\documentclass[epj]{svjour}
\usepackage{graphicx}
\usepackage{color}

\def\mb#1{\setbox0=\hbox{$#1$}\kern-.025em\copy0\kern-\wd0
\kern-0.05em\copy0\kern-\wd0\kern-.025em\raise.0233em\box0}

\begin{document}
   \title{The Brownian Mean Field model}

 \author{Pierre-Henri Chavanis}

\institute{Laboratoire de Physique Th\'eorique (IRSAMC), CNRS and UPS, Universit\'e de Toulouse, F-31062 Toulouse, France\\
\email{chavanis@irsamc.ups-tlse.fr}
}

\titlerunning{The Brownian Mean Field model}

   \date{To be included later }

   \abstract{We discuss the dynamics and thermodynamics of the Brownian Mean
Field (BMF) model which is a system of $N$ Brownian particles moving on a circle
and interacting via a cosine potential. It can be viewed as the canonical
version of the Hamiltonian Mean Field (HMF) model. The BMF model
displays a second order phase transition between a homogeneous phase and an
inhomogeneous phase below a critical temperature $T_c=1/2$. We first complete
the description of this model in the mean field approximation valid for
$N\rightarrow +\infty$. In the strong friction limit, the evolution of the
density towards the mean field Boltzmann distribution is governed by the mean
field Smoluchowski equation. For $T<T_c$, this equation describes a process of
self-organization from a non-magnetized (homogeneous) phase to a magnetized
(inhomogeneous) phase. We obtain an analytical expression for the temporal
evolution of the magnetization close to $T_c$.  Then, we take fluctuations
(finite $N$ effects) into account. The evolution of the density is governed by
the stochastic Smoluchowski equation. From this equation, we derive a stochastic
equation for the magnetization and study its properties both in the homogenous
and inhomogeneous phases. We show that the fluctuations diverge close to the
critical point so that the mean field approximation ceases to be valid.
Actually, the limits $N\rightarrow +\infty$ and $T\rightarrow T_c$ do not
commute. The validity of the mean field approximation requires
$N(T-T_c)\rightarrow +\infty$ so that $N$ must be larger and larger as $T$
approaches $T_c$. We show that the direction of the magnetization changes
rapidly close to $T_c$ but its amplitude takes a long time to relax. We also
indicate  that, for systems with long-range interactions, the lifetime of
metastable states scales as $e^N$ except close to a critical point. The BMF
model shares many analogies with other systems of Brownian particles with long
range interactions such as self-gravitating Brownian particles, the Keller-Segel
model describing the chemotaxis of bacterial populations, and the Kuramoto model
describing the collective synchronization of coupled oscillators.
 }

   \maketitle
%

\section{Introduction}
\label{sec_model}

In the recent years, the  statistical mechanics of systems with long-range
interactions has been a topic of active research
\cite{houches,assisebook,oxford,cdr}. In most papers devoted to this subject,
one assumes that the system is isolated. This corresponds to the {\it
microcanonical ensemble} in which the energy is conserved. This is the proper
description of self-gravitating systems such as galaxies and globular clusters
in astrophysics. Indeed, they can be viewed as isolated Hamiltonian systems of
$N$ point mass stars in gravitational interaction described by the Newton
equations \cite{bt}. This is also the correct approach to the point vortex gas 
in 2D hydrodynamics  described by the Kirchhoff equations \cite{newton}.
Actually, the statistical mechanics of stellar systems and two-dimensional
vortices share many analogies \cite{chavhouches}. Systems with long-range
interactions display very 
interesting properties such as ensemble inequivalence, negative
specific heats, spatially inhomogeneous equilibrium states, violent
collisionless relaxation, slow collisional relaxation, non-Boltzmannian
quasistationary states (QSS) etc. The dynamics and thermodynamics of these
systems is now relatively well-understood \cite{cdr} even if some conceptual
issues remain such as the precise nature of the QSSs.

However, in many situations of physical interest, the system is not
isolated from the surrounding and it is important to take into account
its interaction with the external medium. This interaction usually
results in some effects of forcing and dissipation. In the simplest
situation, the one that we shall consider here, the forcing and the
dissipation satisfy a detailed balance condition so that, formally, the
system can be thought to be in contact with a thermal bath fixing its
temperature $T$. We stress that the thermostat is played by a system
of another nature (physically different from the system under
consideration) which usually has short-range
interactions\footnote{Indeed, it is not possible to define the notion
of thermostat for a purely long-range system (i.e. to divide the
system into a subsystem $+$ a reservoir) since the energy is
non-additive \cite{cdr}.}. To be specific, let us consider a
particular example issued from astrophysics. In the context of
planet formation, one has to study  the motion of dust
particles in gravitational interaction evolving in a gas (the solar
nebula) \cite{aa}. In addition to
the long-range gravitational interaction, the dust particles
experience a friction with the gas and a stochastic force (noise) due
to turbulence or Brownian motion (i.e. short-range collisions with the
molecules of the gas). This situation can be described by $N$
stochastic Langevin equations, one for each particle, coupled together
by the gravitational interaction. This defines the self-gravitating
Brownian model. If we assume a detailed balance condition, the diffusion
coefficient $D$ and the friction coefficient $\xi$ satisfy the Einstein
relation $D=\xi k_B T/m$ where $T$ is the temperature of the bath. In
that case, the proper statistical ensemble is the {\it canonical
ensemble}. The self-gravitating Brownian model has been studied in a
series of papers by Chavanis and Sire (see, e.g.,
\cite{virial1} and references therein) in the strong friction limit
$\xi\rightarrow +\infty$ in which the motion of the particles is
overdamped. Some interesting analogies with the chemotaxis of
bacterial populations, the so-called Keller-Segel model \cite{ks},
have been developed in these papers. Indeed, bacterial
populations may be considered as a system of Brownian particles with long-range
interactions. The bacteria have a diffusive motion (due to their flagella) but
they also secrete a substance (a sort of pheromone) and are collectively
attracted by this substance. Interestingly, it can be shown that the
concentration of the secreted chemical plays the same role as the gravitational
potential. This long-range attraction may result in chemotactic collapse. As a
result, the Smoluchowski-Poisson system and the  Keller-Segel models are
isomorphic \cite{spks}. Furthermore, in biology, the overdamped limit is
justified because inertial effects are generally negligible.

In order to study systems with long-range interactions in a simple setting, toy models have been introduced in statistical mechanics. In particular, the Hamiltonian Mean Field (HMF) model has received a particular
attention [12-37]. This model consists in $N$ particles of unit mass moving
on a circle and interacting via a cosine potential $u=N^{-1}[1-
\cos(\theta_i-\theta_j)]$ where $\theta_i$ denotes the angle that
particle $i$ makes with an axis of reference. Since the energy is conserved, the
fundamental statistical description of the HMF model is the
microcanonical ensemble.   It can be of interest to consider in parallel
the case of a system in which the particles experience, in addition to the cosine interaction,
a friction force and a stochastic force. In that case, their dynamics is
described by coupled Langevin equations. Like for the model of self-gravitating
Brownian particles, we assume that a detailed balance condition holds. We thus
consider a system of Brownian particles with cosine interaction in contact with
a thermal bath. This is the so-called Brownian Mean Field (BMF) model
\cite{cvb,cbo}. Since the temperature is fixed, the fundamental statistical
description of the BMF model is the canonical ensemble. It has been demonstrated
in \cite{bco} that Hamiltonian reservoirs microscopically coupled with the
system \cite{bo1,bo2} and Langevin thermostats \cite{cvb} provide 
equivalent descriptions even out-of-equilibrium. Therefore,
the BMF model has many applications. It is also connected to the Kuramoto model
\cite{kuramoto} describing the collective synchronization in spatially extended
systems of coupled oscillators\footnote{The Kuramoto model is,
however, more complicated since the oscillators usually have different
frequencies.} A generalization of the BMF model, called the $\alpha$-BMF
model, has been considered recently \cite{cr}.

The Brownian Mean Field model was introduced and studied in \cite{cvb}. However,
in this work, the effect of fluctuations was neglected and a mean field
approximation was considered. For systems with long-range interactions the mean
field approximation is usually exact in the thermodynamic limit $N\rightarrow
+\infty$. However, this is no longer true in the vicinity of a critical point.
In that case, the limits $N\rightarrow +\infty$ and $T\rightarrow T_c$ do not
commute and the effect of fluctuations must be properly taken into account. The
objective of the present paper is to go beyond the mean field approximation
considered in \cite{cvb} and study the effect of fluctuations. We shall be
particularly interested in the behavior of the magnetization close to the
critical point. We show that it is described by the usual
phenomenology of second order phase transitions. However, an interest of the BMF
model is that  we can {\it derive} the stochastic equation for the magnetization
${\bf M}(t)$ directly from the $N$-body dynamics by using the stochastic
Smoluchowski equation. As a result, the number of particles $N$ explicitly
enters in the
equations and leads to novel effects such as the non-commutation 
of the limits $N\rightarrow +\infty$ and $T\rightarrow T_c$, and the fact that
the metastable states have very long lifetimes scaling as $e^N$ (except close to
a critical point).

The paper is organized as follows. In Section \ref{sec_inertial} we introduce
the BMF model and present the basic equations. In Section \ref{sec_obmf}, to
simplify the study, we consider the strong friction limit $\xi\rightarrow
+\infty$ in which the inertia of the particles can be neglected. In Section
\ref{sec_mepeq}, we determine the statistical equilibrium state of the BMF model
in the canonical ensemble. It displays a second order phase transition at the
critical temperature $T_c=1/2$.  We solve the thermodynamical stability problem
by different methods and study the equilibrium fluctuations of the
magnetization. We show that they diverge at the critical point.
We also investigate the effect of an external magnetic field and show that the
magnetic susceptibility also diverges at the critical point. In Section
\ref{sec_smoljeans}, we study the dynamical stability of a steady state of the
mean field Smoluchowski equation and show the
equivalence between dynamical and thermodynamical stability. In Section
\ref{sec_linearresponse}, we apply the linear response theory to the BMF model
and study the response of the system to an external perturbation such as a pulse
or a step function. In Section \ref{sec_mmip}, we study the evolution of the
mean magnetization in the inhomogeneous phase and solve the equations
perturbatively close to the critical point. We obtain an
analytical expression for the temporal evolution of the magnetization close to
$T_c$. In Section \ref{sec_scse}, we study the temporal correlations of the
magnetization in the homogeneous phase. We show that they diverge as we approach
the critical point implying  that the mean field approximation ceases to be
valid close to the critical point and that the instability occurs sooner than
predicted by the linear stability analysis. We also explicitly check in this
particular situation the fluctuation-dissipation theorem. In Section
\ref{sec_fmhp} we study the fluctuations of the magnetization in the homogeneous
phase and show that they can be described by an Ornstein-Uhlenbeck process. The
evolution of the probability density of the magnetization is described by a
linear Fokker-Planck equation analogous to the Kramers equation. Finally, in
Section \ref{sec_semip}, we study the fluctuations of the magnetization in the
inhomogeneous phase and solve the problem perturbatively close to the critical
point. As $T\rightarrow T_c$, we show that the direction of the magnetization
changes rapidly while its magnitude takes a long time to relax towards its
equilibrium value.

\section{The inertial BMF model}
\label{sec_inertial}

\subsection{The Langevin equations}
\label{sec_lang}

The BMF model is a system of $N$ Brownian particles of unit mass
moving on a circle and interacting via a cosine binary potential
\cite{cvb}. The dynamics of these particles is governed by the coupled
stochastic Langevin equations
\begin{eqnarray}
{d\theta_{i}\over dt}=v_{i}, \qquad\qquad\qquad\qquad\qquad\qquad\qquad\quad\nonumber\\
{dv_{i}\over dt}=-{\partial\over\partial \theta_{i}}U(\theta_{1},...,\theta_{N})-\xi v_{i}+\sqrt{2D}R_{i}(t),
\label{bkram1}
\end{eqnarray}
where $i=1,...,N$ label the particles. The particles interact
through the potential
$U(\theta_{1},...,\theta_{N})=\frac{1}{N}\sum_{i<j}u(\theta_{i}-\theta_{j})$
where
\begin{equation}
\label{u}
u(\theta-\theta')=1-\cos(\theta-\theta'),
\end{equation}
is the cosine potential. This potential is attractive and the
particles tend to group themselves in order to decrease their potential energy.
This tendency is of course counter-balanced by thermal motion. The Hamiltonian
is $H=\sum_{i=1}^{N}{v_{i}^{2}\over 2}+U(\theta_{1},...,\theta_{N})$. We have
rescaled the potential energy by $1/N$ to make the system extensive. This
corresponds to the Kac prescription \cite{kac}. We note, however, that the
energy remains fundamentally non-additive \cite{cdr}.  $R_i(t)$ is a Gaussian
white noise satisfying $\langle R_i(t)\rangle=0$ and $\langle
R_i(t)R_j(t')\rangle=\delta_{ij}\delta(t-t')$. $D$ and $\xi$ are respectively
the diffusion and friction
coefficients. The former measures the strength of the noise,
whereas the latter quantifies the dissipation to
the external environment. We assume that these two effects
have the same physical origin, like when the system interacts with a
heat bath. In particular, we suppose that the temperature $T$ of the
bath satisfies the Einstein relation $D=\xi T$. The temperature measures the strength of the stochastic force for a given friction coefficient. For $\xi=D=0$, we recover the HMF model \cite{ar} which conserves the energy $H$.

To monitor the evolution of the system, it is convenient to introduce the magnetization ${\bf M}=(M_x,M_y)$ with components
\begin{eqnarray}
M_x=\frac{1}{N}\sum_{i=1}^N \cos\theta_i,\qquad M_y=\frac{1}{N}\sum_{i=1}^N \sin\theta_i.
\label{smf2}
\end{eqnarray}
The magnetization  can serve as an order parameter in the BMF model. In terms of
the magnetization, the potential 
energy is exactly given by $U=N(1-M^2)/2$. The force acting on
particle $i$ is $F_{i}=-{\partial U\over\partial\theta_{i}}$. Using Eqs.
(\ref{u}) and (\ref{smf2}) it can be written as
$F_i=-\frac{1}{N}
\sum_j\sin(\theta_i-\theta_j)=-M_x\sin\theta_i+M_y\cos\theta_i$.

\subsection{The $N$-body Kramers equation}
\label{sec_bkram}

The evolution of the $N$-body distribution function is governed by the Fokker-Planck  equation \cite{hb2}:
\begin{eqnarray}
{\partial P_{N}\over\partial t}+\sum_{i=1}^{N}\biggl (v_{i}{\partial P_{N}\over\partial\theta_{i}}+F_{i}{\partial P_{N}\over\partial v_{i}}\biggr )=\nonumber\\
\sum_{i=1}^{N}{\partial\over\partial v_{i}}\biggl (D{\partial P_{N}\over\partial
v_{i}}+ \xi P_{N}v_{i}\biggr ).
\label{bkram5}
\end{eqnarray}
This is the so-called $N$-body Kramers equation. In the absence of forcing and
dissipation ($\xi=D=0$), it reduces to the Liouville equation. The  $N$-body
Kramers equation satisfies an $H$-theorem for the free energy
\begin{equation}
\label{bkram5b}  F[P_N]=E[P_N]-TS[P_N],
\end{equation}
where $E[P_N]=\int P_N H \, d\theta_1 dv_1...d\theta_N dv_N$ is the energy and  $S[P_N]=-\int P_N \ln P_N \, d\theta_1 dv_1...d\theta_N dv_N$ is the entropy. A simple calculation gives
\begin{eqnarray}
\dot F=-\sum_{i=1}^N\int\frac{\xi}{P_N}\left (T\frac{\partial P_N}{\partial v_i}+P_N v_i\right )^2\, d\theta_1 dv_1...d\theta_N dv_N.\nonumber\\
\label{bkram6}
\end{eqnarray}
Therefore, $\dot F\le 0$ and $\dot F=0$ if, and only, if $P_N$ is the canonical distribution defined by Eq.  (\ref{mcd3}) below. Because of the $H$-theorem, the system converges towards the canonical distribution for $t\rightarrow +\infty$.

\subsection{The canonical distribution}
\label{sec_mcd}

When the system is in contact with a thermal bath, as in the case of the BMF model, the relevant statistical ensemble is the canonical
ensemble. The statistical equilibrium state is described by the canonical distribution
\begin{eqnarray}
P_{N}(\theta_{1},v_{1},...,\theta_{N},v_{N})={1\over Z(\beta)}e^{-\beta H(\theta_{1},v_{1},...,\theta_{N},v_{N})},
\label{mcd3}
\end{eqnarray}
where
\begin{eqnarray}
Z(\beta)=\int e^{-\beta H(\theta_{1},v_{1},...,\theta_{N},v_{N})}\, \prod_{i}d\theta_{i}dv_{i}.
\label{mcd4}
\end{eqnarray}
is the partition function determined by the normalization condition $\int P_N \,
d\theta_1 dv_1....d\theta_N dv_N=1$. The canonical distribution (\ref{mcd3}) is
the steady state of the $N$-body Kramers  
equation (\ref{bkram5}). We note that the velocity distribution
is Gaussian for any $N$.

We define the free energy by $F(T)=-T\ln Z(T)$. We also introduce the Massieu function $J(\beta)=-\beta F(\beta)=\ln Z(\beta)$. In the canonical ensemble, the average energy $E=\langle H\rangle$ is given by $E=\partial (\beta F)/\partial\beta=-\partial J/\partial\beta$. The fluctuations of energy are given by $\langle H^2\rangle-\langle H\rangle^2=T^2 C$ where $C=dE/dT$ is the specific heat. This relation implies that the specific heat is always positive in the canonical ensemble.

We note that the canonical distribution (\ref{mcd3}) is the minimum of $F[P_N]$ respecting the normalization condition. At equilibrium, we get  $F[P_N]=-T\ln Z(T)=F(T)$.

\subsection{The mean field approximation}

In the thermodynamic limit $N\rightarrow +\infty$, we can neglect the correlations between the particles. Therefore, the mean field approximation is exact and the $N$-body distribution function can be factorized in a product of $N$ one-body distribution functions
\begin{equation}
\label{mfa1}
P_N(\theta_1,v_1,...,\theta_N,v_N,t)=\prod_{i=1}^{N}P_1(\theta_i,v_i,t).
\end{equation}
We also have
\begin{eqnarray}
P_N(\theta_1,...,\theta_N,t)=\prod_{i=1}^{N} P_1(\theta_i,t).\label{smf9b}
\end{eqnarray}
We introduce the distribution function $f(\theta,v,t)=P_1(\theta,v,t)$ and the spatial density $\rho(\theta,t)=\int f\, dv=P_1(\theta,t)$. The mean field energy per particle  is given by
\begin{eqnarray}
E=\frac{1}{2}\int f v^2\, d\theta dv+\frac{1}{2}\int \rho\Phi\, d\theta,
\label{mfe1}
\end{eqnarray}
where $\rho(\theta,t)=\int f\, dv$ is the spatial density and
\begin{eqnarray}
\Phi(\theta,t)=\int u(\theta-\theta')\rho(\theta',t)\, d\theta,
\label{smf10}
\end{eqnarray}
is the mean potential. Expanding the cosine function in Eq. (\ref{u}), the mean potential may be written as
\begin{equation}
\label{smf10b}
\Phi(\theta,t)=1-M_x(t)\cos\theta-M_y(t)\sin\theta,
\end{equation}
where
\begin{equation}
\label{magn}
M_x=\int\rho\cos\theta\, d\theta,\qquad M_y=\int\rho\sin\theta\, d\theta,
\end{equation}
are the components of the mean magnetization. In terms of the magnetization the mean field potential energy is given by
\begin{eqnarray}
W=\frac{1}{2}\int \rho\Phi\, d\theta=\frac{1-M^2}{2}.
\label{mfe1w}
\end{eqnarray}
We also note that, in the mean field approximation, the entropy per particle is
\begin{eqnarray}
S=-\int f\ln f\, d\theta dv.
\label{mfe1s}
\end{eqnarray}

\subsection{The equilibrium distribution of the smooth density
and the most probable macrostate}
\label{sec_canof}

We wish to determine the equilibrium distribution of the smooth density $f(\theta,v)$ in phase space. A {\it microstate} is defined by the specification of the
exact positions and velocities $\lbrace \theta_i,v_i\rbrace$ of the $N$ particles. A
{\it macrostate} is defined by  the specification of the (coarse-grained) density
$f(\theta,v)$ of particles in each cell $[\theta,\theta+d\theta]\times
[v,v+dv]$ irrespectively of their  precise position in the
cell.  Let us call $\Omega[f]$ the unconditional number of  microstates $\lbrace \theta_i,v_i\rbrace$
corresponding to the macrostate $f$. The entropy per particle of the macrostate $f(\theta,v)$ is defined  by the Boltzmann formula $S[f]=\frac{1}{N}\ln \Omega[f]$. The unconditional probability density of the distribution $f(\theta,v)$ is therefore $P_0[f]\propto \Omega[f]\propto e^{N S[f]}$. The number of complexions  $\Omega[f]$ can be obtained by a standard combinatorial analysis. For $N\gg 1$, we find that the Boltzmann entropy is given by $S[f]=-\int f \ln f\, d\theta dv$.

To evaluate the partition function (\ref{mcd4}), instead of integrating over the microstates $\lbrace \theta_1,v_1,...,\theta_N,v_N\rbrace$, we can integrate over the macrostates $f(\theta,v)$. Introducing  the unconditional number of microstates $\Omega[f]$ corresponding to the macrostate $f$, and the mean field energy per particle $E[f]$ of the macrostate  $f$, we obtain for $N\gg 1$:
\begin{eqnarray}
Z(\beta) \simeq  \int e^{-N\beta E[f]} \, \Omega[f] \, \delta(I[f]-1)\, {\cal D}f\nonumber\\
\simeq \int  e^{N S[f]-N\beta E[f]} \, \delta(I[f]-1)\, {\cal D}f\nonumber\\
\simeq \int  e^{- N\beta F[f]} \, \delta(I[f]-1)\, {\cal D}f,
\label{c4}
\end{eqnarray}
where $F[f]=E[f]-TS[f]$ is the free energy defined by Eq. (\ref{ffg}), and $I[f]$ is the normalization condition defined by Eq. (\ref{norm}).  The canonical density probability of the distribution $f$ is therefore
\begin{eqnarray}
P[f]=\frac{1}{Z(\beta)}e^{-N\beta F[f]}\delta(I[f]-1).
\label{c5}
\end{eqnarray}
This distribution can be directly obtained by stating that $P[f]\propto W[f] e^{-N\beta  E[f]}\delta(I[f]-1)$ since the microstates with energy $E$ have a probability $\propto e^{-\beta E}$.

For $N\rightarrow +\infty$, we can make the saddle point approximation. We obtain
\begin{eqnarray}
Z(\beta)=e^{-\beta F(\beta)}\simeq e^{-N\beta F[f_*]},
\label{nbh7}
\end{eqnarray}
i.e.
\begin{eqnarray}
\lim_{N\rightarrow +\infty} \frac{1}{N}F(\beta)=F[f_*],
\label{nbh8}
\end{eqnarray}
where $f_*$ is the global minimum of free energy $F[f]$ respecting the normalization condition. This is the most probable macrostate in the canonical ensemble. We are led therefore to solving the minimization problem defined by Eq. (\ref{nff1}). This is a result of large deviations. The critical points of this variational problem are the mean field Maxwell-Boltzmann distributions (\ref{mba1}). They can also be obtained from the canonical distribution (\ref{mcd3}) by writing the first  equation of the Yvon-Born-Green (YBG) hierarchy and using the mean field approximation (\ref{mfa1})(see \cite{hb1,longshort}).

\subsection{The mean field Kramers equation}

In the thermodynamic limit $N\rightarrow +\infty$, the
$N$-body distribution function is a product of $N$ one-body
distribution functions given by Eq. (\ref{mfa1}). Substituting this
factorization in Eq. (\ref{bkram5}) and integrating over $N-1$ variables we find
that the evolution of the distribution function $f(\theta,v,t)$
is governed by the mean field  Kramers equation \cite{hb2}:
\begin{equation}
\label{browbbgky6}
{\partial f\over\partial t}+v{\partial f\over\partial\theta}-{\partial\Phi\over\partial \theta}{\partial f\over\partial v}={\partial \over\partial v}\biggl (D{\partial f\over\partial v}+\xi fv\biggr ),
\end{equation}
where $\Phi(\theta,t)$ is given by Eq. (\ref{smf10}). For $\xi=D=0$, Eq.
(\ref{browbbgky6}) reduces to the Vlasov equation which describes the
collisionless evolution of the HMF model.

Using the Einstein relation, the mean field Kramers equation (\ref{browbbgky6}) may  be rewritten as
\begin{equation}
\frac{\partial f}{\partial t}
+v\frac{\partial f}{\partial\theta}
-\frac{\partial\Phi}{\partial\theta}
\frac{\partial f}{\partial v}
=\xi
\frac{\partial}{\partial v}\left ({T}\frac{\partial f}{\partial v}+f v\right ).
\label{browbbgky7}
\end{equation}
The mean field Kramers equation satisfies an $H$-theorem for the free energy $F[f]$ defined by Eq. (\ref{ffg}) below. Its expression can be obtained from Eq. (\ref{bkram5b}) by using the mean field approximation (\ref{mfa1}). In terms of the free energy, the mean field Kramers equation may be written as a gradient flow
\begin{equation}
\frac{\partial f}{\partial t}
+v\frac{\partial f}{\partial\theta}
-\frac{\partial\Phi}{\partial\theta}
\frac{\partial f}{\partial v}
=\xi\frac{\partial}{\partial v}\left \lbrack f\frac{\partial}{\partial v}\left (\frac{\delta F}{\delta f}\right )\right\rbrack.
\label{browbbgky7b}
\end{equation}
A simple calculation gives
\begin{equation}
{\dot F}=-\xi\int f\left\lbrack\frac{\partial}{\partial v}\left (\frac{\delta F}{\partial f}\right )\right\rbrack^2\, d\theta dv,
\label{browbbgky7bb}
\end{equation}
or equivalently
\begin{equation}
{\dot F}=-\xi\int\frac{1}{f}\left (T\frac{\partial f}{\partial v}+fv\right )^2\, d{\theta}dv.
\label{browbbgky7bbz}
\end{equation}
Therefore, $\dot F\le 0$ and $\dot F=0$ if, and only if, $f$ is the mean field
Maxwell-Boltzmann distribution defined by Eq.  (\ref{mba1}) below with the
temperature of the bath $T$. Because of the $H$-theorem, the system converges,
for $t\rightarrow +\infty$, towards a mean-field Maxwell-Boltzmann distribution
that is a (local) minimum of free energy respecting the normalization condition.
If several minima exist at the same temperature, the selection depends on a
notion of basin of attraction.  
The relaxation time is $t_{B}\sim 1/\xi$.

\section{The overdamped BMF model}
\label{sec_obmf}

\subsection{The Langevin equations}
\label{sec_overlang}

The inertial BMF model has been studied in \cite{cbo,bco}. Here, to simplify the problem, we consider the strong friction
limit $\xi\rightarrow +\infty$ in which  the inertia of the
particles can be neglected. This corresponds to the overdamped BMF model. The stochastic Langevin equations (\ref{bkram1}) reduce to
\begin{eqnarray}
{d\theta_{i}\over dt}=-\mu{\partial\over\partial \theta_{i}}U(\theta_{1},...,\theta_{N})+\sqrt{2D_{*}}R_{i}(t),
\label{smf1}
\end{eqnarray}
where $\mu=1/\xi$ is the mobility and $D_{*}=D/\xi^{2}$ is the
diffusion coefficient in physical space. The Einstein relation may be rewritten as $D_{*}={T}/{\xi}=\mu T$. The temperature measures the strength of the stochastic force (for a given mobility).

\subsection{The $N$-body Smoluchowski equation}
\label{sec_smf}

The evolution of the
$N$-body distribution function $P_N(\theta_1,...,\theta_N,t)$ is governed by the $N$-body
Fokker-Planck equation \cite{hb2}:
\begin{eqnarray}
{\partial P_{N}\over\partial t}=\sum_{i=1}^{N}{\partial\over\partial \theta_{i}}\biggl\lbrack D_{*}{\partial P_{N}\over\partial\theta_{i}}+\mu P_{N}{\partial\over\partial \theta_{i}}U(\theta_{1},...,\theta_{N})\biggr\rbrack.\qquad
\label{smf3}
\end{eqnarray}
This is the so-called $N$-body Smoluchowski equation. It can be derived directly from the stochastic equations (\ref{smf1}). Alternatively, it can be obtained from the $N$-body Kramers equation (\ref{bkram5}) in the strong friction limit $\xi\rightarrow +\infty$ \cite{risken}. In this limit, using the Einstein relation, we find that
\begin{eqnarray}
\label{smf4}
P_N(\theta_1,v_1,...,\theta_N,v_N,t)=\left (\frac{\beta}{2\pi}\right )^{N/2} P_N(\theta_1,...,\theta_N,t) \nonumber\\
\times e^{-\beta \sum_{i=1}^{N}\frac{v_i^{2}}{2}}+O(\xi^{-1}),\qquad\qquad
\end{eqnarray}
where the evolution of $P_N(\theta_1,...,\theta_N,t)$ is governed by Eq. (\ref{smf3}).
The $N$-body Smoluchowski equation satisfies an H-theorem for the free energy
\begin{eqnarray}
\label{smf5}
F[P_N]=\int P_N U\, d\theta_1...d\theta_N+T\int P_N\ln P_N\, d\theta_1...d\theta_N\nonumber\\
-\frac{N}{2}T\ln T-\frac{N}{2}T\ln(2\pi).\qquad
\end{eqnarray}
The expression (\ref{smf5}) can be obtained from the free energy (\ref{bkram5b})  by using Eq. (\ref{smf4}). A simple calculation gives
\begin{eqnarray}
\label{smf6}
\dot F=-\sum_{i=1}^{N}\int \frac{1}{\mu P_N}\left (D_{*}{\partial P_{N}\over\partial\theta_{i}}+\mu P_{N}{\partial U\over\partial \theta_{i}}\right )^2\, d\theta_1...d\theta_N.\nonumber\\
\end{eqnarray}
Therefore, $\dot F\le 0$ and $\dot F=0$ if, and only, if $P_N$ is the canonical distribution in physical space defined by Eq. (\ref{smf7}) below. Because of the $H$-theorem, the system converges towards the canonical distribution (\ref{smf7}) for $t\rightarrow +\infty$.

We note that the free energy may be written as $F[P_N]=E[P_N]-T
S[P_N]$,
where
\begin{eqnarray}
\label{smf5b}
E[P_N]=\frac{1}{2}N T+\int P_N U\, d\theta_1...d\theta_N,
\end{eqnarray}
\begin{equation}
\label{smf5c}
S[P_N]=-\int P_N\ln P_N\, d\theta_1...d\theta_N
+\frac{1}{2}N \ln\left ({2\pi  T}\right )+\frac{1}{2}N
\end{equation}
are the energy and  the entropy.

\subsection{The canonical distribution}
\label{sec_mcdb}

The statistical equilibrium state in configuration space is described by the canonical distribution
\begin{eqnarray}
P_{N}(\theta_{1},...,\theta_{N})={1\over Z_{conf}(\beta)}e^{-\beta U(\theta_{1},...,\theta_{N})},
\label{smf7}
\end{eqnarray}
where
\begin{eqnarray}
Z_{conf}(\beta)=\int e^{-\beta U(\theta_{1},...,\theta_{N})}\, \prod_{i}d\theta_{i},
\label{mcd4b}
\end{eqnarray}
is the configurational partition function determined by the normalization condition $\int P_N \, d\theta_1....d\theta_N=1$. The canonical distribution (\ref{smf7})  is the steady state of the $N$-body Smoluchowski equation (\ref{smf3}). It can also be obtained from Eq. (\ref{mcd3}) by integrating over the velocity. We then find that $Z(\beta)=Z_{conf}(\beta)(2\pi/\beta)^{N/2}$.

We note that the   canonical distribution (\ref{smf7}) is the minimum of $F[P_N]$ respecting the normalization condition. At equilibrium, we get  $F[P_N]=-T\ln Z_{conf}(T)-\frac{N}{2}T\ln(2\pi T)=-T\ln Z(T)=F(T)$.

\subsection{The distribution of the smooth density and the most
probable macrostate}
\label{sec_canorho}

We wish to determine the equilibrium distribution of the smooth density $\rho(\theta)$ in position space. A {\it microstate} is defined by the specification of the exact positions $\lbrace \theta_i\rbrace$ of the $N$ particles. A
{\it macrostate} is defined by  the specification of the (coarse-grained) density
$\rho(\theta)$ of particles in each cell $[\theta,\theta+d\theta]$ irrespectively of their  precise position in the cell.  Let us call $\Omega[\rho]$ the unconditional number of  microstates $\lbrace \theta_i\rbrace$ corresponding to the macrostate $\rho(\theta)$. The unconditional entropy per particle of the macrostate  $\rho(\theta)$  is defined  by the
Boltzmann formula $S_0[\rho]=\frac{1}{N}\ln \Omega[\rho]$.
The unconditional probability density of the density $\rho(\theta)$ is therefore $P_0[\rho]\propto \Omega[\rho]\propto e^{N S_0[\rho]}$. The number of complexions  $\Omega[\rho]$ can be obtained by a standard combinatorial analysis. For $N\gg 1$, we find that the Boltzmann entropy per particle is given by $S_0[\rho]=-\int {\rho}\ln\rho\, d\theta$.

To evaluate the partition function  $Z(\beta)=Z_{conf}(\beta)(2\pi/\beta)^{N/2}$ with Eq. (\ref{mcd4b}), instead of integrating over the microstates $\lbrace \theta_1,...,\theta_N\rbrace$, we can integrate over the macrostates $\rho(\theta)$.  Introducing  the unconditional number of microstates $\Omega[\rho]$ corresponding to the macrostate $\rho$ and the mean field potential energy per particle  $W[\rho]$ of the macrostate  $\rho$, we obtain for $N\gg 1$:
\begin{eqnarray}
Z(\beta)\simeq e^{\frac{N}{2}\ln \left (\frac{2\pi}{\beta} \right )}\int e^{-N\beta  W[\rho]}\Omega[\rho]\, \delta(I[\rho]-1)\, {\cal D}\rho\nonumber\\
 \simeq e^{\frac{N}{2}\ln \left (\frac{2\pi}{\beta} \right )}\int e^{N S_0[\rho]-N \beta W[\rho]}\, \delta(I[\rho]-1)\, {\cal D}\rho\nonumber\\
\simeq \int e^{-N \beta F[\rho]} \, \delta(I[\rho]-1) \, {\cal D}\rho,\qquad
\label{h3}
\end{eqnarray}
where the free energy per particle $F[\rho]$ is given by Eq. (\ref{tsce1zero}). The canonical  probability density of the distribution $\rho$ is therefore
\begin{eqnarray}
P[\rho]=\frac{1}{Z(\beta)}e^{-N \beta F[\rho]}\delta(I[\rho]-1).
\label{h5}
\end{eqnarray}

For $N\rightarrow +\infty$, we can make the saddle point approximation. We obtain
\begin{eqnarray}
Z(\beta)=e^{-\beta F(\beta)}\simeq e^{-N\beta F[\rho_*]},
\label{h7}
\end{eqnarray}
i.e.
\begin{eqnarray}
\lim_{N\rightarrow +\infty} \frac{1}{N}F(\beta)=F[\rho_*],
\label{h8}
\end{eqnarray}
where $\rho_*$ is the global minimum of free energy $F[\rho]$ respecting the normalization condition. This is the most probable macrostate in the canonical ensemble. We are led therefore to solving the minimization problem defined by Eq. (\ref{ff1}). This is a result of large deviations. The critical points of this variational problem are the mean field Boltzmann distributions (\ref{cp4}). They can also be obtained from the canonical distribution (\ref{smf7}) by writing the first  equation of the Yvon-Born-Green (YBG) hierarchy and using the mean field approximation (\ref{smf9b})(see \cite{hb1,longshort}).

\subsection{The mean field Smoluchowski equation}
\label{sec_mfse}

In the thermodynamic limit $N\rightarrow +\infty$, the
$N$-body distribution function is a product of $N$ one-body
distribution functions given by Eq. (\ref{smf9b}). Substituting this
factorization in Eq. (\ref{smf3}) and integrating over $N-1$ angular variables
we find that the evolution of the density $\rho(\theta,t)$ is
governed by the mean field Smoluchowski equation \cite{hb2}:
\begin{eqnarray}
{\partial\rho\over\partial
t}={\partial\over\partial\theta}\biggl\lbrack {1\over\xi}\biggl
(T{\partial \rho\over\partial
\theta}+\rho{\partial\Phi\over\partial\theta}\biggr
)\biggr\rbrack, \label{smf10g}
\end{eqnarray}
where $\Phi(\theta,t)$ is given by Eq. (\ref{smf10}). The mean field
Smoluchowski equation (\ref{smf10g}) can also be obtained from the
mean field Kramers equation (\ref{browbbgky7}) by using an
expansion in power of $1/\xi$ when $\xi\rightarrow +\infty$ \cite{risken}.
In that limit, the distribution function is
close to the Maxwellian
\begin{equation}
\label{smf11}
f(\theta,v,t)={1\over \sqrt{2\pi T}}\rho(\theta,t) e^{-{v^{2}\over 2T}}+O(\xi^{-1}),
\end{equation}
with the temperature of the bath, and the evolution of the density is governed by Eq. (\ref{smf10g}).

The mean field cosine Smoluchowski equation (\ref{smf10g}) may be written in the form of an integro-differential equation
\begin{eqnarray}
\xi{\partial\rho\over\partial t}=T{\partial^{2}\rho\over\partial\theta^{2}}+{\partial\over\partial\theta}\biggl\lbrace \rho\int_{0}^{2\pi}\sin(\theta-\theta')\rho(\theta',t)d\theta'\biggr\rbrace.
\label{dyn1}
\end{eqnarray}
It may also be written as
\begin{eqnarray}
\xi{\partial\rho\over\partial t}=T{\partial^{2}\rho\over\partial\theta^{2}}+{\partial\over\partial\theta}\biggl\lbrack \rho M(t) \sin(\theta-\phi(t))\biggr\rbrack,
\label{dyn1b}
\end{eqnarray}
where $M$ is the  modulus of the magnetization  and $\phi$ is its phase so that ${\bf M}=Me^{i\phi}$.

The mean field Smoluchowski equation satisfies an $H$-theorem for the free
energy $F[\rho]$ defined by Eq. (\ref{tsce1zero}) below. Its expression can be
obtained from Eq. (\ref{smf5}) by using the mean field approximation
(\ref{smf9b}). It can also be 
obtained from Eq. (\ref{ffg}) by using Eq. (\ref{smf11}). It can
be written as $F[\rho]=E[\rho]-T S[\rho]$, where
\begin{eqnarray}
\label{h4b}
E[\rho]=\frac{1}{2}T+\frac{1}{2}\int\rho\Phi\, d\theta,
\end{eqnarray}
\begin{eqnarray}
\label{h4c}
S[\rho]=-\int {\rho}\ln\rho\, d\theta+\frac{1}{2}\ln\left ({2\pi T}\right )+\frac{1}{2}
\end{eqnarray}
are the energy and the entropy. In terms of the free energy, the mean field
Smoluchowski equation may be written as a gradient flow
\begin{eqnarray}
\xi{\partial\rho\over\partial t}=\frac{\partial}{\partial\theta}\left\lbrack\rho\frac{\partial}{\partial\theta}\left (\frac{\delta F}{\delta\rho}\right )\right\rbrack.
\label{gflow}
\end{eqnarray}
A simple calculation gives
\begin{equation}
\label{smf13}
\dot F=-\int\frac{\rho}{\xi}\left\lbrack\frac{\partial}{\partial\theta}\left (\frac{\delta F}{\partial\rho}\right )\right\rbrack^2\, d\theta,
\end{equation}
or equivalently
\begin{equation}
\label{smf13b}
\dot F=-\int \frac{1}{\xi \rho}\left (T{\partial \rho\over\partial\theta}+ \rho {\partial \Phi\over\partial \theta}\right )^2\, d\theta.
\end{equation}
Therefore, $\dot F\le 0$ and $\dot F=0$ if, and only if, $\rho$ is the mean field
Boltzmann distribution defined by Eq.  (\ref{cp4}) below with the temperature of the bath $T$. Because of the $H$-theorem, the system converges, for $t\rightarrow +\infty$, towards a mean-field Boltzmann distribution that is a (local) minimum of free energy respecting the normalization condition\footnote{The steady states of the mean field  Smoluchowski equation are the critical points  (minima, maxima, saddle points) of the free energy $F[\rho]$ respecting the normalization condition. It can be shown \cite{nfp} that a critical point of free energy is dynamically stable with respect to the mean field Smoluchowski equation if, and only, if it is a (local) minimum. Maxima are unstable for all perturbations so they cannot be reached by the system. Saddle points are unstable only for certain perturbations so they can be reached if the system does not spontaneously generate these dangerous perturbations. The same comments apply to the mean field Kramers equation (\ref{browbbgky6}).}. If several minima exist at the same temperature, the selection depends on a notion of basin of attraction.  The relaxation time is $t_{B}\sim 1/\xi$, independent of $N$.

The mean field Smoluchowski equation (\ref{smf10g}) may also be written as
\begin{eqnarray}
{\partial\rho\over\partial
t}={\partial\over\partial\theta}\biggl\lbrack {1\over\xi}\biggl
({\partial p\over\partial
\theta}+\rho{\partial\Phi\over\partial\theta}\biggr
)\biggr\rbrack, \label{smf10gbis}
\end{eqnarray}
where $p(\theta,t)$ is a pressure related to the density by the isothermal equation of state
\begin{eqnarray}
p(\theta,t)=\rho(\theta,t)T.
\label{piso}
\end{eqnarray}
This equation of state can be obtained from the expression of the
local kinetic pressure $p(\theta,t)=\int f(\theta,v,t)
(v-u(\theta,t))^2\, dv$, where $u(\theta,t)=\frac{1}{\rho}\int f v\,
dv$ is the local velocity, combined with the expression (\ref{smf11})
of the distribution function valid in the strong friction limit (see
\cite{nfp} for a generalization of this result). The steady states of
the mean field Smoluchowski equation satisfy the equation
\begin{eqnarray}
T\frac{d\rho}{d\theta}+\rho\frac{d\Phi}{d\theta}=0,
\label{smf16c}
\end{eqnarray}
which may be interpreted as a condition of hydrostatic equilibrium.

{\it Remark:} at $T=0$, the free energy reduces to the potential energy
$W$ and the $H$-theorem
(\ref{smf13}) becomes
$\dot W=-\int (\rho/\xi)(\partial\Phi/\partial\theta)^2\, d\theta\le 0$. In that case, the system relaxes to the ground state $\rho=\delta(\theta)$, $M=1$, $W=0$ (see Section \ref{sec_mepeq}).

\subsection{The stochastic Smoluchowski equation}

The previous equations, which are based on a mean field approximation, ignore fluctuations. However, fluctuations become important close to a critical point. As we shall see, the BMF model displays a critical temperature $T_c=1/2$. As we approach the critical temperature the mean field approximation becomes less and less accurate (or requires a larger and larger number of particles $N$). As a result, the limits $N\rightarrow +\infty$ and $T\rightarrow T_c$ do not commute.

For Brownian particles with long-range interactions, the fluctuations can be taken into account by adapting the theory of fluctuating hydrodynamics developed by Landau and Lifshitz \cite{ll}. Using this theory, we can derive the stochastic Smoluchowski equation (see Appendix B of \cite{hb5}):
\begin{eqnarray}
\xi{\partial\rho\over\partial t}={\partial\over\partial\theta}\biggl
(T{\partial \rho\over\partial
\theta}+\rho{\partial\Phi\over\partial\theta}\biggr
)+\frac{1}{\sqrt{N}}\frac{\partial}{\partial\theta}\left (\sqrt{2\xi T\rho}R(\theta,t)\right ),\nonumber\\
\label{dyn1ma}
\end{eqnarray}
with Eq. (\ref{smf10}) where $R(\theta,t)$ is a Gaussian white noise such that
$\langle R(\theta,t)\rangle=0$ and $\langle
R(\theta,t)R(\theta',t')\rangle=\delta(\theta-\theta')\delta(t-t')$. This
equation applies to the ``smooth'', but still fluctuating, distribution 
of particles $\rho(\theta,t)$. This is the so-called
``coarse-grained'' density. It is usually denoted $\overline{\rho}(\theta,t)$
but we shall omit the bar to simplify the notations. Eq. (\ref{dyn1ma})  is
physically different, but similar in form, to the exact stochastic equation
derived by Dean \cite{dean} for the discrete distribution of particles 
$\rho_d(\theta,t)=\frac{1}{N}\sum_i \delta(\theta-\theta_i(t))$ which is a sum
of Dirac distributions.  We refer to Appendix B of \cite{longshort} (and
references therein) for more details about the domain of validity of these
different equations.

Introducing the  mean field free energy (\ref{tsce1zero}), the
stochastic Smoluchowski equation can  be rewritten as
\begin{eqnarray}
{\partial\rho\over\partial t}=\frac{1}{\xi}\frac{\partial}{\partial\theta}\left\lbrack \rho\frac{\partial}{\partial\theta}
\left (\frac{\delta F}{\delta\rho}\right )\right\rbrack+\frac{\partial}{\partial\theta}\left(\sqrt{\frac{2T\rho}{\xi}}{R}\right ).
\label{sto6}
\end{eqnarray}
Eq. (\ref{sto6}) may be interpreted as a stochastic Langevin equation for the field $\rho(\theta,t)$. The corresponding Fokker-Planck equation for the probability density $P[\rho,t]$ of the density profile $\rho(\theta,t)$ at time $t$ is
\begin{eqnarray}
\label{sto7}
&&\xi\frac{\partial P}{\partial t}[\rho,t]\nonumber\\
&=&-\int\frac{\delta}{\delta\rho(\theta,t)}\left\lbrace \frac{\partial}{\partial\theta} \rho \frac{\partial}{\partial\theta}\left\lbrack {T}\frac{\delta}{\delta\rho}+\frac{\delta F}{\delta\rho}\right\rbrack P[\rho,t]\right\rbrace\, d\theta.\nonumber\\
\end{eqnarray}
Its stationary solution returns the canonical distribution (\ref{h5}) which
shows 
the consistency of our approach. Actually, the form of the noise in Eq.
(\ref{sto6}) may be determined precisely in order to recover the distribution
(\ref{h5}) at statistical equilibrium. We note that the noise is multiplicative
since it depends on $\rho(\theta,t)$ (it vanishes in regions devoid of
particles).

The fluctuations have several effects. First of all, if the number of
particles is small, the fluctuations must be taken into account in all
cases. On the other hand, if the system displays a critical point, the
fluctuations invalidate the mean field theory close to that critical
point as we have explained previously. Finally, when the free energy
$F[\rho]$ has several minima, the fluctuations allow the system to
jump from one minimum to the other. The timescale of the transition
depends on the height of the barrier of free energy that has to be
crossed. If we consider a very long timescale, the system will explore
the free energy landscape. Of course, it will spend more time in the
global minimum of free energy than in a local one. However, for
long-range interactions, local minima (metastable states) have very
long lifetimes scaling as $e^{N \Delta F/k_B T}\sim e^N$ because the
barrier of free energy is proportional to $N$ \cite{meta,cdnew}.
If we use the mean field Smoluchowski equation (\ref{smf10g}), valid for $N\rightarrow
+\infty$, the system will remain ``blocked'' in a minimum of free
energy even if it is not the global minimum. For finite $N$,
fluctuations taken into account in the stochastic Smoluchowski
equation (\ref{dyn1ma}) can ``un-block'' the system by allowing it to
jump into another minimum. Their effect will be particularly important
close to the critical point where the barrier of free energy per
particle $\Delta F$ is small.

{\it Remark:} the stochastic Smoluchowski equation (\ref{sto6})
is different from the stochastic Ginzburg-Landau equation
\begin{eqnarray}
{\partial\rho\over\partial t}=-\Gamma\frac{\delta F}{\delta\rho}+\sqrt{2\Gamma T}\zeta(\theta,t),
\label{gl}
\end{eqnarray}
where $\zeta(\theta,t)$ is a Gaussian white noise, used to describe 
the time-dependent fluctuations about equilibrium. Eq. (\ref{gl}) is a
phenomenological equation because, in general, it is an impossible task to
derive the true equation for the macroscopic variables directly from the
dynamics of the microscopic variables of the system \cite{goldenfeld}. However,
for Brownian particles with long-range interactions, this task is realizable and
leads to the stochastic Smoluchowski equation (\ref{sto6}) instead of Eq.
(\ref{gl}).

\section{Statistical equilibrium states in the canonical ensemble}
\label{sec_mepeq}

\subsection{The equilibrium distribution}
\label{sec_ed}

In the canonical ensemble, the statistical equilibrium state of the inertial BMF
model is determined 
by the minimization problem (see Sec. \ref{sec_canof} and \cite{hmfarxiv}):
\begin{eqnarray}
\label{nff1}
F(T)=\min_{f}\left\lbrace F\lbrack f\rbrack=E[f]-T S[f]\, |\, I\lbrack f\rbrack=1\right\rbrace,
\end{eqnarray}
where
\begin{eqnarray}
\label{ffg}
F[f]=\frac{1}{2}\int f v^2\, d\theta dv+\frac{1}{2}\int\rho\Phi\, d\theta+T\int f\ln f\, d\theta dv,\nonumber\\
\end{eqnarray}
is the free energy per particle and
\begin{eqnarray}
\label{norm}
I[\rho]=\int\rho \, d\theta=1,
\end{eqnarray}
is the normalization condition. The critical points of this minimization
problem are determined by the variational principle $\delta F+\alpha T\delta
I=0$ where $\alpha$ (chemical potential) is a Lagrange multiplier taking the
normalization condition into account.  Performing the variations, we find that
the critical points are given by the mean field Maxwell-Boltzmann distribution
\begin{equation}
\label{mba1}
f(\theta,v)=A\, e^{-\beta \lbrack \frac{v^2}{2}+\Phi(\theta)\rbrack},
\end{equation}
where $A=e^{-1-\alpha}$ and $\Phi(\theta)$ is the mean potential defined by Eq.
(\ref{smf10}).

To solve the minimization problem (\ref{nff1}), we can proceed in two steps \cite{hmfarxiv}. We first minimize $F[f]$ at fixed normalization {\it and} density $\rho(\theta)$. This gives
\begin{eqnarray}
\label{aace2o}
f(\theta,v)=\left (\frac{\beta}{2\pi}\right )^{1/2}\rho(\theta) e^{-\beta v^2/2}.
\end{eqnarray}
Using Eq. (\ref{aace2o}) we can express the free energy $F[f]$ given by Eq.  (\ref{ffg}) as a functional of the density $\rho$. We get
\begin{eqnarray}
\label{tsce1zero}
F[\rho]={1\over 2}\int\rho\Phi \, d\theta+T\int \rho\ln\rho\, d\theta-\frac{1}{2}T\ln T-\frac{T}{2}\ln(2\pi).\nonumber\\
\end{eqnarray}
Finally, the solution of the minimization problem (\ref{nff1}) is given by Eq. (\ref{aace2o}) where $\rho(\theta)$ is the solution of the minimization problem
\begin{eqnarray}
\label{ff1}
F(T)=\min_{\rho}\left\lbrace F\lbrack \rho\rbrack\, |\, I\lbrack \rho\rbrack=1\right\rbrace.
\end{eqnarray}
It can be shown that the minimization problems (\ref{nff1}) and (\ref{ff1}) are
equivalent for global and local minimization \cite{hmfarxiv}. If we consider the
overdamped BMF model, its statistical equilibrium state is directly determined 
by the minimization problem (\ref{ff1}) (see Sec.
\ref{sec_canorho} and \cite{hmfarxiv}).  The critical points of this
minimization problem are determined by the variational principle $\delta
F+\alpha' T\delta I=0$. Performing the variations, we find that the critical
points are given by the mean field Boltzmann distribution
\begin{eqnarray}
\label{cp4}
\rho(\theta)=A'\, e^{-\beta\Phi(\theta)},
\end{eqnarray}
where $A'=e^{-1-\alpha'}$ and $\Phi(\theta)$ is the mean potential defined by Eq. (\ref{smf10}). This distribution may also be obtained by integrating Eq. (\ref{mba1}) over the velocity. Using the expression (\ref{smf10b}) of the potential, the density (\ref{cp4}) can be rewritten as
\begin{eqnarray}
\label{cp5}
\rho(\theta)=A'\, e^{-\beta\left (1-M_x\cos\theta-M_y\sin\theta\right )}.
\end{eqnarray}
It is convenient to write $M_x=M\cos\phi$ and $M_y=M\sin\phi$ where $M=(M_x^2+M_y^2)^{1/2}$ is the modulus of the magnetization and $\phi$ its phase. In that case, the foregoing expression takes the form
\begin{eqnarray}
\label{cp7}
\rho(\theta)=\frac{1}{2\pi I_{0}(\beta M)}\, e^{\beta M\cos(\theta-\phi)},
\end{eqnarray}
where we have used the normalization condition (\ref{norm}) to determine the amplitude. Here
\begin{equation}
I_n(x)=\frac{1}{2\pi}\int_0^{2\pi}e^{z\cos\theta}\cos(n\theta)\, d\theta,
\label{cp8}
\end{equation}
is the modified Bessel function of order $n$. If $M=0$, the density is uniform. This defines the homogeneous phase. If $M>0$, the equilibrium state is inhomogeneous with one cluster centered about
$\theta=\phi$.

\begin{figure}
\begin{center}
\includegraphics[clip,scale=0.3]{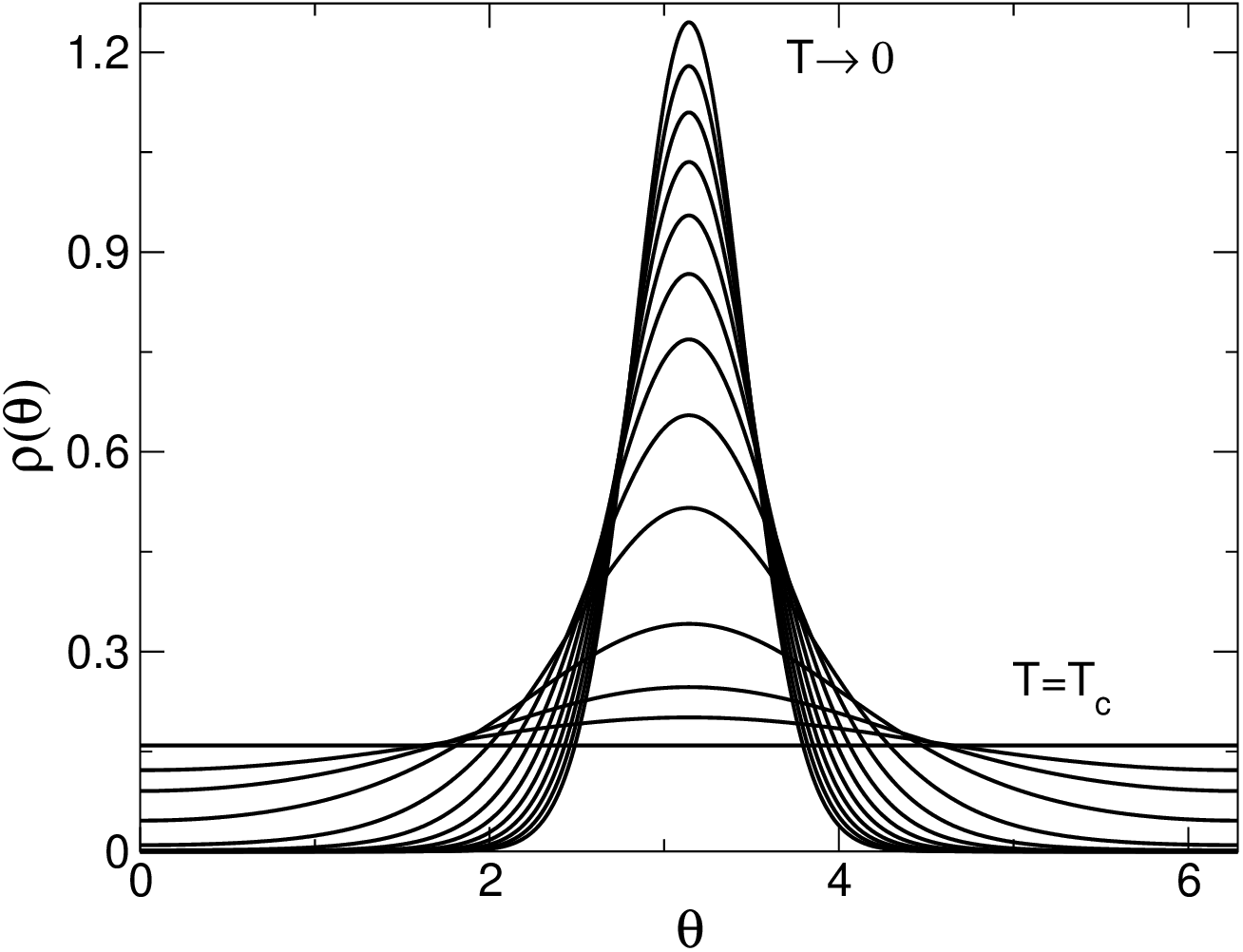}
\caption{Evolution of the density profile as temperature is decreased (from bottom to top).}
\label{rho}
\end{center}
\end{figure}

The magnetization $M$ is determined by substituting Eq.  (\ref{cp7}) in Eq.  (\ref{magn}). This yields the self-consistency relation
\begin{equation}
M=\frac{I_{1}(\beta M)}{I_{0}(\beta M)}.
\label{cp9}
\end{equation}
Equation (\ref{cp9}) determines the magnetization $M$ as a function of the temperature $T$. Then, the density profile  is given by Eq. (\ref{cp7}). The critical points are degenerate. There exist an infinity of critical points which differ only by their phase $\phi$, i.e. by the position of the maximum of the density profile. They have the same value of free energy (see below). We can take $\phi=0$ without loss of generality. In that case, $M_x=M$ and $M_y=0$. Then, the density can be written as
\begin{eqnarray}
\rho(\theta)=\frac{1}{2\pi I_0(\beta M)} e^{\beta M\cos\theta},
\label{cp11}
\end{eqnarray}
where $M$ is determined in terms of $T$ by Eq.  (\ref{cp9}).
Some density profiles are plotted in Figure \ref{rho}.

In the canonical ensemble, we have to select free energy {\it minima} and discard
free energy maxima and saddle points. In a first step, we shall
determine all the critical points of free energy. The thermodynamical stability of these solutions will be studied in a second step.

\subsection{The mean magnetization at equilibrium}
\label{sec_mme}

\begin{figure}
\begin{center}
\includegraphics[clip,scale=0.3]{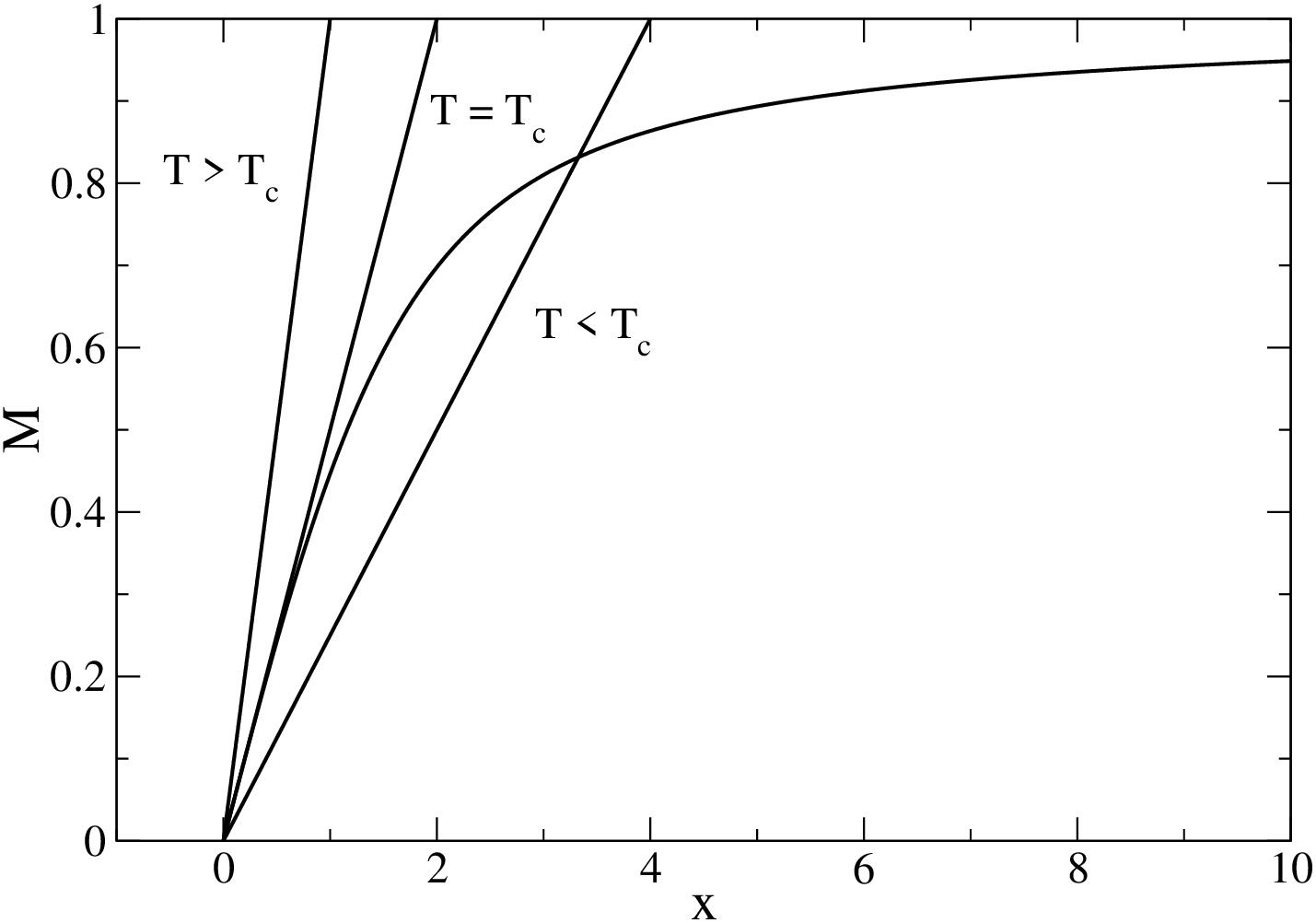}
\caption{Graphical construction determining the solutions of the self-consistency relation (\ref{cp9}). The critical points of free energy respecting the normalization condition are given by the intersection(s) between the curve $M=M(x)$ defined by Eq.  (\ref{tp6}) and the straight line $M=Tx$. There is one solution $M=0$ for $T>T_c=1/2$ and two solutions $M=0$ and $M(T)>0$ for $T<T_c$. It can be shown \cite{hmfarxiv} that a critical point of free energy is a minimum (resp. maximum) if the slope of the curve $M(x)$ at that point is smaller (resp. larger) than the slope of the straight line $M=Tx$. Therefore, the inhomogeneous states ($M>0$) are always stable while the homogeneous states ($M=0$) are stable for $T>T_c$ and unstable for $T<T_c$.}
\label{blambda}
\end{center}
\end{figure}

Using Eqs. (\ref{smf10b}), (\ref{magn}) and (\ref{cp11}), the equilibrium free energy (\ref{tsce1zero}) is given by
\begin{eqnarray}
\label{tsce1}
F(T)=\frac{1-M^2}{2}-T\ln I_0(\beta M)+ M^2\nonumber\\
-\frac{1}{2}T\ln T-\frac{3T}{2}\ln(2\pi).
\end{eqnarray}
Using the self-consistency relation (\ref{cp9}), we can obtain the curves $M(T)$ and $F(T)$. To that purpose, we can proceed as follows. Introducing the parameter
\begin{eqnarray}
\label{tp5}
x=\beta M,
\end{eqnarray}
we can rewrite the self-consistency relation as
\begin{eqnarray}
\label{tp6}
M=M(x)\equiv \frac{I_1(x)}{I_0(x)}.
\end{eqnarray}
The function $M(x)$ is plotted in Figure \ref{blambda}. The self-consistency relation may be solved by a simple graphical construction explained in the Figure caption. For future reference, we note the identity
\begin{eqnarray}
M'(x)=1-\frac{M(x)}{x}-M(x)^2,
\label{idint}
\end{eqnarray}
which can be obtained from the standard properties of the Bessel functions. From Eqs. (\ref{tp5}) and (\ref{tp6}), the temperature may be expressed in terms of $x$ as
\begin{eqnarray}
\label{tp6bis}
T=\frac{M(x)}{x}.
\end{eqnarray}
From the relations $M=M(x)$, $T=T(x)$ and $F=F(x)$, we can obtain the curves $M(T)$ and $F(T)$ in parametric form with the parameter $x$ going from $0$ to $+\infty$.
These parametric equations apply to the inhomogeneous phase ($M\neq 0$).  For the homogeneous phase, we have
\begin{eqnarray}
\label{tp12b}
M(T)=0, \quad F(T)=-\frac{1}{2}T\ln T-\frac{3}{2}T \ln (2\pi)+\frac{1}{2}.
\end{eqnarray}
The homogeneous phase exists for $T\ge 0$. The inhomogeneous  phase bifurcates from the homogeneous phase at $x=0$, corresponding to
\begin{eqnarray}
\label{tp12c}
T_c=\lim_{x\rightarrow 0} \frac{M(x)}{x}=\frac{1}{2}.
\end{eqnarray}
The inhomogeneous phase exists for $0\le T\le T_c=1/2$.

\begin{figure}
\begin{center}
\includegraphics[clip,scale=0.3]{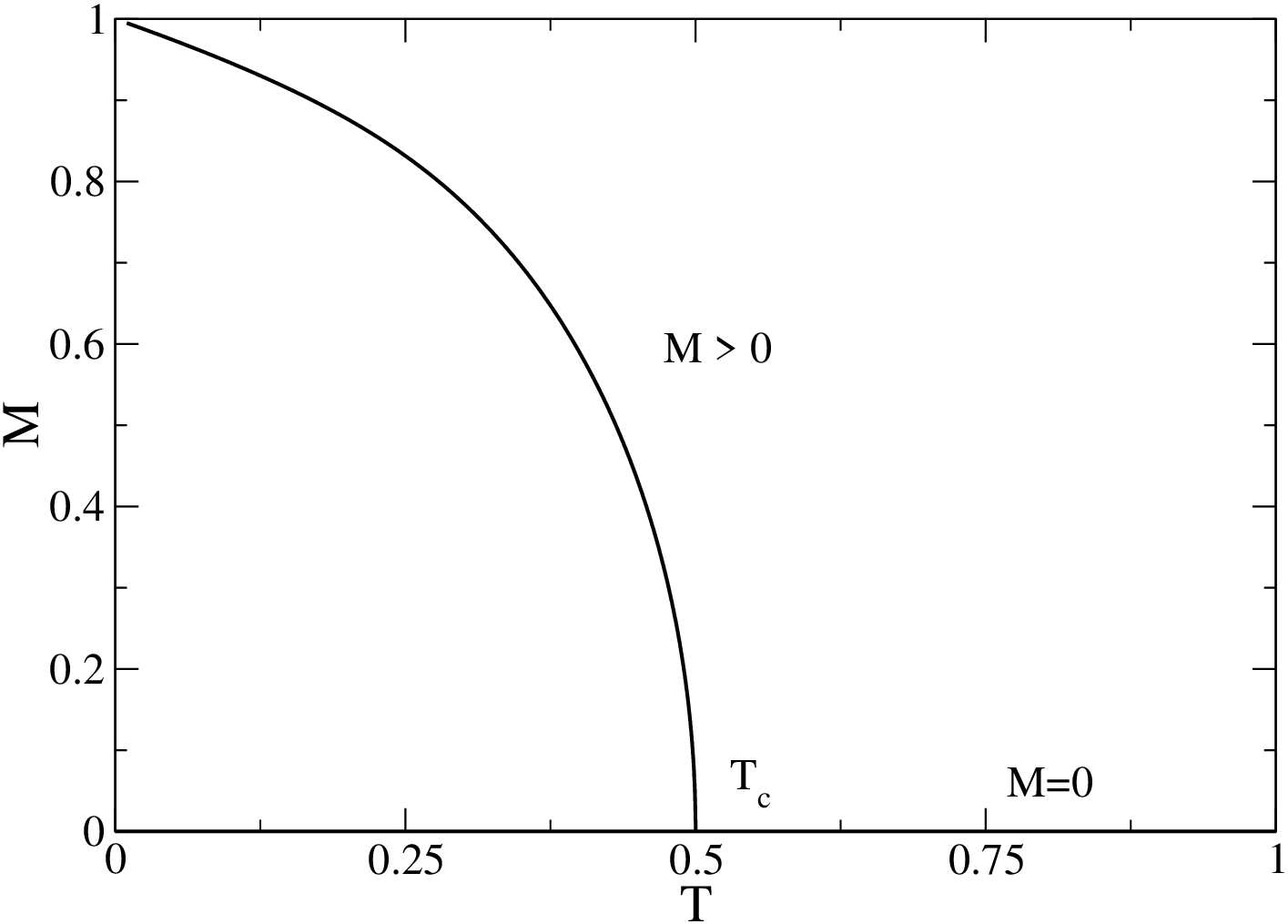}
\caption{Magnetization (order parameter) as a function of the temperature. The system is magnetized (inhomogeneous) for $T<T_c$ and non-magnetized (homogeneous) for $T>T_c$.}
\label{beta}
\end{center}
\end{figure}

\begin{figure}
\begin{center}
\includegraphics[clip,scale=0.3]{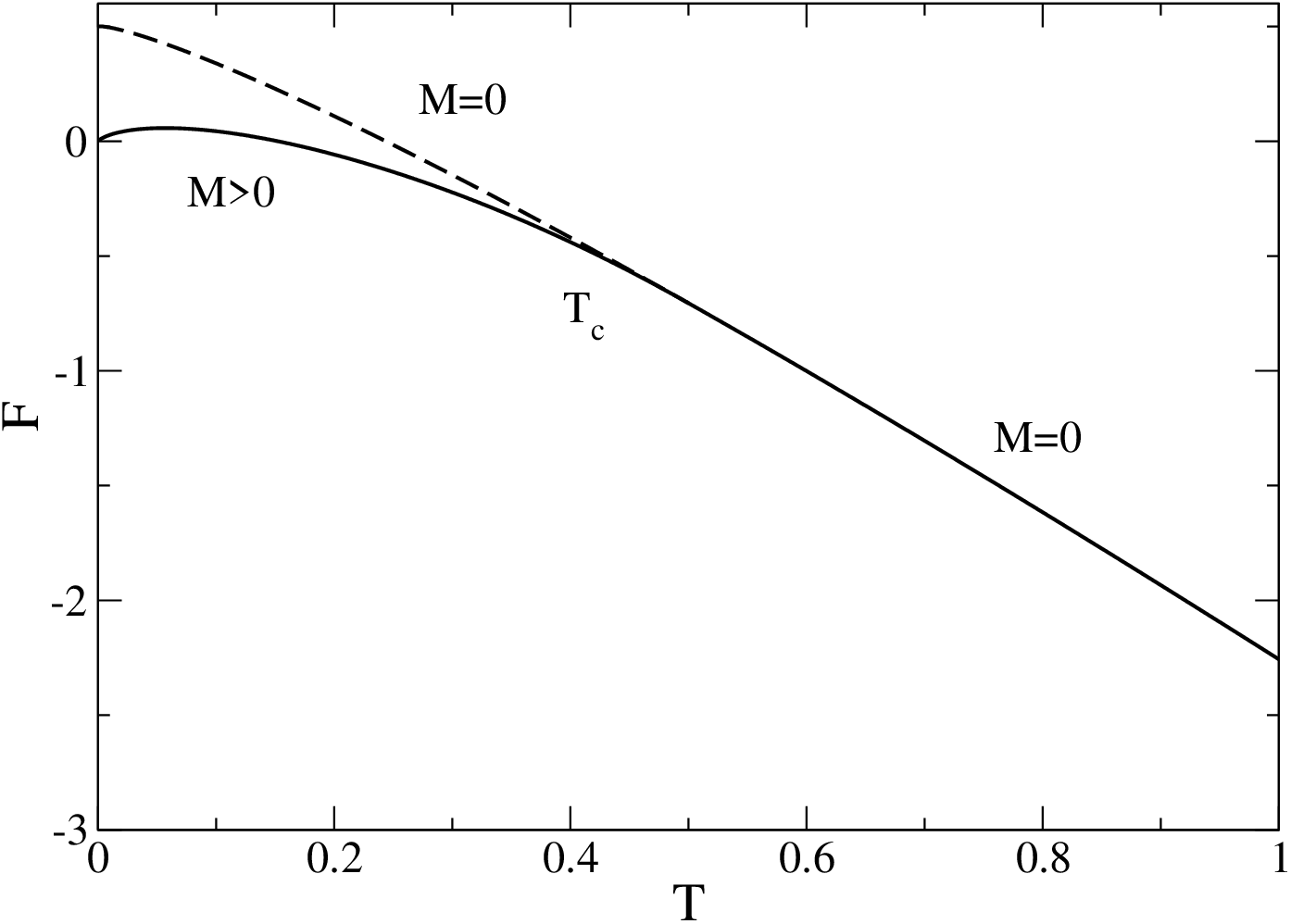}
\caption{Free energy as a function of the temperature.}
\label{tf}
\end{center}
\end{figure}

At $T=0$ (ground state), all the particles are at $\theta=0$. The density profile $\rho(\theta)=\delta(\theta)$ is a Dirac peak and the magnetization is $M=1$. Close to the ground state ($x\rightarrow +\infty$, $T\rightarrow 0$):
\begin{eqnarray}
\label{tp13}
M\simeq 1-\frac{T}{2}-\frac{3T^2}{8},\qquad F\sim -T\ln (2\pi T).
\end{eqnarray}
At $T=T_c$ (critical temperature), the density profile is spatially homogeneous $\rho(\theta)=1/(2\pi)$ and the magnetization is $M=0$. Close to the bifurcation point ($x\rightarrow 0$, $T\rightarrow T_c^-$):
\begin{eqnarray}
\label{tp15}
M\sim 2(T_c-T)^{1/2},
\end{eqnarray}
\begin{eqnarray}
\label{tp17}
F-F_c\sim \frac{1}{2}\lbrack 1-\ln 2+3\ln (2\pi)\rbrack (T_c-T),
\end{eqnarray}
where $F_c=1/2+(1/4)\ln 2-(3/4)\ln (2\pi)\simeq -0.705$ is the value of the free energy at the critical point.

The curves  $M(T)$ and $F(T)$  are plotted  in Figures \ref{beta} and \ref{tf}.
These curves contain all the critical points of free energy respecting the normalization condition. They reveal a second order phase transition between the homogeneous phase and the inhomogeneous phase at the critical temperature $T_c$. It is marked by the discontinuity of the second derivatives of the free energy. This is equivalent to the discontinuity of the derivative of the energy and the discontinuity of the derivative of the magnetization.

\subsection{The eigenvalue equation for thermodynamical stability}
\label{sec_tsee}

Among the critical points of free energy respecting the normalization condition, we have to select minima and discard maxima and saddle points.

The second order variations of free energy are given by
\begin{eqnarray}
\label{tsce4}
\delta^{2} F={1\over 2}\int \delta\rho\delta\Phi d\theta+{1\over 2}T\int {(\delta\rho)^{2}\over \rho}d\theta,
\end{eqnarray}
with
\begin{equation}
\label{ejl3}
\delta\Phi(\theta)=\int_{0}^{2\pi} u(\theta-\theta') \delta\rho(\theta')\, d\theta'.
\end{equation}
The Boltzmann distribution is a (local) minimum of free energy respecting the  normalization condition if, and only, if $\delta^2 F>0$ for all perturbations satisfying $\int \delta\rho\, d\theta=0$.

Let us first consider the homogeneous phase where $\rho=1/(2\pi)$. We decompose the perturbation $\delta\rho$ in Fourier modes according to
\begin{equation}
\label{def5}
\delta\rho(\theta)=\sum_{n=-\infty}^{+\infty}\, e^{i n\theta}\delta\hat\rho_n,\qquad \delta\hat\rho_n=\int_0^{2\pi}\frac{d\theta}{2\pi}\, e^{-i n\theta}\delta\rho(\theta).
\end{equation}
We use a similar decomposition for $\delta\Phi$ and $u$. According to Eq. (\ref{ejl3}), we have
\begin{eqnarray}
\label{tsce4b}
\delta\hat{\Phi}_n=2\pi \hat{u}_n \delta\hat{\rho}_n,
\end{eqnarray}
with
\begin{equation}
\label{jh4}
\hat{u}_n=\frac{1}{2}(2\delta_{n,0}-\delta_{n,1}-\delta_{n,-1}).
\end{equation}
From Eq. (\ref{def5}), we obtain
\begin{eqnarray}
\label{tsce4a}
\int (\delta\rho)^{2}\, d\theta=2\pi \sum_n |\delta\hat{\rho}_n|^2.
\end{eqnarray}
On the other hand, using Eq. (\ref{tsce4b}), we get
\begin{eqnarray}
\label{tsce4c}
\int \delta\rho\delta\Phi \, d\theta=2\pi \sum_n \delta\hat{\rho}_{n}\delta\hat{\Phi}_{-n}=(2\pi)^2 \sum_n \hat{u}_n |\delta\hat{\rho}_n|^2.
\end{eqnarray}
Substituting these relations in Eq. (\ref{tsce4}), we find that
\begin{eqnarray}
\delta^2 F=(2\pi)^2 \sum_{n=1}^{+\infty} (T+\hat{u}_n)|\delta\hat{\rho}_n|^2.
\label{dyn1a}
\end{eqnarray}

This equation is valid for a general potential of interaction. If $\hat{u}_n>0$ for all $n$ (repulsive interaction),  the homogeneous phase is always thermodynamically stable. If  $\hat{u}_{n}<0$ for some mode(s) $n$ (attractive interaction), the homogeneous phase is thermodynamically  stable when $T>T_c=\max_n|\hat{u}_n|$ and thermodynamically unstable (with respect to the modes such that $T+\hat{u}_n<0$) when $T<T_c$. For the cosine potential (\ref{u}), using Eq. (\ref{jh4}), we obtain
\begin{eqnarray}
\delta^2 F=(2\pi)^2 T \sum_{n=2}^{+\infty} |\delta\hat{\rho}_n|^2+(2\pi)^2 (T-T_c)|\delta\hat{\rho}_1|^2.
\label{dyn1bb}
\end{eqnarray}
For $T>T_c$, we clearly have $\delta^2F>0$ so that the homogeneous phase is thermodynamically stable. For $T<T_c$, taking $\delta\hat{\rho}_n=0$ for $n\neq \pm 1$ and  $\delta\hat{\rho}_{\pm 1}\neq 0$, we see that $\delta^2F<0$ for these particular perturbations  so that the homogeneous phase is thermodynamically unstable.

To treat the general case where the system may be spatially inhomogeneous, we introduce the notation
\begin{eqnarray}
q(\theta)=\int_0^{\theta}\delta\rho(\theta',t)\, d\theta',\qquad \delta\rho=\frac{dq}{d\theta},
\label{dyn1c}
\end{eqnarray}
which corresponds to the mass perturbation in the interval $[0,\theta]$. The conservation of mass implies $q(0)=q(2\pi)=0$. Substituting Eq. (\ref{dyn1c}) in Eq. (\ref{tsce4}) and making simple integrations by parts, we can put the second order variations of free energy in the quadratic form
\begin{eqnarray}
\label{ka1}
\delta^{2} F=\int_0^{2\pi}\int_0^{2\pi}d\theta d\theta' q(\theta)K(\theta,\theta')q(\theta'),
\end{eqnarray}
with
\begin{eqnarray}
\label{ka2}
K(\theta,\theta')=-\frac{1}{2}\cos(\theta-\theta')-\frac{1}{2}T\delta(\theta-\theta')
\frac{d}{d\theta}\left (\frac{1}{\rho}\frac{d}{d\theta}\right ).\nonumber\\
\end{eqnarray}
We are led therefore to considering the eigenvalue problem
\begin{eqnarray}
\label{ka3}
\int_0^{2\pi} K(\theta,\theta')q(\theta')\, d\theta'=\lambda q(\theta),
\end{eqnarray}
or, more explicitly,
\begin{eqnarray}
{d\over d\theta}\biggl ({1\over\rho}{dq\over d\theta}\biggr )+{1\over  T }\int_{0}^{2\pi}q(\theta')\cos(\theta-\theta')d\theta'=-2\lambda q.\nonumber\\
\label{smf17a}
\end{eqnarray}
The Boltzmann distribution is a (local) minimum of free energy respecting the normalization condition if all the eigenvalues $\lambda$ are positive,  and a saddle point of free energy if at least one
of the eigenvalues is negative.

In the homogeneous phase ($\rho=1/(2\pi)$ and $\Phi=1$), Eq. (\ref{smf17a}) reduces to
\begin{equation}
\label{see1}
2\pi{d^{2}q\over d\theta^{2}}+{1\over T}\int_{0}^{2\pi}q(\theta')\cos(\theta-\theta')d\theta'=-2\lambda q.
\end{equation}
The eigenmodes  of Eq. (\ref{see1}) are $q_{n}=A_{n}\cos
(n\theta)$ and $q_{n}=B_{n}\sin
(n\theta)$.  For $n\neq \pm 1$, the eigenvalues are $\lambda_{n\neq \pm 1}={\pi n^{2}}>0$,
so that these modes do not induce instability. For $n=\pm 1$, the (degenerate) eigenvalues are
$\lambda_{\pm 1}=-\pi (T_c/T-1)$.
Therefore, the uniform phase is thermodynamically stable when $T>T_{c}$
and thermodynamically unstable when $T<T_{c}$, as discussed previously.

\begin{figure}
\begin{center}
\includegraphics[clip,scale=0.3]{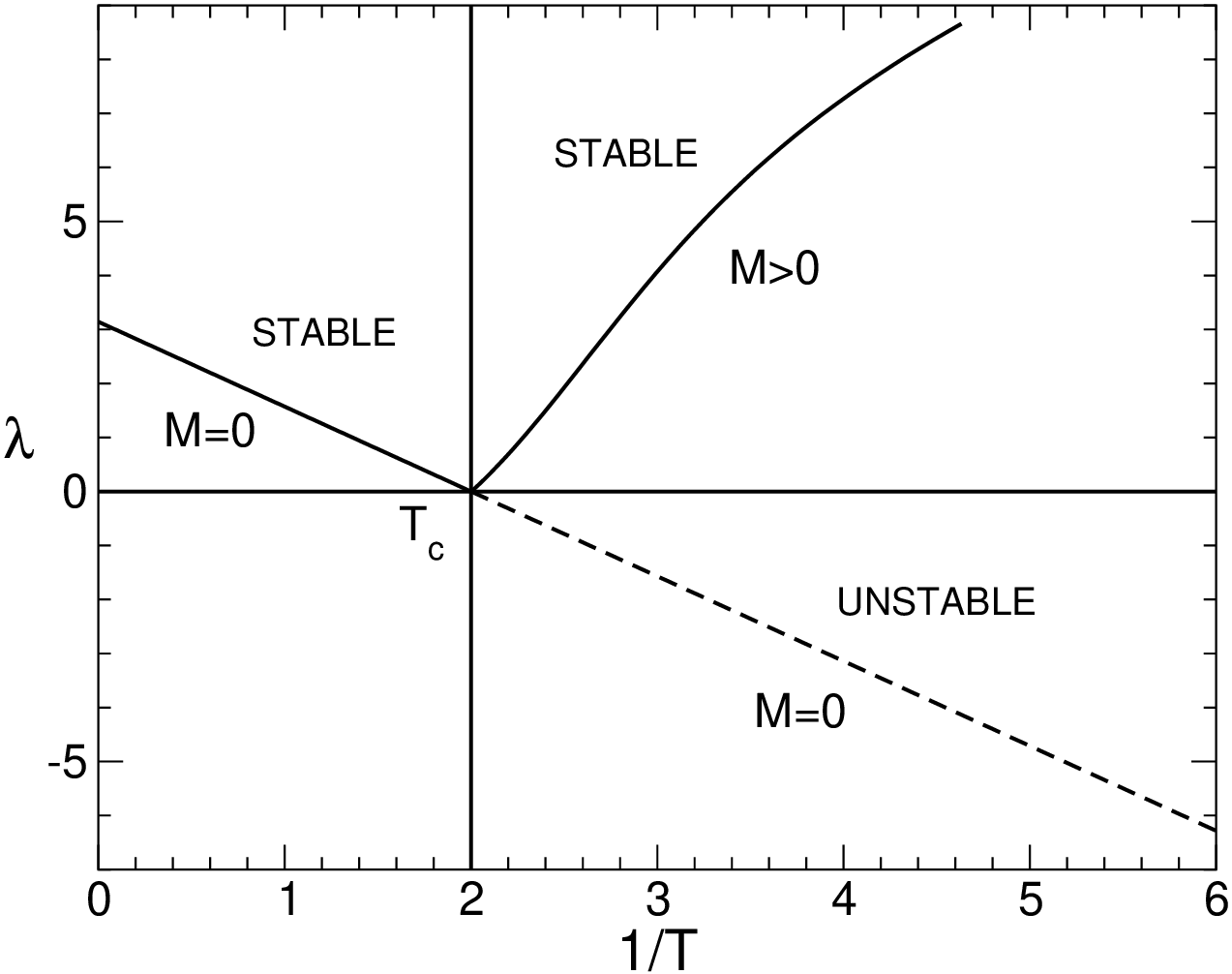}
\caption{Dependence of the smallest eigenvalue $\lambda$ with the temperature. A positive value of $\lambda$ corresponds to stability ($\delta^{2}F>0$) and a negative value of $\lambda$ corresponds to instability.}
\label{stabthermo}
\end{center}
\end{figure}

In the inhomogeneous phase, it is possible to solve Eq.  (\ref{smf17a}) analytically  close to the bifurcation point $T_{c}$ where $M\rightarrow 0$ \cite{cvb}. The smallest eigenvalue is $\lambda\simeq 2\pi ({T_c/T}-1)$. More generally, the smallest eigenvalue $\lambda$ obtained by solving  Eq.  (\ref{smf17a}) numerically, is plotted as a function of the inverse temperature in Figure \ref{stabthermo}.  Since $\lambda>0$, the inhomogeneous phase is always stable.

{\it Remark:} it is possible to determine analytically the eigenfunction $q_0(\theta)$ corresponding to the neutral mode $\lambda=0$ in Eq. (\ref{smf17a}). This is done in Appendix F of \cite{cd} (see also Appendix E of \cite{cc} and Appendix C of \cite{hmfmagn} for generalizations). With the notations of the present paper, assuming that the equilibrium state is symmetric with respect to the $x$-axis, the expression of the neutral mode is (see Eq. (F.5) of \cite{cd}):
\begin{equation}
\label{margf}
\delta\rho_0=\frac{dq_0}{d\theta}=\beta\rho(\theta)\left\lbrack \delta M_x(\cos\theta-M)+\delta M_y\sin\theta\right\rbrack,
\end{equation}
where $\delta M_x$ and $\delta M_y$ are determined self-consistently by the relations $\delta M_x=\int \delta\rho_0 \cos\theta\, d\theta$ and $\delta M_y=\int \delta\rho_0\sin\theta\, d\theta$. Using the properties of the Bessel functions, they lead to the trivial identity $\delta M_y=\delta M_y$ and to the condition $\delta M_x=\beta \delta M_x M'(\beta M)$. For $\delta M_x\neq 0$, this condition is satisfied only at the critical point ($T=T_c$, $M=0$, $x=0$) \cite{cd}. For $\delta M_x=0$, the neutral mode is $\delta\rho_0=\beta\rho(\theta)\delta M_y\sin\theta$ and it corresponds to a mere rotation of the system\footnote{Using Eq. (\ref{cp11}), we note that $\delta\rho_0\propto \rho'(\theta)$. This result can be understood as follows. Since the system is invariant by rotation, if $\rho(\theta)$ is an equilibrium state, then $\rho(\theta+\phi)$ is also an equilibrium state for any $\phi$ (see Section \ref{sec_ed}). Considering $\phi\ll 1$ and using $\rho(\theta+\phi)\simeq \rho(\theta)+\rho'(\theta)\phi$, we conclude that $\rho'(\theta)$ is a neutral mode ($\lambda=0$ or $\delta^2 S=0$).} \cite{hmfmagn}. In conclusion, for $T<T_c$, there is no solution of Eq. (\ref{smf17a}) for which $\lambda=0$ (except a trivial rotation of the system) \cite{cd}. The neutral mode occurs only at the bifurcation point $T=T_c$ in agreement with the Poincar\'e theorem \cite{hmfarxiv}. This is in agreement with the preceding results showing that $\lambda>0$ in the inhomogeneous phase ($T<T_c$). These results have been generalized to the $\alpha$-BMF model in Ref. \cite{cr}.

\subsection{The free energy $F(M)$}
\label{sec_fm}

To solve the minimization problem (\ref{ff1}), we can proceed in two steps \cite{hmfarxiv}. We first minimize $F[\rho]$ at fixed normalization {\it and} magnetization ${\bf M}$. This gives
\begin{eqnarray}
\label{aace2}
\rho(\theta)= \frac{1}{2\pi I_{0}(\lambda)} e^{\lambda\cos(\theta-\phi)},
\end{eqnarray}
where $\phi$ is an arbitrary phase and $\lambda$ is determined by the modulus of the magnetization according to
\begin{equation}
M=M(\lambda)=\frac{I_{1}(\lambda)}{I_{0}(\lambda)}.
\label{aace3}
\end{equation}
Using Eq. (\ref{aace2}) we can express the free energy $F[\rho]$ given by Eq.  (\ref{tsce1zero}) as a function of the magnetization $M$. We get
\begin{eqnarray}
F(M)=\frac{1-M^2}{2}+T \lambda M-T\ln I_0(\lambda)\nonumber\\
-\frac{1}{2}T\ln T-\frac{3}{2}T\ln(2\pi),
\label{aace4}
\end{eqnarray}
where $\lambda(M)$ is obtained by inverting equation (\ref{aace3}). Finally, the solution of the minimization problem (\ref{ff1}) is given by Eq. (\ref{aace2}) where $M$ is the solution of the minimization problem
\begin{eqnarray}
\label{aace5}
F(T)=\min_M\left\lbrace F(M)\right\rbrace.
\end{eqnarray}
It can be shown that the minimization problems (\ref{ff1}) and (\ref{aace5}) are
equivalent for global 
and local minimization \cite{hmfarxiv}. The minimization problem
(\ref{aace5}) can also be directly derived from the canonical distribution in
the $N\rightarrow +\infty$ limit \cite{hmfarxiv}. This is a result of large
deviations.

Using the identity  $I'_0(\lambda)=I_1(\lambda)$, we can check that the condition $F'(M)=0$ gives $\lambda=x=\beta M$ leading to the self-consistency relation (\ref{cp9}).

In the homogeneous phase ($M=0$), computing the second derivatives of the free energy (\ref{aace4}), we obtain
\begin{eqnarray}
F''(0)=2(T-T_c).
\label{aace14}
\end{eqnarray}
From this analytical formula, we immediately conclude that the homogeneous states are stable ($F''(0)>0$) for $T>T_c$ and unstable ($F''(0)<0$) for $T<T_c$ as found previously by other methods.

In the inhomogeneous phase, using Eq. (\ref{idint}), we find that the second derivatives of the free energy (\ref{aace4}) at a critical point can be written as
\begin{eqnarray}
F''(M)=-1+\frac{T}{1-T-M^2},
\label{aace10b}
\end{eqnarray}
where $M$ and $T$ are related by the self-consistency relation (\ref{cp9}). By studying the sign of the second derivative of $F(M)$ we can show (see \cite{hmfarxiv} and the caption of Figure \ref{blambda}) that the inhomogeneous states are always stable ($F''(M)>0$).

\begin{figure}
\begin{center}
\includegraphics[clip,scale=0.3]{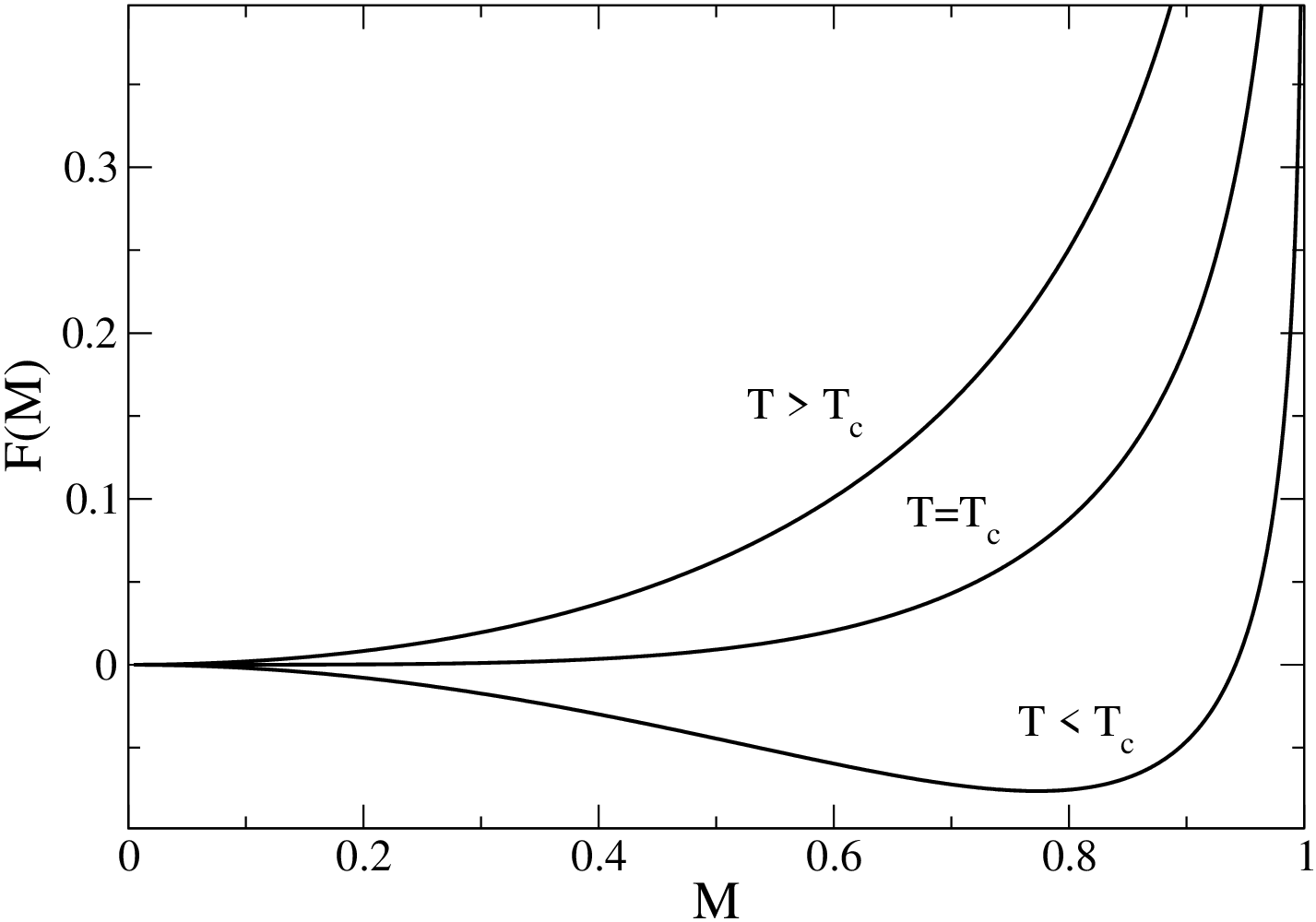}
\caption{Free energy $F(M)$ as a function of the magnetization $M$ for a given value of the temperature $T$ (for clarity, we have subtracted the free energy  $F_{0}(T)$ of the homogeneous phase $M=0$). For $T>T_c$, this curve has a unique (global)  minimum at $M=0$. For $T<T_c$, this curve has a local maximum at $M=0$ and a global minimum at $M(T)>0$. Specifically, we have taken $T=0.3$, 0.5\ and $0.7$. For  $M\rightarrow 1$, we have $F(M)\sim -(1/2)T\ln (1-M)$.}
\label{mfTfixe}
\end{center}
\end{figure}

We can plot the function $F(M)$ for a prescribed temperature $T$. It
is defined in parametric form (with parameter $\lambda$) by
Eqs. (\ref{aace3}) and (\ref{aace4}). This function displays the two
behaviors described above, as illustrated in Figure \ref{mfTfixe}.  For
$T>T_c$ the free energy $F(M)$ has a unique minimum at $M=0$
(homogeneous phase). For $M\rightarrow 0$, we can make the
approximation
\begin{eqnarray}
F(M)\simeq F_{0}(T)+(T-T_c)M^2,
\label{fbzab}
\end{eqnarray}
where $F_{0}(T)$ is the equilibrium free energy of the homogeneous phase $(M=0)$ given by Eq. (\ref{tp12b}). From this formula, we explicitly check that the minimum of free energy is $M=0$ and that  $F''(0)=2(T-T_c)>0$ in agreement with Eq. (\ref{aace14}). For $T<T_c$, the free energy $F(M)$ has a maximum at $M=0$ and a minimum at $M(T)>0$ (inhomogeneous phase). Close to the critical point $T\rightarrow T_c^{-}$, the equilibrium magnetization $M(T)$ tends to zero. For $M\rightarrow 0$, we can make the approximation
\begin{eqnarray}
F(M)\simeq F_{0}(T)+(T-T_c)M^2+\frac{1}{8}M^4.
\label{fbza}
\end{eqnarray}
From this formula, we explicitly check that the minimum of free energy is given by Eq. (\ref{tp15}) and that $F''(M)=4(T_c-T)>0$ at that point.

{\it Remark:} In addition to
the approach developed previously, the thermodynamical stability analysis of
the cosine model may also be performed by determining the minimizer of the free
energy \cite{ms}, by evaluating the partition function using the
Hubbard-Stratonovich transformation and the saddle point approximation
\cite{ar}, by using the Poincar\'e theory of linear series of equilibria
\cite{hmfarxiv,inagaki},  by applying the theory of large deviations
\cite{largedev}, or by determining the minimum value of the second order
variations of free energy \cite{ccstab}. The advantage of the
approach developed in the present paper, based on the minimization of the free
energy $F[f]$, $F[\rho]$, or $F(M)$, is to be simple and physical.

\subsection{The equilibrium fluctuations of the magnetization}
\label{sec_efluc}

The distribution of the magnetization in the canonical ensemble  is given, for $N\rightarrow +\infty$, by \cite{hmfarxiv}:
\begin{eqnarray}
P({\bf M})=\frac{1}{Z(\beta)}e^{-\beta N F(M)},
\label{magn1}
\end{eqnarray}
where $Z(\beta)=\int  e^{-\beta N F(M)}\, d{\bf M}$ is the partition function and $F(M)$ is the free energy defined by Eq. (\ref{aace4}). We note that the average value of the magnetization vector ${\bf M}$ is always zero. This is due to the rotational invariance of the system.

In the homogeneous phase $(T>T_c)$ the particles are uniformly distributed on the circle so that the equilibrium magnetization vanishes ($M=0$). For $N\rightarrow +\infty$, $P({\bf M})$ is strongly peaked around the minimum of $F(M)$ that is $M=0$. Therefore, we can make the Gaussian approximation
\begin{eqnarray}
P({\bf M})=\frac{1}{\pi \langle M^2\rangle}e^{-\frac{M^2}{\langle M^2\rangle}}.
\label{magn3}
\end{eqnarray}
The variance of the magnetization is given by
\begin{eqnarray}
N\langle  M^2\rangle=\frac{2}{\beta F''(0)}.
\label{magn4a}
\end{eqnarray}
Using Eq. (\ref{aace14}), we obtain
\begin{equation}
\label{magn6a}
N\langle M^2\rangle=\frac{1}{1-T_c/T}.
\end{equation}

In the inhomogeneous phase $(T<T_c)$ the particles are concentrated around a certain point so that the equilibrium magnetization has a modulus $M\neq 0$ and a phase $\phi$. However, due to the degeneracy of the equilibrium states (see Section \ref{sec_ed}), the phase $\phi$ changes from realization to realization (or in the course of time when we consider a long timescale). This is why the  average value of the magnetization vector ${\bf M}$ vanishes in that case. In order to avoid this degeneracy, it may be useful to {\it impose} the direction of the magnetization. For example, we can impose $M_y=0$ (i.e. $\phi=0$). The distribution of the $x$-component of the magnetization is therefore given by
\begin{eqnarray}
P(M_x)=\frac{1}{Z(\beta)}e^{-\beta N F(M_x)}.
\label{magn1cd}
\end{eqnarray}
For $N\rightarrow +\infty$, $P(M_x)$ is strongly peaked around the minimum of $F(M_x)$. Therefore, we can make the Gaussian approximation
\begin{eqnarray}
P(M_x)=\frac{1}{\sqrt{2\pi \langle (\Delta M_x)^2\rangle}}e^{-\frac{(\Delta M_x)^2}{2\langle (\Delta M_x)^2\rangle}},
\label{magn3cd}
\end{eqnarray}
where $\Delta M_x$ is the fluctuation of the magnetization around its equilibrium value $M_x$. The variance of the magnetization is given by
\begin{eqnarray}
N\langle (\Delta M_x)^2\rangle=\frac{1}{\beta F''(M_x)},
\label{magn4acd}
\end{eqnarray}
Using Eq. (\ref{aace10b}), we get
\begin{eqnarray}
N\langle (\Delta M_x)^2\rangle=\frac{1}{\frac{1}{1-T-M_x^2}-\frac{1}{T}},
\label{magn7a}
\end{eqnarray}
where the magnetization  $M_x$ is related to the temperature by Eq. (\ref{cp9}). Close to the bifurcation point $T\rightarrow T_c^{-}$, using Eq. (\ref{fbza}), we obtain
\begin{eqnarray}
N\langle (\Delta M_x)^2\rangle\sim\frac{1}{8(T_c-T)}.
\label{magn9a}
\end{eqnarray}
Close to the ground state $T\rightarrow 0$, using Eq. (\ref{tp13}), we get
\begin{eqnarray}
N\langle (\Delta M_x)^2\rangle\sim\frac{T^2}{2}.
\label{magn11a}
\end{eqnarray}

\begin{figure}
\begin{center}
\includegraphics[clip,scale=0.3]{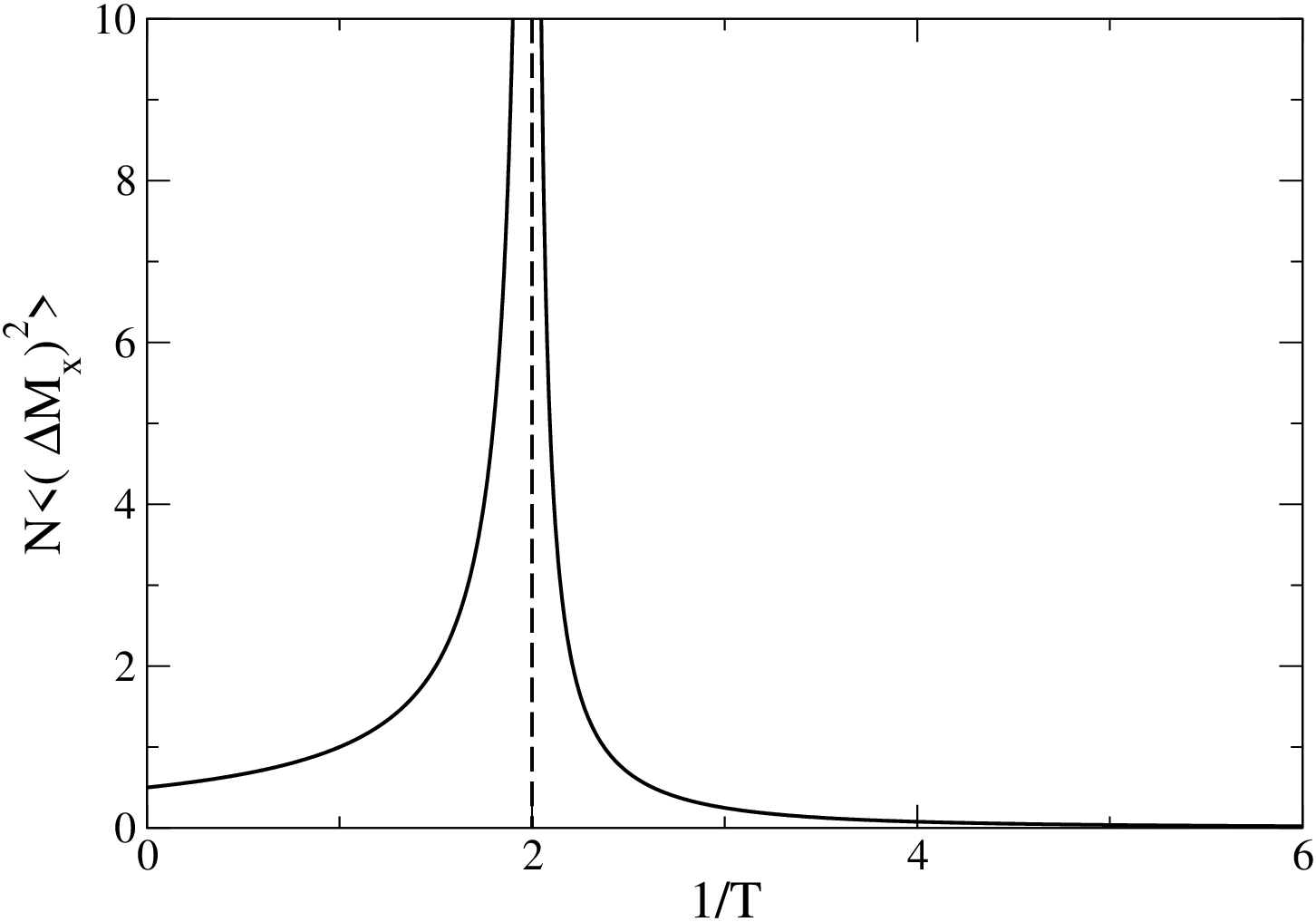}
\caption{Variance of the magnetization as a function of 
the inverse temperature in the canonical ensemble.  At the
critical point $T_c=1/2$, the variance of the magnetization diverges  like
$N\langle M_x^2\rangle=1/[2(1-T_c/T)]$ for $T>T_c$ and like $N\langle (\Delta
M_x)^2\rangle\sim 1/[4(T_c/T-1)]$ for $T\rightarrow T_c^{-}$. These expressions 
differ by a factor $2$.}
\label{varianceeta}
\end{center}
\end{figure}

The variance $\langle (\Delta M_x)^2\rangle$ of the magnetization is
plotted as a function of the inverse temperature in Figure
\ref{varianceeta} (to obtain this curve, it is convenient to write
Eq. (\ref{magn7a}) in parametric form with parameter $x$ and use the
results of Section \ref{sec_mme}).  According to Eq. (\ref{magn4acd}),
Figure \ref{varianceeta} also gives the evolution of $F''(M_x)$ at a
critical point ($F'(M_x)=0$) as a function of the temperature $T$. It
shows, in particular, that the inhomogeneous phase is always a minimum
of free energy ($F''(M_x)>0$). On the other hand, we can directly read
from Eq. (\ref{magn6a}) that the homogeneous phase is stable ($F''(0)>0$) for
$T>T_c$ and unstable ($F''(0)<0$) for $0\le T<T_c$. This is of course
equivalent to the results of Section \ref{sec_fm}.

\subsection{The effect of an external magnetic field}
\label{sec_magn}

The statistical equilibrium state of the cosine model under a magnetic field  has been studied in detail in \cite{hmfmagn}. We give here a few complements that will be needed in the sequel. The effect of an external magnetic field $h$ pointing in the $x$ direction can be taken into account by adding a term $-hM_x$ in the Hamiltonian $H/N$, hence in the free energy $F$. The statistical equilibrium state is determined by the Boltzmann distribution (\ref{cp4}) where $\Phi(\theta)$ is replaced by $\Phi(\theta)-h\cos\theta$. The equilibrium distribution may be rewritten as in Eq. (\ref{cp11}) where $M$ is replaced by $M_x+h$. As a result, the relation between the magnetization, the temperature, and the magnetic field is
\begin{eqnarray}
M_x=M(x)
\label{sun1}
\end{eqnarray}
with
\begin{eqnarray}
x=\beta(M_x+h).
\label{sun2}
\end{eqnarray}
For a fixed temperature $T$, the magnetization $M_x$ is related to the magnetic field $h$ by eliminating $x$ between Eq. (\ref{sun1}) and
\begin{eqnarray}
h=Tx-M(x),
\label{sun3}
\end{eqnarray}
issued from Eq. (\ref{sun2}). From Eqs. (\ref{sun1}) and (\ref{sun3}), the magnetic susceptibility $\chi_M=dM_x/dh$ is  given by
\begin{eqnarray}
\chi_M=\frac{1}{\frac{T}{M'(x)}-1}.
\label{sun4}
\end{eqnarray}
For a fixed magnetic field $h$, the magnetic susceptibility $\chi_M$ is related to the temperature $T$ by eliminating $x$ between Eq. (\ref{sun4}) and
\begin{eqnarray}
T=\frac{M(x)+h}{x},
\label{sun5}
\end{eqnarray}
issued from Eq. (\ref{sun2}). The case of an arbitrary magnetic field $h$ is considered in \cite{hmfmagn}. Here, we restrict ourselves to a weak field. For $T>T_c$ and $h\rightarrow 0$ we can take $x\rightarrow 0$ in Eqs. (\ref{sun1}) and (\ref{sun3}). We obtain $M\sim x/2$ and $h\sim (T-T_c)x$ leading to
\begin{eqnarray}
M_x=\frac{h}{2(T-T_c)},\qquad \chi_M=\frac{1}{2(T-T_c)}.
\label{sun6}
\end{eqnarray}
For $T<T_c$ and $h\rightarrow 0$, the magnetic susceptibility is related to the temperature by eliminating $x$ between Eq. (\ref{sun4}) and Eq. (\ref{sun5}) with $h=0$. These equations can be rewritten as
\begin{eqnarray}
\chi_M=\frac{1}{\frac{T}{1-T-M(x)^2}-1}, \qquad T=\frac{M(x)}{x},
\label{sun7}
\end{eqnarray}
where we have used the identity (\ref{idint}). For $T\rightarrow T_c^{-}$, we can take $x\rightarrow 0$ and  we obtain $T_c-T\sim x^2/16$ and  $\chi_M\sim 4/x^2$ leading to
\begin{eqnarray}
\chi_M\sim \frac{1}{4(T_c-T)},\qquad (T\rightarrow T_c^{-}).
\label{sun8}
\end{eqnarray}
Actually, we can obtain these asymptotic results directly from the normal form of the free energy close to the critical point $T\rightarrow T_c$ in the weak field limit $h\rightarrow 0$:
\begin{eqnarray}
F(M_x)=F_{0}(T)+(T-T_c)M_x^2+\frac{1}{8}M_x^4-hM_x.
\label{sun9}
\end{eqnarray}
It is obtained from Eq. (\ref{fbza}) by adding $-hM_x$. The minimum of free energy is determined by
\begin{eqnarray}
2(T-T_c)M_x+\frac{1}{2}M_x^3-h=0.
\label{sun10a}
\end{eqnarray}
For $T>T_c$, we get $M_x\simeq h/[2(T-T_c)]$ returning Eq. (\ref{sun6}) and for $T<T_c$, we get $M_x\simeq 2(T_c-T)^{1/2}+h/[4(T_c-T)]$ returning Eq. (\ref{sun8}).  We also note that the magnetization at the critical point $T=T_c$ behaves as $M_x=(2h)^{1/3}$ when the magnetic field $h\rightarrow 0$ (critical isotherm).

\begin{figure}
\begin{center}
\includegraphics[clip,scale=0.3]{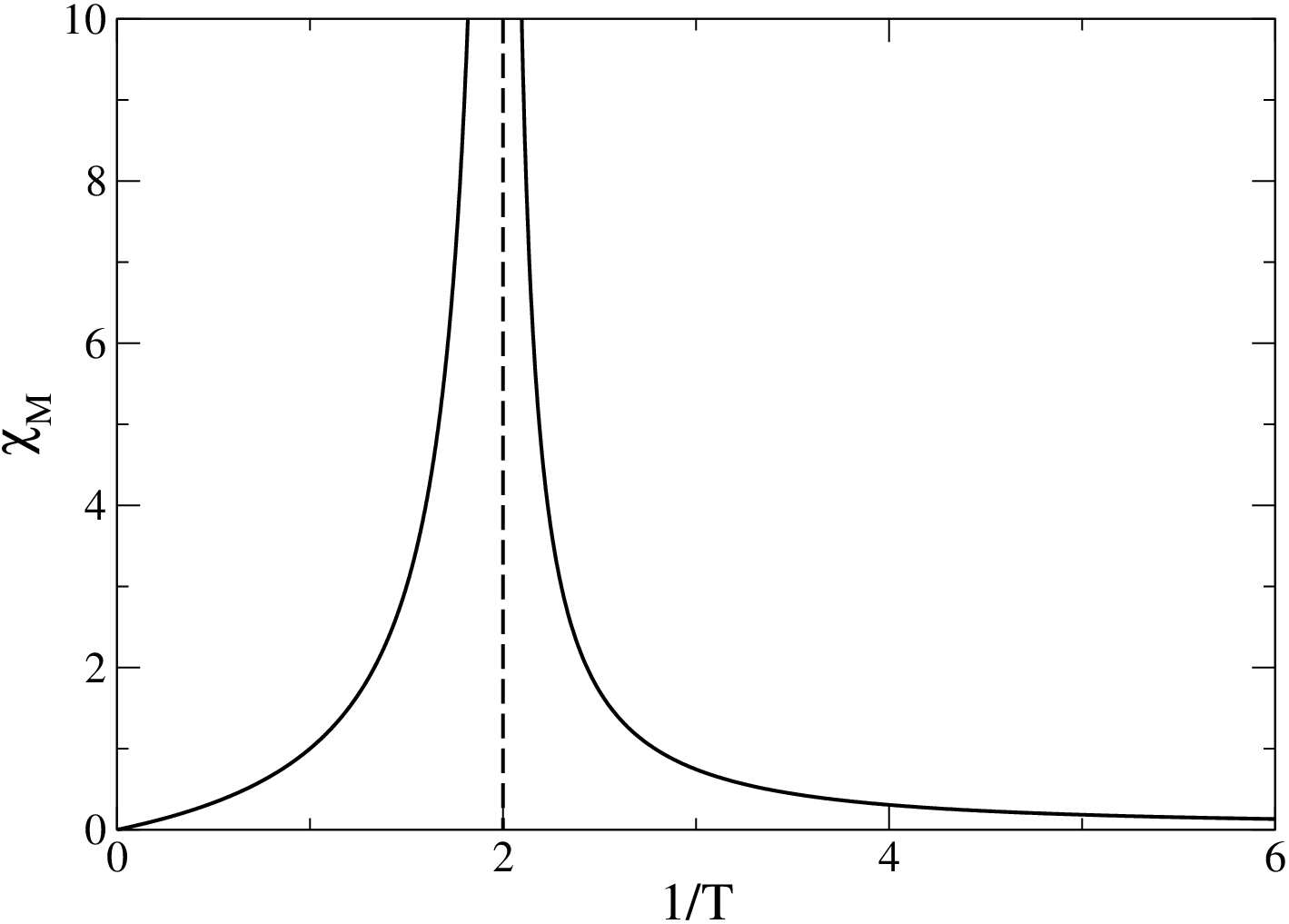}
\caption{Magnetic susceptibility $\chi_M=dM_x/dh$ as a function of the temperature $T$ in the weak field limit $h\rightarrow 0$. At the critical point $T_c=1/2$, the susceptibility diverges like $\chi_M=T_c/(T-T_c)$ for $T>T_c$ and like $\chi_M\sim T_c/[2(T_c-T)]$ for $T\rightarrow T_c^{-}$. These expressions differ by a factor $2$.}
\label{chi}
\end{center}
\end{figure}

The magnetic susceptibility in the weak field limit $h\rightarrow 0$ is plotted as a function of the temperature in Figure \ref{chi}. Comparing Eqs. (\ref{magn6a}) and (\ref{magn7a}) with Eqs. (\ref{sun6}) and (\ref{sun7}) we find that
\begin{eqnarray}
\chi_M=\beta N\langle (\Delta M_x)^2\rangle.
\label{sun10}
\end{eqnarray}
This is the well-known fluctuation-dissipation theorem (see Appendix
\ref{sec_fdtm} and Section \ref{sec_fdt}). This relation is actually valid for
an arbitrary magnetic 
field as explicitly checked in \cite{hmfmagn}. In the Gaussian
approximation, $\chi_M=1/F''(M_x)$ but Eq. (\ref{sun10}) is valid beyond the
Gaussian approximation (see Appendix \ref{sec_fdtm}).

\section{Dynamical stability of the steady states of the mean field Smoluchowski equation}
\label{sec_smoljeans}

The steady states of the mean field Smoluchowski equation (\ref{smf10g}) correspond to the mean field Boltzmann distribution (\ref{cp4}). They are the critical points of the free energy (\ref{tsce1zero}) respecting the normalization condition. Using general arguments based on the fact that the free energy is the Lyapunov functional of the mean field Smoluchowski equation, we can show that dynamical and thermodynamical stability coincide \cite{nfp}: the mean field Boltzmann distribution is dynamically stable with respect to the Smoluchowski equation if, and only, if it is a (local) minimum of free energy respecting the normalization condition (thermodynamical stability). We shall confirm this result in this section by a direct calculation. This study will provide in addition the explicit growth rate or damping rate of the perturbation.

\subsection{Spectral stability of the homogeneous phase: Jeans-like instability}
\label{sec_jh}

We first study the spectral stability of a spatially homogeneous steady state of the mean field  Smoluchowski equation: $\rho(\theta)=\rho=1/(2\pi)$ and $\Phi(\theta)=1$. Considering a small perturbation about this steady solution, the linearized mean field Smoluchowski equation is
\begin{eqnarray}
\xi{\partial\delta \rho\over\partial t}={\partial\over\partial\theta}\biggl (T{\partial\delta\rho \over\partial \theta}+\rho{\partial\delta\Phi\over\partial\theta}\biggr ),
\label{jh1}
\end{eqnarray}
with
\begin{equation}
\label{jh2}
\delta\Phi(\theta,t)=\int u(\theta-\theta') \delta\rho(\theta',t)\, d\theta'.
\end{equation}
We look for solutions in the form of plane waves $\delta\rho(\theta,t)=\delta\hat\rho_n e^{i(n\theta-\omega t)}$ and $\delta\Phi(\theta,t)=\delta\hat\Phi_n e^{i(n\theta-\omega t)}$.
The first equation gives
\begin{equation}
\label{jh5}
-i\xi\omega \delta\hat\rho_n=-Tn^2\delta\hat\rho_n-\rho n^2\delta\hat\Phi_n,
\end{equation}
and the second equation gives Eq. (\ref{tsce4b}).
Eliminating $\delta\hat\Phi_n$ between Eqs. (\ref{tsce4b}) and (\ref{jh5}), we obtain the dispersion relation
\begin{eqnarray}
i\xi\omega=n^2(T+\hat{u}_n).
\label{jh6}
\end{eqnarray}
We note that the pulsation is purely imaginary: $\omega=i\omega_i$. In the linear regime, the different modes of the density perturbation behave as
\begin{eqnarray}
\delta\hat{\rho}_n(t)=\delta\hat{\rho}_n(0)e^{-n^2(T+\hat{u}_n)t/\xi}.
\label{jh6dd}
\end{eqnarray}

The foregoing equations are valid for a general potential of interaction. The neutral mode ($\omega=0$) is determined by the condition
\begin{eqnarray}
T+\hat{u}_n=0.
\label{jh6ddfd}
\end{eqnarray}
If $\hat{u}_n>0$  for all $n$ (repulsive interaction), the homogeneous phase is always dynamically stable. If  $\hat{u}_{n}<0$ for some mode(s) $n$ (attractive interaction), the homogeneous phase is dynamically stable when $T>T_c=\max_n|\hat{u}_n|$ and dynamically unstable (for the modes such that $T+\hat{u}_n<0$) when $T<T_c$. For the cosine potential (\ref{u}), using Eq.  (\ref{jh4}), we find that the modes $n\neq \pm 1$ are damped  with a damping rate
\begin{eqnarray}
\omega_i=-\frac{1}{\xi}T n^2<0.
\label{jh7}
\end{eqnarray}
On the other hand, the complex pulsation of the modes $n=\pm 1$ is
\begin{eqnarray}
\omega_i=-\frac{1}{\xi}(T-T_c).
\label{jh8}
\end{eqnarray}
The homogeneous phase is dynamically stable when $T>T_c=1/2$ and
dynamically unstable (with respect to the modes $n=\pm 1$) when
$T<T_c$. In that case, the system is expected to become spatially
inhomogeneous and form clusters. This is similar to the Jeans
instability in astrophysics \cite{bt}. When $T>T_c$, all the modes
decay. When $T<T_c$, the modes $n\neq \pm 1$ decay while the modes
$n=\pm 1$ grow. Therefore, contrary to the Jeans instability in
astrophysics where the gravitational potential is scale invariant
resulting in several clusters, for the cosine potential the linear
instability is expected to generate a {\it single} cluster
corresponding to the growth of the modes $n=\pm 1$ (see
Sec. \ref{sec_mmip}).

{\it Remark:} we note that the 
relaxation time $t_R=1/\omega_i$ diverges when $T\rightarrow T_c^+$. This
corresponds to a critical slowing down \cite{goldenfeld}. As $T\rightarrow T_c$,
it takes longer and longer to equilibriate the system.

\subsection{Spectral stability of the inhomogeneous phase}
\label{sec_smoljeansinho}

We now consider a steady state of the mean field Smoluchowski equation that may be spatially inhomogeneous. The linearized mean field Smoluchowski equation is
\begin{eqnarray}
\xi{\partial\delta \rho\over\partial t}={\partial\over\partial\theta}\biggl (T{\partial\delta\rho \over\partial \theta}+\delta\rho{\partial\Phi\over\partial\theta}+\rho{\partial\delta\Phi\over\partial\theta}\biggr ),
\label{smf16a}
\end{eqnarray}
with Eqs. (\ref{smf10}) and (\ref{jh2}). Considering a perturbation of the form $\delta\rho\sim e^{\omega_i t} g(\theta)$, we obtain the eigenvalue equation
\begin{eqnarray}
{d\over d\theta}\biggl (T{d\delta\rho \over d \theta}+\delta\rho{d\Phi\over d\theta}+\rho{d\delta\Phi\over d\theta}\biggr )=\xi \omega_i \delta \rho.
\label{smf16b}
\end{eqnarray}
Introducing the notation (\ref{dyn1c}), and using the condition of hydrostatic equilibrium (\ref{smf16c}), we can put the eigenvalue equation (\ref{smf16b}) in the form
\begin{eqnarray}
{d\over d\theta}\biggl ({1\over\rho}{dq\over d\theta}\biggr )+{1\over  T }\int_{0}^{2\pi}q(\theta')\cos(\theta-\theta')d\theta'={\omega_i\xi\over T\rho} q.\qquad
\label{smf17}
\end{eqnarray}
This equation is similar to the eigenvalue equation (\ref{smf17a}) obtained by studying the sign of $\delta^2 F$. In particular, these two equations coincide at the point of marginal stability ($\omega_i=0$).

In the uniform phase, the destabilizing perturbations are $\delta\rho\sim \cos\theta \ e^{\omega_i t}$ and $\delta\rho\sim \sin\theta \ e^{\omega_i t}$, corresponding to the modes $n=\pm 1$. The corresponding eigenvalue is given by Eq. (\ref{jh8}). When $T<T_{c}$ the perturbation grows exponentially rapidly while it is damped exponentially rapidly when $T>T_{c}$. This returns the results of Section \ref{sec_jh}.

\begin{figure}
\begin{center}
\includegraphics[clip,scale=0.3]{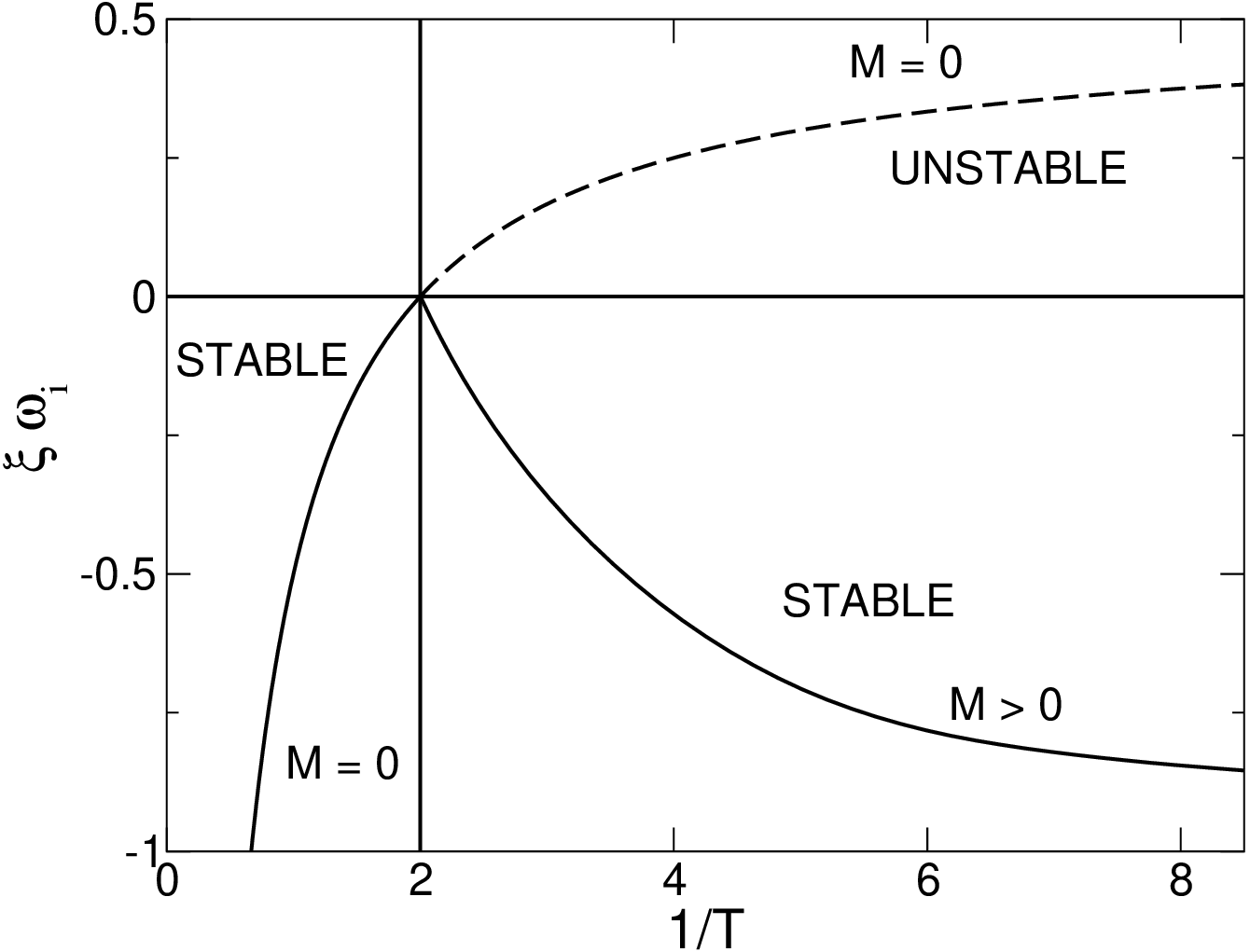}
\caption{Dependence of the largest eigenvalue $\omega_i$ with the temperature. A negative value of $\omega_i$ corresponds to stability and a positive value of $\omega_i$ corresponds to instability.}
\label{lambdaEulernew}
\end{center}
\end{figure}

Considering now the inhomogeneous phase when $T<T_c$, and using a perturbative approach valid for $T\rightarrow T_{c}^{-}$ (the calculations are similar to those reported in Appendix A of \cite{cvb}), we find that the largest eigenvalue is
\begin{eqnarray}
\omega_i=-{2\over\xi} (T_{c}-T).
\label{smf19}
\end{eqnarray}
Since $\omega_i<0$, the perturbation is damped exponentially rapidly. More generally, by solving Eq. (\ref{smf17}) numerically (see Figure \ref{lambdaEulernew}), we find that $\omega_i$ is always negative in the inhomogeneous phase. This implies that the inhomogeneous phase is always dynamically stable.

In conclusion, we find that dynamical and thermodynamical stability coincide. We note that the modes $n=\pm 1$ are associated with the magnetization (see Appendix \ref{sec_app}). Therefore, Eqs. (\ref{jh8}) and (\ref{smf19}) give the growth rate or the damping rate of the magnetization (see also Section \ref{sec_mmip}). Starting from an unstable homogeneous distribution  below the critical temperature ($T<T_{c}$), there is first an exponential growth of the magnetization on a timescale $\sim \xi (T_{c}-T)^{-1}$. Then,
nonlinear terms come into play. Finally, the system relaxes towards a stable clustered state on a timescale $\sim (1/2)\xi (T_{c}-T)^{-1}$. The fact that only the modes $n=\pm 1$ grow during the linear regime explains why we observe just {\it one} cluster in the numerical simulation of the mean field cosine Smoluchowski equation (see Figure \ref{NEWrhoteta2} in Section \ref{sec_mmip}). We also note that the relaxation is very slow close to the critical point. In fact, close to the critical point, the mean field approximation is not valid anymore due to the enhancement of fluctuations (see Section \ref{sec_semip}).

\section{The linear response theory applied to the mean field Smoluchowski equation}
\label{sec_linearresponse}

In this section, we study the linear response of the BMF model to an
external perturbation. Other applications of the linear response
theory to systems with long-range interactions have been developed in
Refs. \cite{lr1,lr2,response}.

\subsection{The mean field Smoluchowski equation with an external potential}
\label{sec_de}

In the presence of an external potential $\Psi(\theta,t)$, the mean field Smoluchowski equation becomes
\begin{equation}
\label{de1}
\xi\frac{\partial\rho}{\partial t}=\frac{\partial}{\partial\theta}\left (T\frac{\partial\rho}{\partial\theta}+
\rho\frac{\partial\Phi}{\partial\theta}+\rho\frac{\partial\Psi}{\partial\theta}\right ),
\end{equation}
with Eq. (\ref{smf10}). We assume that the system has a homogeneous distribution $\rho=1/(2\pi)$ and we examine its response to a small external potential $\Psi(\theta,t)\ll 1$. Since the perturbation is small, we can develop a linear response theory. The linearized Smoluchowski equation writes
\begin{equation}
\label{de2}
\xi\frac{\partial\delta \rho}{\partial t}=\frac{\partial}{\partial\theta}\left (T\frac{\partial\delta\rho}{\partial\theta}+
\rho\frac{\partial\delta\Phi}{\partial\theta}+\rho\frac{\partial\Psi}{\partial\theta}\right ),
\end{equation}
with Eq. (\ref{jh2}). Since the external potential is introduced at $t=0$ (say), it is convenient to use Laplace transforms in time and Fourier transforms in space. The Fourier-Laplace transform of the perturbed density $\delta\rho(\theta,t)$ is defined by
\begin{equation}
\label{de3}
\delta\tilde\rho_n(\omega)=\int_0^{2\pi}\frac{d\theta}{2\pi}\int_0^{+\infty}dt\, e^{-i(n\theta-\omega t)}\delta\rho(\theta,t).
\end{equation}
This expression for the Laplace transform is valid for ${\rm Im}(\omega)$ sufficiently large. For the remaining part of the complex $\omega$ plane, it is defined by an analytic continuation. The inverse
Fourier-Laplace transform is
\begin{equation}
\label{de4}
\delta\rho(\theta,t)=\sum_n\int_{\cal C}\frac{d\omega}{2\pi}\, e^{i(n\theta-\omega t)}\delta\tilde\rho_n(\omega),
\end{equation}
where the Laplace contour ${\cal C}$ in the complex $\omega$ plane must pass above all poles of the integrand. Similar definitions are introduced for the Fourier-Laplace transform of the perturbed potential.

Taking the Fourier-Laplace transform of Eqs. (\ref{jh2}) and (\ref{de2}), and using Eq. (\ref{tsce4b}), we obtain
\begin{equation}
\label{de5}
\delta\tilde{\Phi}_n(\omega)=\frac{n^2\hat{u}_n}{i\xi \omega-n^2 (T+\hat{u}_n)}\tilde\Psi_n(\omega),
\end{equation}
where we have assumed $\delta\Phi(\theta,t)=0$ at $t=0$. The response function is defined by
\begin{equation}
\label{de6}
\delta\tilde{\Phi}_n(\omega)=R_n(\omega)\tilde\Psi_n(\omega).
\end{equation}
Therefore, we get
\begin{equation}
\label{de7}
R_n(\omega)=\frac{n^2\hat{u}_n}{i\xi \omega-n^2(T+\hat{u}_n)}.
\end{equation}
The polarization function is defined by
\begin{equation}
\label{de8}
\delta\tilde{\Phi}_n(\omega)=P_n(\omega)\lbrack \tilde\Psi_n(\omega)+\delta\tilde{\Phi}_n(\omega)\rbrack,
\end{equation}
yielding
\begin{equation}
\label{de9}
P_n(\omega)=\frac{R_n(\omega)}{1+R_n(\omega)}=\frac{n^2\hat{u}_n}{i\xi \omega-T n^2}.
\end{equation}
Finally, the dielectric function is defined by
\begin{equation}
\label{de10}
\epsilon_n(\omega)=1-P_n(\omega)=1-\frac{n^2\hat{u}_n}{i\xi \omega-T n^2}.
\end{equation}
We note the relations
\begin{equation}
\label{de11}
R_n(\omega)=\frac{P_n(\omega)}{1-P_n(\omega)}=\frac{1-\epsilon_n(\omega)}{\epsilon_n(\omega)}.
\end{equation}
The pure modes ($\Psi=0$) correspond to $1/R_n(\omega)=\epsilon_n(\omega)=0$. This returns the dispersion relation (\ref{jh6}). The physical meaning of the response, polarization, and dielectric functions is further discussed in \cite{bt,response}.

\subsection{The density response function}
\label{sec_dere}

Another important quantity is the density response function defined by
\begin{equation}
\label{de12}
\delta\tilde{\rho}_n(\omega)=\chi_n(\omega)\tilde\Psi_n(\omega).
\end{equation}
According to Eqs. (\ref{tsce4b}) and (\ref{de6}) it is related to the response function by $R_n(\omega)=2\pi \hat{u}_n\chi_n(\omega)$. From Eq. (\ref{de7}) we get
\begin{equation}
\label{de13}
\chi_n(\omega)=\frac{1}{2\pi}\frac{n^2}{i\xi \omega-n^2(T+\hat{u}_n)}.
\end{equation}

The previous relation gives the dynamical density response function. We now consider the static case. We start from the mean field Boltzmann distribution (\ref{cp4}), introduce an external field, and consider the weak field limit. The  static response function is defined by
\begin{equation}
\label{de12b}
\delta\tilde{\rho}_n=\chi_n\tilde\Psi_n.
\end{equation}
Since the Boltzmann distribution is the steady state of the mean field Smoluchowski equation, the previous study remains valid provided that we set $\omega=0$. Therefore $\chi_n=\chi_n(0)$ yielding
\begin{equation}
\label{de13b}
\chi_n=-\frac{1}{2\pi}\frac{1}{T+\hat{u}_n}.
\end{equation}
Similar results hold for $R_n$, $P_n$, and $\epsilon_n$. These results can also be obtained as in Appendix \ref{sec_gene}.

\subsection{The magnetization}
\label{sec_detr}

We assume that the external potential is of the form
\begin{equation}
\label{pr14mar1}
\Psi(\theta,t)=-h(t)\cos\theta,
\end{equation}
where $h$ may be interpreted as a magnetic field acting in the $x$-direction (see Section \ref{sec_magn}). Its Fourier-Laplace transform is
\begin{equation}
\label{pr14mar2}
\tilde\Psi_n(\omega)=-\frac{1}{2}\tilde{h}(\omega)(\delta_{n,1}+\delta_{n,-1}).
\end{equation}
On the other hand, from Eq. (\ref{smf10b}), the fluctuations of the potential can be expressed in terms of the fluctuations of the magnetization as
\begin{equation}
\label{pr14mar3}
\delta\Phi(\theta,t)=-M_x(t)\cos\theta- M_y(t)\sin\theta.
\end{equation}
Taking the Fourier-Laplace transform of this equation, we obtain
\begin{eqnarray}
\label{pr14mar4}
\delta\tilde{\Phi}_n(\omega)=-\frac{1}{2}(\tilde{M}_x+i\tilde{M}_y)(\omega)\delta_{n,-1}
\nonumber\\
-\frac{1}{2}(\tilde{M}_x-i\tilde{M}_y)(\omega)\delta_{n,1}.
\end{eqnarray}
From Eqs. (\ref{de5}) and (\ref{pr14mar2}), we see that $\delta\tilde\Phi_1(\omega)=\delta\tilde\Phi_{-1}(\omega)$. According to Eq. (\ref{pr14mar4}), this implies that $\tilde{M}_y(\omega)=0$. Therefore, Eq. (\ref{pr14mar4}) reduces to
\begin{equation}
\label{pr14mar5}
\delta\tilde{\Phi}_n(\omega)=-\frac{1}{2}\tilde{M}_x(\omega)(\delta_{n,1}+\delta_{n,-1}).
\end{equation}
We can rewrite Eq. (\ref{de6}) as
\begin{equation}
\label{pr14mar6}
\tilde{M}_x(\omega)=R(\omega)\tilde{h}(\omega),
\end{equation}
with
\begin{equation}
\label{pr14mar7}
R(\omega)=-\frac{T_c}{i\xi \omega-(T-T_c)}.
\end{equation}
We also have
\begin{equation}
\label{pr14mar8}
\tilde{M}_x(\omega)=P(\omega)\lbrack \tilde{h}(\omega)+\tilde{M}_x(\omega)\rbrack,
\end{equation}
with
\begin{equation}
\label{pr14mar9}
P(\omega)=\frac{R(\omega)}{1+R(\omega)}=-\frac{T_c}{i\xi \omega-T}.
\end{equation}
The dielectric function is
\begin{equation}
\label{pr14mar11}
\epsilon(\omega)=1-P(\omega)=1+\frac{T_c}{i\xi \omega-T}.
\end{equation}
We note the relations
\begin{equation}
\label{pr14mar10}
R(\omega)=\frac{P(\omega)}{1-P(\omega)}=\frac{1-\epsilon(\omega)}{\epsilon(\omega)}.
\end{equation}

\subsection{The response to a pulse}
\label{sec_pe}

We consider the response of the system to a magnetic ``pulse'' localized at $t=0$. It can be represented by the Dirac distribution
\begin{equation}
\label{pr14mar12}
h(t)=\delta(t).
\end{equation}
The Laplace transform of the magnetic field is
\begin{equation}
\label{pr14mar13}
\tilde{h}(\omega)=1.
\end{equation}
According to Eq. (\ref{pr14mar6}), the perturbation caused by a pulse is equal to the response function
\begin{equation}
\label{pr14mar14}
\tilde{M}_x(\omega)=R(\omega).
\end{equation}
Taking the inverse Laplace transform of Eq. (\ref{pr14mar7}), we obtain
\begin{equation}
\label{pe4a}
{M}_x(t)=R(t)=-\int_{\cal C}\frac{d\omega}{2\pi}e^{-i\omega t}\frac{T_c}{i\xi \omega-(T-T_c)}.
\end{equation}
The integral can be easily calculated with the residue theorem leading to
\begin{equation}
\label{pe4b}
{M}_x(t)=R(t)=\frac{T_c}{\xi}e^{-(T-T_c)t/\xi}.
\end{equation}
The polarization function can be calculated similarly yielding
\begin{equation}
\label{pe4c}
P(t)=\frac{T_c}{\xi}e^{-Tt/\xi}.
\end{equation}
In the stable case $T>T_c$, the perturbation $M_x(t)$ is damped exponentially rapidly (it tends to $0$ for $t\rightarrow +\infty$) and in the unstable case $T<T_c$, the perturbation grows exponentially rapidly. This is of course consistent with the results of Section \ref{sec_smoljeans}. The linear response theory is another manner to study the dynamical stability of a system.

\subsection{The response to a step function}

We consider the response of the system to a constant magnetic field that is ``switched on'' suddenly at $t=0$. It can be represented by a step function
\begin{equation}
\label{pr14mar15}
h(t)=H(t)h,
\end{equation}
where $H(t)=0$ for $t<0$ and $H(t)=1$ for $t>0$ (Heaviside function). The Laplace transform of the magnetic field is
\begin{equation}
\label{pr14mar16}
\tilde{h}(\omega)=\frac{i}{\omega}h.
\end{equation}
According to Eq. (\ref{pr14mar6}), the perturbation caused by a step function is
\begin{equation}
\label{pr14mar17}
\tilde{M}_x(\omega)=R(\omega)\frac{i}{\omega}h.
\end{equation}
Taking the inverse Laplace transform of this equation and using Eq. (\ref{pr14mar7}), we obtain
\begin{equation}
\label{pe4d}
{M}_x(t)=-h\int_{\cal C}\frac{d\omega}{2\pi}e^{-i\omega t}\frac{T_c}{i\xi \omega-(T-T_c)}\frac{i}{\omega}.
\end{equation}
The integral  can be easily calculated with the residue theorem leading to
\begin{equation}
\label{pe4e}
{M}_x(t)=\frac{T_c}{T-T_c}\left \lbrack 1-e^{-(T-T_c)t/\xi}\right \rbrack h.
\end{equation}
In the stable case $T>T_c$,  the perturbation $M_x(t)$  tends to the asymptotic value
\begin{equation}
\label{pe4f}
({M}_x)_{\infty}=\frac{T_c}{T-T_c}h,
\end{equation}
which corresponds to the equilibrium magnetization of the BMF model under a weak magnetic field (see Section \ref{sec_magn}). It arises here as the pole of the integral (\ref{pe4d}) at $\omega=0$. We note that the magnetic susceptibility is
\begin{equation}
\label{pe4fb}
\chi_M=R=-\pi\chi=\frac{T_c}{T-T_c}.
\end{equation}
In the unstable case $T<T_c$, the perturbation grows
exponentially rapidly. Of course, the linear response theory ceases to
be valid when the perturbation has grown significatively, so the
expressions (\ref{pe4b}) and (\ref{pe4e}) are only valid for
sufficiently ``short'' times in the unstable case.

\subsection{The evolution of the magnetization in the presence of a magnetic field}

The previous results may be obtained in a more synthetic manner as
follows. If we decompose the density perturbation
$\delta\rho(\theta,t)$ in Fourier modes according to Eq. (\ref{app1})
and substitute this decomposition in the linearized Smoluchowski
equation (\ref{de2}), we obtain the modal equations
\begin{eqnarray}
\xi\frac{d\delta\hat{\rho}_n}{dt}+n^2(T+\hat{u}_n)\delta\hat{\rho}_n=-n^2\rho\hat{\Psi}_n(t).
\label{mon1}
\end{eqnarray}
The different modes evolve as
\begin{eqnarray}
\delta\hat{\rho}_n(t)=-\frac{n^2\rho}{\xi}\int_0^t \hat{\Psi}_n(s) e^{-\frac{n^2}{\xi}(T+\hat{u}_n)(t-s)}\, ds,
\label{mon2}
\end{eqnarray}
where we have assumed that $\delta\hat{\rho}_n(0)=0$. When the external potential is a step function $\hat{\Psi}_n(t)=H(t)\hat{\Psi}_n$, we get
\begin{eqnarray}
\delta\hat{\rho}_n(t)=-\frac{\rho\hat{\Psi}_n}{T+\hat{u}_n}\left\lbrack 1-e^{-n^2(T+\hat{u}_n)t/\xi}\right\rbrack,
\label{mon3}
\end{eqnarray}
which tends to
\begin{eqnarray}
\delta\hat{\rho}_n^{\infty}=-\frac{\rho\hat{\Psi}_n}{T+\hat{u}_n},
\label{mon4}
\end{eqnarray}
for $t\rightarrow +\infty$ (for the stable modes). We recover the expression (\ref{de13b}) of the static density response function. These relations generalize the preceding results.

In the case of the cosine potential (\ref{u}), and for an external potential of the form (\ref{pr14mar1}), the modal equations become
\begin{eqnarray}
\xi \frac{d\delta \hat{\rho}_n}{dt}+Tn^2\delta \hat{\rho}_n=0, \qquad (n\neq \pm 1)
\label{dyn1f}
\end{eqnarray}
\begin{eqnarray}
\xi \frac{d\delta \hat{\rho}_{\pm 1}}{dt}+(T-T_c)\delta \hat{\rho}_{\pm 1}=\frac{1}{4\pi} h(t).
\label{dyn1g}
\end{eqnarray}
We see that the modes $n\neq \pm 1$ are damped exponentially rapidly as $e^{-Tn^2t/\xi}$. Only the modes $n=\pm 1$ have a non trivial evolution. Recalling that the modes $\delta \hat{\rho}_{\pm 1}$ are related to the magnetization (see Appendix \ref{sec_app}), we can rewrite Eq. (\ref{dyn1g}) in the form
\begin{eqnarray}
\xi \frac{d{\bf M}}{dt}+(T-T_c){\bf M}=T_c {\bf h},
\label{dyn1i}
\end{eqnarray}
where ${\bf M}=M_x+i M_y$ and ${\bf h}=h$. From this equation, we can establish the results of Eqs. (\ref{pe4b}), (\ref{pe4e}), and (\ref{pe4f}).

\section{The evolution of the mean magnetization in the inhomogeneous phase}
\label{sec_mmip}

In the previous sections, we have considered the linear dynamical
stability of a steady state of the mean field cosine Smoluchowski
equation, and the linear response of the system to a weak external
potential.  We now turn to the nonlinear evolution of the mean field
cosine Smoluchowski equation.  This equation exhibits an interesting
process of self-organization. Indeed, for $T<T_c$, the spatially
homogeneous phase is unstable and the system evolves towards an
equilibrium state with a spatially inhomogeneous distribution
(clustered phase). This process of self-organization is illustrated on
Figure \ref{NEWrhoteta2} for $T=1/4$. In this section, we analytically
study the evolution of the magnetization close to the critical point.

\subsection{The modal decomposition of the mean field cosine Smoluchowski equation}

\begin{figure}[!h]
\begin{center}
\includegraphics[clip,scale=0.3]{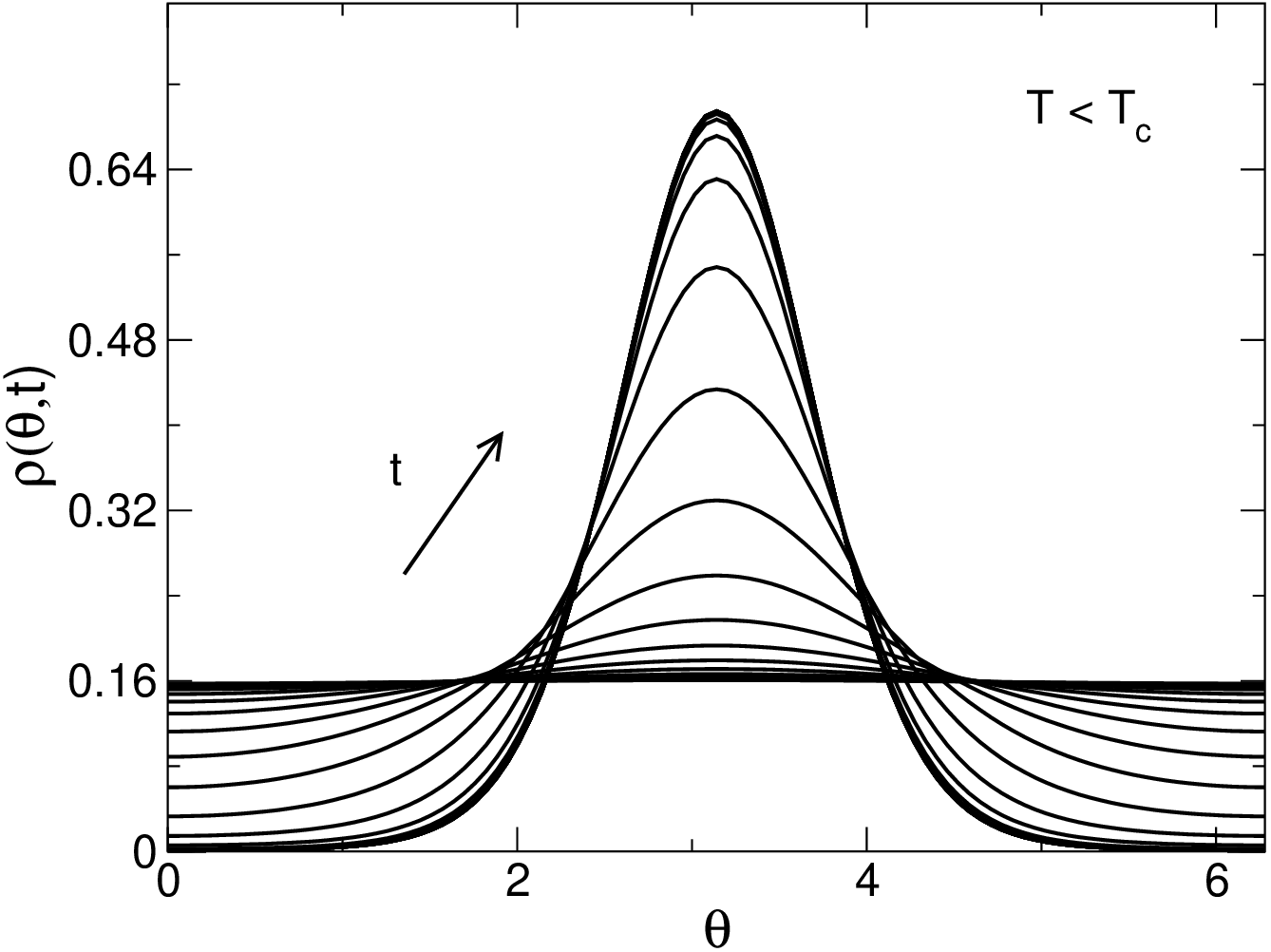}
\caption{Evolution of the density profile according to the cosine
Smoluchowski equation. For $T<T\sc$ (specifically $T=1/4$), the homogeneous
solution is unstable and the system  forms a cluster (from \cite{cvb}).}
\label{NEWrhoteta2}
\end{center}
\end{figure}

In the mean field approximation, the evolution of the density profile $\rho(\theta,t)$ is given by the cosine Smoluchowski equation (\ref{smf10g}). Decomposing the density profile in Fourier modes according to Eq. (\ref{app1}),  and using the identities of Appendix \ref{sec_app},  we obtain
a hierarchy of coupled ordinary differential equations
\begin{eqnarray}
\xi \frac{d\hat{\rho}_n}{dt}+Tn^2\hat{\rho}_n=-2\pi n \sum_m m \hat{\rho}_m \hat{u}_m \hat{\rho}_{n-m}.
\label{me1}
\end{eqnarray}
For the cosine potential, using Eq. (\ref{jh4}), the hierarchy of equations (\ref{me1}) takes the form
\begin{eqnarray}
\xi \frac{d\hat{\rho}_n}{dt}+Tn^2\hat{\rho}_n=\pi n (\hat{\rho}_1 \hat{\rho}_{n-1}-\hat{\rho}_{-1}\hat{\rho}_{n+1}).
\label{me2}
\end{eqnarray}
The modes $\hat{\rho}_{\pm 1}$ are directly related to the components
of the magnetization ${\bf M}$ (see Appendix \ref{sec_app}). This
infinite hierarchy of equations is equivalent to the mean field cosine
Smoluchowski equation (\ref{smf10g}). A good approximation of the
solution can be obtained by taking a sufficient number of modes ${\cal
N}\gg 1$ and closing the hierarchy by imposing the condition
$\hat{\rho}_{\pm {\cal N}}=0$.  The density profile can then be
reconstructed from Eq. (\ref{app1}). This is the numerical procedure
used in \cite{cvb} to obtain the result of Figure
\ref{NEWrhoteta2}.

At equilibrium, the hierarchy of equations (\ref{me2}) reduces to
\begin{eqnarray}
Tn^2\hat{\rho}_n=\pi n (\hat{\rho}_1 \hat{\rho}_{n-1}-\hat{\rho}_{-1}\hat{\rho}_{n+1}).
\label{me3}
\end{eqnarray}
Using $\hat{\rho}_n=I_n(\beta M)/[2\pi I_0(\beta M)]$ according to Eqs. (\ref{cp11}) and (\ref{app2}), we find that Eq. (\ref{me3}) is equivalent to the recursive relation $2nI_n(x)/x=I_{n-1}(x)-I_{n+1}(x)$ satisfied by the Bessel functions. On the other hand, considering a small perturbation $\delta\hat{\rho}_n\propto e^{\omega_i t}$ around the equilibrium state (assumed to be symmetric with respect to the $x$-axis), linearizing Eq. (\ref{me2}), and using the identities of Appendix \ref{sec_app}, we obtain the eigenvalue equations
\begin{eqnarray}
\xi\omega_i \delta M_x^{(n)}&=&-Tn^2\delta M_x^{(n)}+\frac{1}{2}n M_x [\delta M_x^{(n-1)}-\delta M_x^{(n+1)}]\nonumber\\
&+&\frac{1}{2}n \delta M_x [M_x^{(n-1)}-M_x^{(n+1)}],
\label{me4}
\end{eqnarray}
\begin{eqnarray}
\xi\omega_i \delta M_y^{(n)}&=&-Tn^2\delta M_y^{(n)}+\frac{1}{2}n M_x [\delta M_y^{(n-1)}-\delta M_y^{(n+1)}]\nonumber\\
&+&\frac{1}{2}n \delta M_y [M_x^{(n-1)}+M_x^{(n+1)}].
\label{me5}
\end{eqnarray}
These equations are equivalent to the eigenvalue problem written in the form of a differential equation in Section \ref{sec_smoljeansinho}.

\subsection{The closure of the hierarchy close to the critical point}

When $T<T_c$, the equilibrium distribution is spatially inhomogeneous ($M\neq 0$) but close to the critical point $T\rightarrow T_c^+$ the magnetization $M\rightarrow 0$. We assume that the initial magnetization is small so that $M(t)$ remains small during all the evolution. In that case, analytical results can be obtained. Indeed, according to Eq. (\ref{me2}),  the density modes scale as $\hat{\rho}_{n}\sim M^{n}$. Therefore, when $M\ll 1$, the modes of higher and higher order become less and less important. After a transient regime of duration $2\xi/n^{2}$,   the modes $n>0$ are given by
\begin{eqnarray}
\hat{\rho}_{n}\sim \frac{2\pi}{n}\hat{\rho}_{1}\hat{\rho}_{n-1}\sim {(2\pi)^{n-1}}\frac{1}{n!}\hat{\rho}_{1}^n,
\label{mod3}
\end{eqnarray}
and the modes $n<0$ by
\begin{eqnarray}
\hat{\rho}_{n}\sim -\frac{2\pi}{n}\hat{\rho}_{-1}\hat{\rho}_{n+1}\sim {(2\pi)^{|n|-1}}\frac{1}{|n|!}\hat{\rho}_{-1}^{|n|}.
\label{mod4}
\end{eqnarray}
In particular, for $n=2$, we get
\begin{eqnarray}
\hat{\rho}_{\pm 2}\sim {\pi}\hat{\rho}_{\pm 1}^2.
\label{mod5}
\end{eqnarray}
This shows that the second mode is slaved to the first (this corresponds to an adiabatic approximation). In that case, we obtain the closed equations
\begin{eqnarray}
\xi \frac{d\hat{\rho}_{\pm 1}}{dt}+(T-T_c)\hat{\rho}_{\pm 1}=-{\pi^2} \hat{\rho}_{\mp 1} \hat{\rho}_{\pm 1}^2,
\label{mod6}
\end{eqnarray}
Using the results of Appendix \ref{sec_app}, they can be rewritten in terms of the components $M_x$ and $M_y$ of the magnetization as
\begin{eqnarray}
\xi \frac{dM_x}{dt}+(T-T_c)M_x=-\frac{M^2}{4}M_x,
\label{mod8}
\end{eqnarray}
\begin{eqnarray}
\xi \frac{dM_y}{dt}+(T-T_c)M_y=-\frac{M^2}{4}M_y,
\label{mod9}
\end{eqnarray}
where $M=(M_x^2+M_y^2)^{1/2}$. Introducing the complex magnetization ${\bf M}=M_x+iM_y$, Eqs. (\ref{mod8}) and (\ref{mod9}) may be combined into a single equation
\begin{eqnarray}
\xi \frac{d{\bf M}}{dt}=-(T-T_c){\bf M}-\frac{M^2}{4}{\bf M}.
\label{mod10}
\end{eqnarray}
We note that
\begin{eqnarray}
\xi \frac{d{\bf M}}{dt}=-\frac{1}{2}\frac{\partial F}{\partial {\bf M}},
\label{mod11}
\end{eqnarray}
where $F(M)$ is the approximate expression (\ref{fbza}) of the free energy close to equilibrium for $T\rightarrow T_c^{-}$. For $T>T_c$, the evolution of the magnetization close to equilibrium is given by Eq. (\ref{mod10}) without the cubic term. In that case, the free energy is given by Eq. (\ref{fbzab}).

The steady states of Eq. (\ref{mod11}) correspond to extrema of
free energy ($F'(M)=0$). Since
\begin{eqnarray}
\dot F=-\frac{1}{2\xi}\left (\frac{\partial F}{\partial {\bf M}}\right )^2\le 0,
\label{mod11v}
\end{eqnarray}
the magnetization ${\bf M}(t)$ relaxes towards a minimum of 
free energy (maxima of free energy are unstable). Considering a small
perturbation about a steady state of Eq. (\ref{mod11}), we find that the
perturbation evolves as $\delta M\propto e^{\omega_i t}$ with
$\omega_i=-\frac{1}{2\xi}F''(M)$.

\subsection{The evolution of the mean magnetization close to the critical point}
\label{sec_meanclose}

\begin{figure}[!h]
\begin{center}
\includegraphics[clip,scale=0.3]{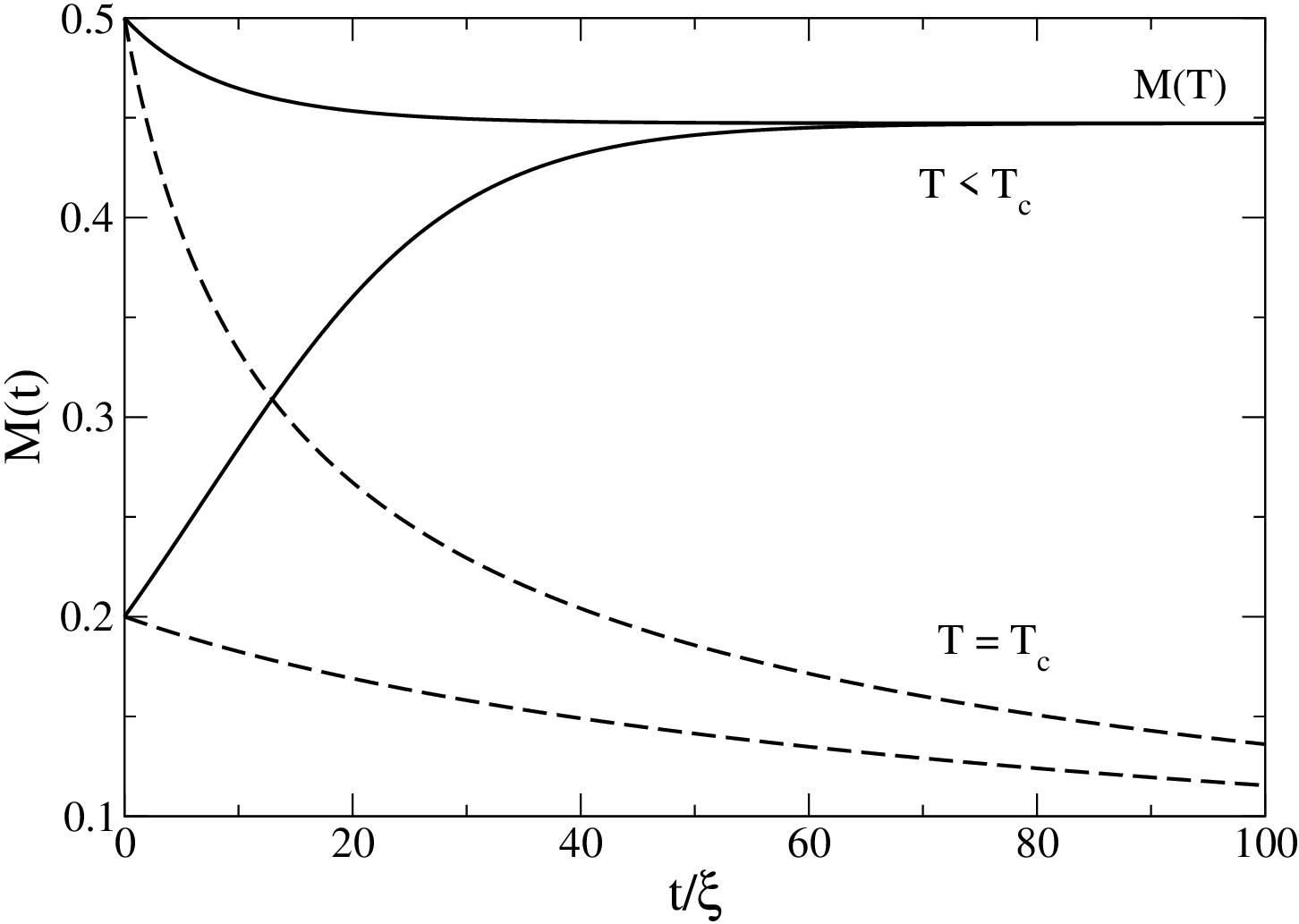}
\caption{Evolution of the mean magnetization close ($T<T_c$) and at ($T=T_c$) 
the critical point according to the mean field theory. These curves 
correspond to Eqs. (\ref{mod13}) and (\ref{mod20}) with $T=0.45$ and
$T=T_c=1/2$. We have taken $M_0=0.2$ and $M_0=0.5$.}
\label{magnfig}
\end{center}
\end{figure}

For sufficiently short times, and for sufficiently small initial magnetization, we can neglect the cubic term in Eq. (\ref{mod10}). This corresponds to the linear regime. The resulting equation
\begin{eqnarray}
\xi \frac{d{\bf M}}{dt}=-(T-T_c){\bf M},
\label{mod10b}
\end{eqnarray}
can be integrated into
\begin{eqnarray}
M(t)\simeq M_0 e^{(T_c-T)t/\xi},
\label{mod17}
\end{eqnarray}
returning the exponential rate (\ref{jh8}). For $T>T_c$, the perturbation 
is damped at a rate $\omega_i=(T_c-T)/\xi<0$ and the
magnetization tends to zero (homogeneous phase). In that case,
the solution (\ref{mod17}) is valid for all times. For $T<T_c$, the
perturbation grows at a rate $\omega_i=(T_c-T)/\xi>0$. This
growth is limited by nonlinear effects represented by the cubic term in Eq.
(\ref{mod10}) so that a magnetized state is finally reached (inhomogeneous
phase). A simple analytical solution describing this saturation may be obtained
\cite{cvb}. From Eq. (\ref{mod10}), we find that the evolution of the modulus of
the magnetization is governed by the equation
\begin{eqnarray}
\xi \frac{dM}{dt}+(T-T_c)M=-\frac{M^3}{4}.
\label{mod12}
\end{eqnarray}
This equation is readily solved (it may be convenient to use $M^{2}$ as a variable) with the initial condition $M(0)=M_0$. We obtain
\begin{eqnarray}
M(t)=\frac{M}{\sqrt{1+\left ( \frac{M^2}{M_0^2}-1\right )e^{-2(T_{c}-T)t/\xi}}},
\label{mod13}
\end{eqnarray}
where
\begin{eqnarray}
M=2\sqrt{T_{c}-T},
\label{mod14}
\end{eqnarray}
is the asymptotic value of the magnetization reached for $t\rightarrow +\infty$. This returns the equilibrium value of the magnetization  (\ref{tp15}) close to the critical point. The magnetization relaxes towards its equilibrium value as
\begin{eqnarray}
M(t)\simeq M\left\lbrack 1-\frac{1}{2}\left (\frac{M^2}{M_0^2}-1\right ) e^{-2(T_{c}-T)t/\xi}\right\rbrack,
\label{mod15}
\end{eqnarray}
returning the damping rate (\ref{smf19}).  On the other hand, for $t\rightarrow 0$, one has
\begin{eqnarray}
M(t)\simeq M_0 \left\lbrack 1+\frac{1}{4\xi}(M^2-M_0^2){t}\right\rbrack.
\label{mod16}
\end{eqnarray}
The magnetization increases if $M_0<M$ and decreases if $M_0>M$. When
$M_0\ll M$, we recover the result of Eq. (\ref{mod17}). The analytical solution (\ref{mod13})
describes the complete evolution of the magnetization close to the critical point $T_c$, from its initial growth to its convergence towards its equilibrium value (see Figure \ref{magnfig}).

Solving Eqs. (\ref{mod8}) and (\ref{mod9}) with Eq. (\ref{mod13}), we
get
\begin{eqnarray}
M_x(t)=\frac{\frac{M_x(0)}{M_0}M}{\sqrt{1+\left ( \frac{M^2}{M_0^2}-1\right )e^{-2(T_{c}-T)t/\xi}}},
\label{mod18}
\end{eqnarray}
\begin{eqnarray}
M_y(t)=\frac{\frac{M_y(0)}{M_0}M}{\sqrt{1+\left ( \frac{M^2}{M_0^2}-1\right )e^{-2(T_{c}-T)t/\xi}}}.
\label{mod19}
\end{eqnarray}
We see that if $M_x(0)=M_y(0)$ initially, then  $M_x(t)=M_y(t)$ for all times. Similarly, if  $M_y(0)=0$ initially then  $M_y(t)=0$ for all times. More generally, writing $M_x(t)=M(t)\cos\phi(t)$ and $M_y(t)=M(t)\sin\phi(t)$, we see that $\tan\phi(t)=M_y(0)/M_x(0)$ so that the phase is conserved. As a result, we can take $\phi=0$ without loss of generality. In that case, $M_y(t)=0$ and $M_x(t)=M(t)$. The density profile close to the critical point can be written as
\begin{eqnarray}
\rho(\theta,t)=\frac{1}{2\pi}+\frac{1}{\pi}M(t)\cos\theta,
\label{mod19b}
\end{eqnarray}
where $M(t)$ is given by Eq. (\ref{mod13}).

At $T=T_{c}$, the solution of Eq. (\ref{mod10}) is
\begin{eqnarray}
M(t)={M_0\over \sqrt{1+{M_0^2 t\over 2\xi}}}.
\label{mod20}
\end{eqnarray}
The magnetization tends to zero {\it algebraically} as $t^{-1/2}$ for $t\rightarrow +\infty$ (assuming that the mean field approximation is correct which is not the case close to $T_c$ as we shall see shortly). Solving Eqs. (\ref{mod8}) and (\ref{mod9}) with Eq. (\ref{mod20}), we get
\begin{eqnarray}
M_x(t)={M_x(0)\over \sqrt{1+{M_0^2 t\over 2\xi}}},\qquad M_y(t)={M_y(0)\over \sqrt{1+{M_0^2 t\over 2\xi}}}.
\label{mod21}
\end{eqnarray}

{\it Remark:} We have truncated the hierarchy of equations (\ref{me2}) by using the approximation (\ref{mod5}). This is valid close to the critical point $T_c$. Another approach would be to truncate the hierarchy of equations at the level of $\hat{\rho}_{\pm 2}$ by taking $\hat{\rho}_{\pm 3}=0$. In that case, we get the coupled equations
\begin{eqnarray}
\xi \frac{d\hat{\rho}_{\pm 1}}{dt}+(T-T_c)\hat{\rho}_{\pm 1}=-\pi \hat{\rho}_{\mp 1} \hat{\rho}_{\pm 2},
\label{mod23}
\end{eqnarray}
\begin{eqnarray}
\xi \frac{d\hat{\rho}_{\pm 2}}{dt}+4T\hat{\rho}_{\pm 2}=2\pi \hat{\rho}_{\pm 1}^2.
\label{mod24}
\end{eqnarray}
This system of equations is closed but it does not seem to admit a simple analytical solution. If we replace Eq. (\ref{mod24}) by its asymptotic expression $\hat{\rho}_{\pm 2}=\pi \hat{\rho}_{\pm 1}^2$ and substitute this relation  in Eq. (\ref{mod23}), we recover Eq. (\ref{mod6}). However, Eqs. (\ref{mod23})-(\ref{mod24}) are more general that Eq. (\ref{mod6}) because they do not rely on an adiabatic assumption.

\section{The temporal correlations of the magnetization in the homogeneous phase}
\label{sec_scse}

\subsection{The density fluctuation spectrum and the structure factor}
\label{sec_gscse}

In the previous sections, we have used a mean field approximation which amounts to neglecting fluctuations. For large values of $N$, this is a good approximation for systems with long-range interactions, except close to a critical point. In this section, we study the correlations of the magnetization in the homogeneous phase and show that they diverge as $T\rightarrow T_c^+$. To that purpose, we return to the stochastic Smoluchowski equation (\ref{dyn1ma}). Since the noise is weak when $N\gg 1$, we can consider small fluctuations $\delta\rho(\theta,t)$ and $\delta\Phi(\theta,t)$ about the homogeneous steady state $\rho=1/(2\pi)$ and $\Phi=1$.  The linearized equation for the fluctuations is
\begin{eqnarray}
\label{corr1}
\xi\frac{\partial\delta\rho}{\partial t}
=T\frac{\partial^2\delta\rho}{\partial\theta^2}+\rho\frac{\partial^2\delta\Phi}{\partial\theta^2}+ \sqrt{\frac{2\xi T\rho}{N}}\frac{\partial R}{\partial\theta}(\theta,t),
\end{eqnarray}
with Eq. (\ref{jh2}). We introduce the Fourier transform of the density fluctuations in space and time
\begin{equation}
\label{corr3}
\delta\hat\rho_n(\omega)=\int_0^{2\pi}\frac{d\theta}{2\pi}\int_{-\infty}^{+\infty}\frac{dt}{2\pi}\, e^{-i(n\theta-\omega t)}\delta\rho(\theta,t).
\end{equation}
The inverse Fourier transform is
\begin{equation}
\label{corr4}
\delta\rho(\theta,t)=\sum_{n=-\infty}^{+\infty}\int_{-\infty}^{+\infty}{d\omega}\, e^{i(n\theta-\omega t)}\delta\hat\rho_n(\omega).
\end{equation}
We use similar notations for the fluctuations of the potential $\delta\Phi(\theta,t)$ and for the noise $R(\theta,t)$.

Taking the Fourier transform of Eqs. (\ref{jh2}) and (\ref{corr1}), and using Eq. (\ref{tsce4b}), we find that the fluctuations of the density induced by the noise are given by
\begin{eqnarray}
\label{corr5}
\delta\hat{\rho}_{n}(\omega)=in\sqrt{\frac{2\xi T\rho}{N}} \frac{1}{Z_n(\omega)}\hat{R}_{n}(\omega),
\end{eqnarray}
where
\begin{eqnarray}
\label{ZZ}
Z_n(\omega)=Tn^2+\hat{u}_n n^2-i\xi\omega.
\end{eqnarray}
Without noise (${R}={0}$), we must have $Z_n(\omega)=0$ and we recover the dispersion relation (\ref{jh6}) corresponding to a pure mode.

For a Gaussian white noise, $\langle \hat{R}_n(\omega)\rangle=0$ and
\begin{eqnarray}
\label{corr7}
\langle \hat{R}_n(\omega)\hat{R}_{n'}(\omega')\rangle =\frac{1}{(2\pi)^2}\delta_{n,-n'}\delta(\omega+\omega').
\end{eqnarray}
Therefore, the correlations of the fluctuations are
\begin{eqnarray}
\label{corr8}
\langle \delta\hat{\rho}_{n}(\omega)\delta\hat{\rho}_{n'}(\omega')\rangle=\frac{1}{(2\pi)^2}\frac{2\xi T\rho}{N} \frac{n^2}{|Z_n(\omega)|^{2}}\delta_{n,-n'}\delta(\omega+\omega').\nonumber\\
\end{eqnarray}
The density fluctuation spectrum is defined by
\begin{eqnarray}
\label{corr8b}
\langle \delta\hat{\rho}_{n}(\omega)\delta\hat{\rho}_{n'}(\omega')\rangle=\frac{1}{2\pi}
S_n(\omega)\delta_{n,-n'}\delta(\omega+\omega').
\end{eqnarray}
According to  Eqs. (\ref{ZZ}) and (\ref{corr8}), we find that the density fluctuation spectrum is the Lorenzian
\begin{eqnarray}
\label{corr8bb}
S_n(\omega)=\frac{1}{2\pi^2}\frac{\xi T n^2}{\xi^2\omega^2+n^4(T+\hat{u}_n)^2}.
\end{eqnarray}
We note that it becomes more and more narrow as we approach the neutral mode defined by Eq. (\ref{jh6ddfd}).
The temporal correlation function of the Fourier components of the
density fluctuations is given by
\begin{eqnarray}
\label{corr9}
\langle \delta\hat{\rho}_{n}(t)\delta\hat{\rho}_{n'}(t')\rangle=\frac{n^2}{(2\pi)^2}\frac{2\xi T \rho}{N}\delta_{n,-n'}\int_{-\infty}^{+\infty} d\omega\,  \frac{e^{-i\omega (t-t')}}{|Z_n(\omega)|^{2}}.\nonumber\\
\end{eqnarray}
The integral over $\omega$ can be easily performed by using the Cauchy residue
theorem yielding
\begin{eqnarray}
\label{corr10}
\langle \delta\hat{\rho}_{n}(t)\delta\hat{\rho}_{n'}(t')\rangle=\frac{1}{N}\frac{T\rho^2}{T+ \hat{u}_n}\delta_{n,-n'} e^{-n^2(T+\hat{u}_n) |t-t'|/\xi}.\nonumber\\
\end{eqnarray}
The equal time  correlation function is
\begin{eqnarray}
\label{corr11}
\langle \delta\hat{\rho}_{n}\delta\hat{\rho}_{n'}\rangle=\frac{1}{N}\frac{T\rho^2 }{T+ \hat{u}_n}
\delta_{n,-n'}.
\end{eqnarray}
The structure factor is defined by
\begin{eqnarray}
\label{corr8c}
\langle \delta\hat{\rho}_{n}\delta\hat{\rho}_{n'}\rangle=\frac{1}{2\pi}
S_n\delta_{n,-n'}.
\end{eqnarray}
According to Eq. (\ref{corr11}), it is given by
\begin{eqnarray}
\label{corr11b}
S_n=\frac{1}{2\pi}\frac{T}{T+ \hat{u}_n}.
\end{eqnarray}
We note that the structure factor diverges at the neutral mode defined by Eq. (\ref{jh6ddfd}). This is manifested by a peak in the spectrum.

The results of this section can be obtained in different manners as shown in Appendices \ref{sec_gene} and \ref{sec_corr}.

\subsection{Application to the cosine potential}
\label{sec_apco}

The foregoing expressions can be simplified for the cosine potential (\ref{u}) using Eq. (\ref{jh4}). The density fluctuation spectrum (\ref{corr8bb}) can be written as
\begin{eqnarray}
\label{corr8s}
S_n(\omega)=\frac{1}{2\pi^2}\frac{\xi T n^2}{\xi^2\omega^2+n^4T^2}\qquad (n\neq \pm 1),
\end{eqnarray}
\begin{eqnarray}
\label{corr8ss}
S_{\pm 1}(\omega)=\frac{1}{2\pi^2}\frac{\xi T}{\xi^2\omega^2+(T-T_c)^2}.
\end{eqnarray}
The Lorenzian $S_{\pm 1}(\omega)$ becomes more and more narrow as we approach the critical temperature $T\rightarrow T_c$. The temporal correlation function of the density fluctuations for the
stable modes $n\neq \pm 1$ is
\begin{eqnarray}
\label{exa1}
\langle\delta\hat{\rho}_{n}(t)\delta\hat{\rho}_{n'}(t')\rangle=\frac{1}{N}\rho^2e^{-Tn^{2}|t-t'|/\xi}\delta_{n,-n'}.
\end{eqnarray}
For the unstable modes $n=\pm 1$, we get
\begin{eqnarray}
\label{exa2q}
\langle\delta\hat{\rho}_{\pm 1}(t)\delta\hat{\rho}_{n'}(t')\rangle=\frac{1}{N}\frac{T\rho^2}{T-T_{c}}e^{-(T-T_{c})|t-t'|/\xi}\delta_{n',\mp 1}.\nonumber\\
\end{eqnarray}
The equal time correlation function is
\begin{eqnarray}
\label{exa1b}
\langle\delta\hat{\rho}_{n}\delta\hat{\rho}_{n'}\rangle=\frac{1}{N}\rho^2\delta_{n,-n'}
\end{eqnarray}
for the stable modes $n\neq \pm 1$ and
\begin{eqnarray}
\label{exa2b}
\langle\delta\hat{\rho}_{\pm 1}\delta\hat{\rho}_{n'}\rangle=\frac{1}{N}\frac{T\rho^2}{T-T_{c}}\delta_{n',\mp 1}
\end{eqnarray}
for the unstable modes $n=\pm 1$. The structure factor is
\begin{eqnarray}
\label{exa2}
S_{n\neq \pm 1}=\frac{1}{2\pi},\qquad S_{\pm 1}=\frac{1}{2\pi}\frac{T}{T-T_c}.
\end{eqnarray}
It diverges at the critical point $T_c$. Using the relations of Appendix \ref{sec_app}, we can express these results in terms of the magnetization. We obtain $\langle M_x(t) M_y(t')\rangle=0$ and $\langle M_x(t) M_x(t')\rangle=\langle M_y(t) M_y(t')\rangle$ with
\begin{eqnarray}
\label{exa5}
\langle M_x(t) M_x(t')\rangle=\frac{T}{2N(T-T_{c})}e^{-(T-T_{c})|t-t'|/\xi}.
\end{eqnarray}
Taking $t'=t$, we find that the equal time correlation function of the magnetization is given by Eq. (\ref{magn6a}).

The physical content of Eq. (\ref{exa5}) is very instructive.
Considering the temporal factor in Eq. (\ref{exa5}), we see that the
correlations decay for $T>T_{c}$ with the rate (\ref{jh8}) given by
the mean field theory, i.e. by the deterministic mean field
Smoluchowski equation (\ref{smf10g}) without noise. However, as we
approach the critical temperature $T_{c}$, the amplitude of the
fluctuations diverges like $(T-T_{c})^{-1}$ so that the phase
transition should occur for $T$ {\it strictly} above
$T_{c}$. Indeed, the fluctuations become large before
ordinary stability theory predicts growth. We had previously reached
this conclusion from the YBG hierarchy
\cite{cvb,hb1}. These results imply that the mean field approximation
breaks down close to the critical point\footnote{Indeed, the limits
$N\rightarrow +\infty$ (mean field) and $T\rightarrow T_{c}$ do not
commute (see Appendix \ref{sec_corr}).} and that the instability
triggering the phase transition occurs {\it sooner} than what is
predicted by the mean field theory (i.e. by the stability analysis of
the mean field Smoluchowski equation). Similar results have been
reported for self-gravitating systems \cite{meta,monaghan,ko}.

\subsection{The fluctuation-dissipation theorem}
\label{sec_fdt}

There exist an important relation between the correlation function and the response of the system to an external perturbation. This is the so-called fluctuation-dissipation theorem \cite{kubo}. It can be derived at a very general level but it is interesting to obtain it explicitly in the present model.

Comparing Eqs. (\ref{de13}) and (\ref{corr8bb}), we find that  the density fluctuation spectrum is related to the density response function  by the relation
\begin{eqnarray}
\label{corr9b}
S_n(\omega)=-\frac{T}{\pi\omega}{\rm Im}\chi_n(\omega).
\end{eqnarray}
Similarly, comparing Eqs. (\ref{de13b}) and (\ref{corr11b}), we find that the structure factor is related to the susceptibility by
\begin{eqnarray}
\label{fdt}
S_n=-T\chi_n.
\end{eqnarray}
This identity can also be derived from Eq. (\ref{corr9b}) by using $S_n=\int S_n(\omega)\, d\omega$.

For the cosine potential, we can obtain the relation (\ref{corr9b}) with $n=\pm 1$ by comparing Eqs. (\ref{pr14mar7}) and (\ref{corr8ss}) and recalling that $R(\omega)=-\pi\chi(\omega)$.  Similarly, we can obtain the relation (\ref{fdt}) with $n=\pm 1$ by comparing Eq. (\ref{pe4fb}) and (\ref{exa2}-b). Finally, using $N\langle M^2\rangle=2\pi S_{\pm 1}$ and $\chi_M=R_{\pm 1}=-\pi\chi_{\pm 1}$, we check the equivalence between Eqs. (\ref{sun10}) and (\ref{fdt}).

\section{The fluctuations of the magnetization in the homogeneous phase}
\label{sec_fmhp}

\subsection{The modal decomposition of the linearized stochastic cosine Smoluchowski equation}

When $T>T_c$ (homogeneous phase) and $N\gg 1$, the fluctuations of the density $\delta\rho(\theta,t)$ about the equilibrium distribution $\rho=1/(2\pi)$ are small. Their evolution is described by the linearized stochastic cosine Smoluchowski equation (\ref{corr1}) with Eq. (\ref{jh2}). Decomposing the density fluctuations $\delta\rho(\theta,t)$ in Fourier modes according to Eq. (\ref{app1}), and using Eq. (\ref{tsce4b}), we find that the evolution of the different modes is given by
\begin{eqnarray}
\xi\frac{d\delta\hat{\rho}_n}{dt}+n^2 (T+\hat{u}_n)\delta\hat{\rho}_n=i n \sqrt{\frac{\xi T}{\pi N}}\, \hat{R}_n(t),
\label{fm10}
\end{eqnarray}
where $\hat{R}_n(t)$ is the Fourier transform of $R(\theta,t)$. This is a Gaussian white noise with zero mean $\langle \hat{R}_n(t)\rangle=0$ and  correlator
\begin{eqnarray}
\langle \hat{R}_n(t)\hat{R}_{n'}(t')\rangle=\frac{1}{2\pi} \delta_{n,-n'}\delta(t-t').
\label{fm8}
\end{eqnarray}
For the cosine potential (\ref{u}), using Eq. (\ref{jh4}), we get
\begin{eqnarray}
\xi \frac{d \delta \hat{\rho}_n}{dt}+n^2 T\delta \hat{\rho}_n=in\sqrt{\frac{\xi T}{\pi N}}\, \hat{R}_n(t),\quad (n\neq \pm 1)
\label{fm4}
\end{eqnarray}
\begin{eqnarray}
\xi \frac{d \delta \hat{\rho}_{\pm 1}}{dt}+(T-T_c) \delta \hat{\rho}_{\pm 1}=\pm i\sqrt{\frac{\xi T}{\pi N}}\, \hat{R}_{\pm 1}(t).
\label{fm5}
\end{eqnarray}

\subsection{The Langevin equation for the fluctuations of the magnetization}

From Eq. (\ref{fm5}), using the relations of Appendix \ref{sec_app}, we find that the evolution of the fluctuations of the magnetization is given by
\begin{eqnarray}
\xi \frac{d M_x}{dt}+(T-T_c) M_x=\sqrt{\frac{\xi T}{N}}R_x(t),
\label{fm10v}
\end{eqnarray}
\begin{eqnarray}
\xi \frac{d M_y}{dt}+(T-T_c) M_y=\sqrt{\frac{\xi T}{N}}R_y(t),
\label{fm11}
\end{eqnarray}
where $R_x(t)=i\sqrt{\pi} (\hat{R}_1-\hat{R}_{-1})$ and  $R_y(t)=-\sqrt{\pi} (\hat{R}_1+\hat{R}_{-1})$ are Gaussian white noises with $\langle {R}_i(t)\rangle={0}$ and $\langle R_i(t)R_{j}(t')\rangle=\delta_{ij}\delta(t-t')$. The magnetization vector ${\bf M}=M_x+iM_y$  satisfies a Langevin equation of the form
\begin{eqnarray}
\xi\frac{d {\bf M}}{dt}+(T-T_c) {\bf M}=\sqrt{\frac{\xi T}{N}}{\bf R}(t),
\label{fm15l}
\end{eqnarray}
where ${\bf R}=R_x+iR_y$. Equation (\ref{fm15l})  can be rewritten as
\begin{eqnarray}
\xi\frac{d {\bf M}}{dt}=-\frac{1}{2}\frac{\partial F}{\partial {\bf M}}+\sqrt{\frac{\xi T}{N}}{\bf R}(t),
\label{fm16}
\end{eqnarray}
where $F({\bf M})$ is the approximate expression (\ref{fbzab}) of the free energy in the homogeneous phase when $M\ll 1$.

\subsection{The Fokker-Planck equation for the fluctuations of the magnetization}
\label{sec_fpfluc}

The Langevin equation (\ref{fm15l}) for the fluctuations of the magnetization defines an Ornstein-Uhlenbeck process. The Fokker-Planck equation governing the evolution of the distribution $P({\bf M},t)$ of the magnetization is
\begin{eqnarray}
\xi\frac{\partial P}{\partial t}=\frac{\partial}{\partial {\bf M}}\cdot \left \lbrack\frac{T}{2N}\frac{\partial P}{\partial {\bf M}}+P(T-T_c){\bf M}\right\rbrack.
\label{fm17}
\end{eqnarray}
It can be written as
\begin{eqnarray}
\xi\frac{\partial P}{\partial t}=\frac{1}{2}\frac{\partial}{\partial {\bf M}}\cdot \left (\frac{T}{N}\frac{\partial P}{\partial {\bf M}}+P\frac{\partial F}{\partial {\bf M}}\right ),
\label{fm18}
\end{eqnarray}
where $F({\bf M})$ is given by Eq. (\ref{fbzab}). The equilibrium distribution of the magnetization is the Gaussian
\begin{eqnarray}
P({\bf M})=\frac{N(T-T_c)}{\pi T}e^{-N(T-T_c)M^2/T}.
\label{fm19}
\end{eqnarray}
It can be written as Eq. (\ref{magn1}) with Eq. (\ref{fbzab}). These results are valid in the Gaussian approximation where the fluctuations of the magnetization are very much peaked around the equilibrium value ${\bf M}={\bf 0}$. This corresponds to the weak noise limit valid when $N\gg 1$.

The probability of observing the fluctuation ${\bf M}$ at time $t$ provided that the system has the magnetization ${\bf M}'$ at time $t'$ is
\begin{eqnarray}
P({\bf M},t|{\bf M}',t')=\frac{N(T-T_c)}{\pi T \left\lbrack 1-e^{-2(T-T_c)(t-t')/\xi}\right\rbrack}\nonumber\\
\times e^{-\frac{N(T-T_c)\left\lbrack {\bf M}-e^{-(T-T_c)(t-t')/\xi}{\bf M}'\right\rbrack^2}{T \left\lbrack 1-e^{-2(T-T_c)(t-t')/\xi}\right\rbrack}}.
\label{fm21}
\end{eqnarray}
At equilibrium, the joint probability density
\begin{eqnarray}
P_2({\bf M},t|{\bf M}',t')=P({\bf M},t|{\bf M}',t')P({\bf M}'),
\label{fm22}
\end{eqnarray}
is
\begin{eqnarray}
P_2({\bf M},t|{\bf M}',t')=\frac{N^2(T-T_c)^2}{\pi^2 T^2 \lbrack 1-e^{-2(T-T_c)|t-t'|/\xi}\rbrack}\nonumber\\
\times e^{-\frac{N(T-T_c)\left (M^2+{M'}^2-2{\bf M}\cdot {\bf M}' e^{-(T-T_c)|t-t'|/\xi}\right )}{T \left\lbrack 1-e^{-2(T-T_c)|t-t'|/\xi}\right\rbrack}}.
\label{fm23}
\end{eqnarray}
The temporal correlation function of the $x$-component of the magnetization at equilibrium is
\begin{eqnarray}
\langle M_x(t) M_x(t')\rangle=\int P_2(M_x,t|M_x',t') M_x M_x' \, dM_x dM_x'.\nonumber\\
\label{fm24}
\end{eqnarray}
Similar expressions hold for $\langle M_y(t) M_y(t')\rangle$ and $\langle M_x(t) M_y(t')\rangle$.
Using Eq. (\ref{fm23}), we recover Eqs. (\ref{magn6a}) and (\ref{exa5}). More generally, the relaxation of the temporal correlation function may be calculated from Eq. (\ref{fm21}) or directly from the Langevin equation (\ref{fm15l}). In that case we find that
\begin{eqnarray}
\langle M_x(t) M_x(t')\rangle=M_x(0)^2 e^{-(T-T_c)(t+t')/\xi}\nonumber\\
+\frac{T}{2N(T-T_c)}\left\lbrack e^{-(T-T_c)|t-t'|/\xi}-e^{-(T-T_c)(t+t')/\xi}\right\rbrack.\nonumber\\
\label{fm26}
\end{eqnarray}
The temporal evolution of the variance of the magnetization is
\begin{eqnarray}
\langle M_x^2(t)\rangle=M_x(0)^2 e^{-2(T-T_c)t/\xi}\nonumber\\
+\frac{T}{2N(T-T_c)}\left\lbrack 1-e^{-2(T-T_c)t/\xi}\right\rbrack.
\label{fm27}
\end{eqnarray}
We have similar expressions for the correlations of $M_y(t)$. The crossed correlation functions are simply $\langle M_x(t) M_y(t')\rangle=M_x(0)M_y(0)e^{-(T-T_c)(t+t')/\xi}$ and $\langle M_x(t) M_y(t)\rangle=M_x(0)M_y(0)e^{-2(T-T_c)t/\xi}$. For large times, we recover Eqs. (\ref{magn6a}) and (\ref{exa5}).

\section{The stochastic evolution of the magnetization in the inhomogeneous phase}
\label{sec_semip}

\subsection{The modal decomposition of the stochastic cosine Smoluchowski equation}

The general evolution of the density, taking the fluctuations into account, is governed by the stochastic cosine Smoluchowski equation (\ref{dyn1ma}) with Eq. (\ref{smf10}). Substituting the Fourier decomposition (\ref{app1}) of the density in Eq. (\ref{dyn1ma}), and using the identities of Appendix \ref{sec_app}, we obtain the infinite hierarchy of equations
\begin{eqnarray}
\xi \frac{d\hat{\rho}_n}{dt}+Tn^2\hat{\rho}_n=-2\pi n \sum_m m \hat{\rho}_m \hat{u}_m \hat{\rho}_{n-m}+\hat{Q}_n(t),\nonumber\\
\label{se4}
\end{eqnarray}
where we have defined
\begin{eqnarray}
\hat{Q}_n(t)=\frac{1}{\sqrt{N}}\frac{in}{2\pi}\int_0^{2\pi} d\theta\, e^{-in\theta}\sqrt{2\xi T\rho(\theta,t)}R(\theta,t).\nonumber\\
\label{se5}
\end{eqnarray}
This is a noise with zero mean $\langle \hat{Q}_n(t)\rangle=0$ and correlator
\begin{eqnarray}
\langle \hat{Q}_n(t)\hat{Q}_{n'}(t')\rangle=-\frac{\xi T}{N\pi} n n' \hat{\rho}_{n+n'}\delta(t-t').
\label{se6}
\end{eqnarray}
For the cosine potential (\ref{u}), using Eq. (\ref{jh4}), the hierarchy of equations (\ref{se4}) takes the form
\begin{eqnarray}
\xi \frac{d\hat{\rho}_n}{dt}+Tn^2\hat{\rho}_n=\pi n (\hat{\rho}_1 \hat{\rho}_{n-1}-\hat{\rho}_{-1}\hat{\rho}_{n+1})+\hat{Q}_n(t).\nonumber\\
\label{se4b}
\end{eqnarray}
If we linearize Eqs. (\ref{se4})-(\ref{se4b}) about a homogeneous state, we recover Eqs. (\ref{fm10})-(\ref{fm5}).

\subsection{The stochastic evolution of the  magnetization close to the critical point}

Close to the critical point ($T\rightarrow T_c^{-}$), we can make the approximation (\ref{mod5}) and we can replace $\rho(\theta,t)$ by $\rho=1/(2\pi)$ in the expression (\ref{se5}) of the noise.
With these approximations, we obtain the closed equations
\begin{eqnarray}
\xi \frac{d\hat{\rho}_{\pm 1}}{dt}+(T-T_c)\hat{\rho}_{\pm 1}=-\pi^2 \hat{\rho}_{\mp 1} \hat{\rho}_{\pm 1}^2\pm i\sqrt{\frac{\xi T}{\pi N}} \hat{R}_{\pm 1}(t),\nonumber\\
\label{se10}
\end{eqnarray}
where $\hat{R}_n(t)$ has been defined in Section \ref{sec_fmhp}. Using the relations given in Appendix \ref{sec_app}, these equations can be rewritten in terms of the magnetization as
\begin{eqnarray}
\xi \frac{dM_x}{dt}+(T-T_c)M_x=-\frac{M^2}{4}M_x+\sqrt{\frac{\xi T}{N}}R_x(t),
\label{se12}
\end{eqnarray}
\begin{eqnarray}
\xi \frac{dM_y}{dt}+(T-T_c)M_y=-\frac{M^2}{4}M_y+\sqrt{\frac{\xi T}{N}}R_y(t).
\label{se13}
\end{eqnarray}

{\it Remark:} We could also close the hierarchy by assuming that $a_n=0$ for $|n|\ge 3$. In that case, we get
\begin{eqnarray}
\xi \frac{d\hat{\rho}_{\pm 1}}{dt}+(T-T_c)\hat{\rho}_{\pm 1}=-\pi \hat{\rho}_{\mp 1} \hat{\rho}_{\pm 2}+i\sqrt{\frac{\xi T}{\pi N}} \hat{R}_1(t),\nonumber\\,
\label{mod17a}
\end{eqnarray}
\begin{eqnarray}
\xi \frac{d\hat{\rho}_{\pm 2}}{dt}+4T\hat{\rho}_{\pm 2}=2\pi \hat{\rho}_{\pm 1}^2\pm 2i\sqrt{\frac{\xi T}{\pi N}} \hat{R}_{\pm 2}(t),
\label{mod18a}
\end{eqnarray}
where we have again replaced $\rho(\theta,t)$ by $\rho=1/(2\pi)$ in the expression (\ref{se5}) of the noise. These equations are more general than Eq. (\ref{se10}) but they are also more complicated.

\subsection{The Fokker-Planck equation for the magnetization}

Introducing the complex magnetization ${\bf M}=M_x+iM_y$, we can rewrite  the Langevin equations (\ref{se12}) and (\ref{se13}) as
\begin{eqnarray}
\xi\frac{d{\bf M}}{dt}+(T-T_c){\bf M}=-\frac{M^2}{4}{\bf M}+\sqrt{\frac{\xi T}{N}}{\bf R}(t),
\label{se19}
\end{eqnarray}
or, equivalently, as
\begin{eqnarray}
\xi \frac{d{\bf M}}{dt}=-\frac{1}{2}\frac{\partial F}{\partial {\bf M}}+\sqrt{\frac{\xi T}{N}}{\bf R}(t),
\label{se20}
\end{eqnarray}
where $F({\bf M})$ is the approximate expression (\ref{fbza}) of the free energy close to the critical point.

The Fokker-Planck equation governing the evolution of the distribution $P({\bf M},t)$ of the magnetization is
\begin{eqnarray}
\xi\frac{\partial P}{\partial t}=\frac{\partial}{\partial {\bf M}}\cdot \left\lbrack \frac{T}{2N}\frac{\partial P}{\partial {\bf M}}+P(T-T_c){\bf M}+P\frac{M^2}{4}{\bf M}\right\rbrack.\nonumber\\
\label{se21}
\end{eqnarray}
It can be written as
\begin{eqnarray}
\xi\frac{\partial P}{\partial t}=\frac{1}{2}\frac{\partial}{\partial {\bf M}}\cdot \left ( \frac{T}{N}\frac{\partial P}{\partial {\bf M}}+P\frac{\partial F}{\partial {\bf M}}\right ),
\label{se22}
\end{eqnarray}
where $F({\bf M})$ is given by Eq. (\ref{fbza}). The equilibrium distribution of the magnetization is
\begin{eqnarray}
P({\bf M})=A e^{-\frac{N}{T}\left\lbrack (T-T_c)M^2+\frac{M^4}{8}\right\rbrack}.
\label{se23}
\end{eqnarray}
It can be written as Eq. (\ref{magn1}) with Eq. (\ref{fbza}).

The magnetization vector may be written as ${\bf M}(t)=M(t)e^{i\phi(t)}$ where $M$ is its modulus and $\phi$ is its phase. The Fokker-Planck equation governing the evolution of the distribution $P(M,\phi,t)$ is
\begin{eqnarray}
\xi\frac{\partial P}{\partial t}&=&\frac{1}{M}\frac{\partial}{\partial {M}}\left\lbrace M \left\lbrack \frac{T}{2N}\frac{\partial P}{\partial {M}}+P(T-T_c){M}+P\frac{M^3}{4}\right\rbrack\right\rbrace\nonumber\\
&+&\frac{1}{M^2}\frac{T}{2N}\frac{\partial^2P}{\partial\phi^2}.
\label{se25}
\end{eqnarray}
This equation shows that the modulus of the magnetization relaxes towards its equilibrium value on a typical timescale
\begin{eqnarray}
t_R\sim \frac{\xi}{T_c-T},
\label{se26}
\end{eqnarray}
in agreement with the results of Section \ref{sec_mmip}.
On the other hand, the phase diffuses with a diffusion coefficient
\begin{eqnarray}
D_{\phi}\sim \frac{T}{N M^2}\sim \frac{T}{N(T_c-T)},
\label{se27}
\end{eqnarray}
where we have used Eq. (\ref{mod14}) to evaluate the equilibrium magnetization.
This defines a timescale
\begin{eqnarray}
t_{\phi}\sim \frac{2\pi}{D_{\phi}}\sim \frac{N}{T}(T_c-T),
\label{se28}
\end{eqnarray}
determining the spread of the phase. For given $T<T_c$ and $N\rightarrow +\infty$ we see that $t_R\sim 1$ and $t_{\phi}\sim N\rightarrow +\infty$. This shows that the modulus of the magnetization relaxes on a timescale of order $O(1)$ and that the direction of the magnetization (phase) changes slowly on a timescale $O(N)$. Actually, if we fix the interval of time $t$ (any) and let $N\rightarrow +\infty$, the direction of the magnetization does not change (see Section \ref{sec_mmip}).  On the other hand, for fixed $N$ and $T\rightarrow T_c$, we see that the scalings are reversed:  $t_R\rightarrow +\infty$ and $t_{\phi}\rightarrow 0$. Close to the critical point, the direction of the magnetization changes rapidly (it diffuses) and its magnitude takes a long time to relax.

In conclusion, the limits
$T\rightarrow T_c$ and $N\rightarrow +\infty$ do not commute. For fixed $T<T_c$
and $N\rightarrow +\infty$, the particles rapidly form a cluster ($t_R\sim 1$)
and the position of this cluster slowly diffuses ($t_{\phi}\sim N$). For fixed
$N$ and $T\rightarrow T_c$, the fluctuations are very important (the diffusion
dominates: $t_{\phi}\rightarrow 0$) and the formation of a cluster is hardly
visible ($t_R\rightarrow +\infty$).

{\it Remark:} If we impose $M_y=0$, the Langevin equation (\ref{se19}) reduces to
\begin{eqnarray}
\xi\frac{d{M}_x}{dt}+(T-T_c){M}_x=-\frac{M_x^3}{4}+\sqrt{\frac{\xi T}{N}}{R}_x(t).
\label{se29}
\end{eqnarray}
When $T<T_c$, the free energy $F(M_x)$ has two symmetric minima at $M_x=\pm
2(T_c-T)^{1/2}$ separated by a maximum at $M_x=0$. In that case, the
magnetization 
undergoes random changes between the two minima (metastable
states). These random changes can be analyzed with standard technics
\cite{risken}. In particular, the probability of transition from
a minimum to the other scales like $e^{-N\Delta F/T}$ where $\Delta
F=2(T_c-T)^2$ is the barrier of free energy (per particle) between the
minimum and the maximum computed from Eq. (\ref{fbza}). For
fixed $T<T_c$ and $N\rightarrow +\infty$ the system remains in one of the minima
for a very long time scaling like $e^N$. For fixed $N$ and $T\rightarrow T_c$,
the barrier of free energy is reduced and the random transitions between the two
minima (bistability) should be observed. This
type of random transitions has been recently studied for a model of
self-gravitating Brownian particles and chemotaxis \cite{cdnew}. Similar
results should be obtained for the BMF model and for other models with
long-range interactions presenting a phenomenon of bistability.

\section{Conclusion}

We have provided a detailed analysis of the BMF model in the overdamped
limit $\xi\rightarrow +\infty$ improving and extending the study of
\cite{cvb}. We have considered the mean field approximation generally
valid when $N\rightarrow +\infty$ and we have studied the process of
self-organization from an unstable homogeneous state to a stable
inhomogenous state when $T<T_c$. Interestingly, this process of
self-organization can be described analytically close to the critical
point where the magnetization is small. Indeed, in that limit, we can
approximate the free energy by its normal form close to a second order
phase transition. We have indicated that the mean field approximation
becomes incorrect as we approach the critical point $T\rightarrow T_c$
due to the enhancement of fluctuations. In particular, the limits
$N\rightarrow +\infty$ and $T\rightarrow T_c$ do not commute. We have
studied the stochastic evolution of the magnetization close to the
critical point and we have shown that the correlation functions
diverge at the critical point.

The BMF model may be viewed as the canonical counterpart of the HMF
model. The HMF model evolves at fixed energy $E$ while the BMF model
dissipates the energy and evolves instead at fixed temperature
$T$. There has been a lot of studies dedicated to the HMF model
\cite{cdr}.  The HMF model also displays a process of
self-organization\footnote{The statistical equilibrium states of the
HMF and BMF models are both described by the mean field
Maxwell-Boltzmann distribution with $E$ and $T$ as a control parameter
respectively. Furthermore, the microcanonical and canonical ensembles are
equivalent for the cosine interaction. However, the relaxation
towards these equilibrium states is very different in the HMF model
(fixed $E$) and in the BMF model (fixed $T$).} but this process is
difficult to describe analytically because we do not know the explicit
kinetic equation governing the relaxation of the system towards the
microcanonical equilibrium state. Indeed, for the HMF model, the
relaxation towards the Boltzmann distribution (with a temperature
$T(E)$ determined by the energy) is due to finite $N$ effects and we
must therefore take correlations into account. Unfortunately, the
Lenard-Balescu collision term which takes correlations into account at
the order $1/N$ vanishes for one dimensional systems and the kinetic
equation valid at the next order is not explicitly known. By contrast,
for the BMF model, the relaxation towards the canonical equilibrium
state is due to the coupling with the bath, not to finite $N$
effects. As a result, for $N\rightarrow +\infty$, the relaxation
towards the Boltzmann distribution (with the temperature $T$ of the
bath) is explicitly described by the mean field Kramers equation, or
by the mean field Smoluchowski equation in the strong friction
limit. This makes the study of the BMF model much easier than that of
the HMF model\footnote{This remark concerns only the relaxation
towards the canonical distribution. Of course, the intermediate
dynamics of the inertial BMF model is extremely rich and complex since
the properties of the HMF model should be recovered for
$\xi\rightarrow 0$. This intermediate dynamics has been studied in
\cite{bco}. Depending on the relative importance of $\xi$ and $N$, the
inertial BMF model may display Vlasov QSSs and microcanonical QSSs
before reaching the canonical equilibrium state. For $N\rightarrow
+\infty$, the system generally exhibits a dynamical phase transition
between a Vlasov QSS and the canonical distribution
\cite{cbo}. In the overdamped limit $\xi\rightarrow +\infty$, the QSSs
are destroyed and only the canonical equilibrium state
remains.}. Indeed, an almost complete description of the relaxation
process can be given for the BMF model and the influence of the
fluctuations can be taken into account by using the stochastic Kramers
or Smoluchowski equations.

\appendix

\section{Fourier decomposition of the density and of the magnetization}
\label{sec_app}

It is convenient to decompose the density in Fourier modes as
\begin{eqnarray}
\rho(\theta,t)=\sum_{n=-\infty}^{+\infty} \hat{\rho}_n(t) e^{in\theta},
\label{app1}
\end{eqnarray}
where
\begin{eqnarray}
\hat{\rho}_n(t)=\frac{1}{2\pi}\int_{0}^{2\pi} \rho(\theta,t) e^{-in\theta}\, d\theta.
\label{app2}
\end{eqnarray}
We note that $\hat{\rho}_n^*=\hat{\rho}_{-n}$.  For $n=0$, we have
\begin{eqnarray}
\hat{\rho}_0=\frac{1}{2\pi}.
\label{app3}
\end{eqnarray}
We define
\begin{eqnarray}
M_x^{(n)}(t)=\int\rho(\theta,t)\cos(n\theta)\, d\theta,
\label{app3b}
\end{eqnarray}
\begin{eqnarray}
M_y^{(n)}(t)=\int\rho(\theta,t)\sin(n\theta)\, d\theta.
\label{app3bb}
\end{eqnarray}
For $n=1$, we recover the mean magnetization: $M_x^{(1)}=M_x$ and $M_y^{(1)}=M_y$. For $n=0$, we have  $M_x^{(0)}=M_y^{(0)}=1$. According to Eq. (\ref{app2}), we have
\begin{eqnarray}
{\bf M}_n=M_x^{(n)}+iM_y^{(n)}=2\pi \hat{\rho}_{-n},
\label{app5}
\end{eqnarray}
\begin{eqnarray}
{\bf M}_n^*=M_x^{(n)}-iM_y^{(n)}=2\pi \hat{\rho}_{n}.
\label{app5b}
\end{eqnarray}
Inversely, we obtain
\begin{eqnarray}
M_x^{(n)}=\pi (\hat{\rho}_{n}+\hat{\rho}_{-n}),\quad M_y^{(n)}=i \pi (\hat{\rho}_{n}-\hat{\rho}_{-n}).
\label{app6}
\end{eqnarray}
In particular, the modes $n=\pm 1$ of the density are related to the components (\ref{magn}) of the mean magnetization. We note the identities
\begin{eqnarray}
\hat{\rho}_n^2+ \hat{\rho}_{-n}^2=\frac{1}{2\pi^2}({M_x^{(n)}}^2-{M_y^{(n)}}^2),
\label{app7}
\end{eqnarray}
\begin{eqnarray}
\hat{\rho}_n^2-\hat{\rho}_{-n}^2=-\frac{i}{\pi^2}M_x^{(n)} M_y^{(n)},
\label{app7b}
\end{eqnarray}
\begin{eqnarray}
\hat{\rho}_n \hat{\rho}_{-n}=\frac{M_n^2}{4\pi^2},
\label{app8}
\end{eqnarray}
where $M_n=\sqrt{{M_x^{(n)}}^2+{M_y^{(n)}}^2}$ (for $n=1$, this is the modulus of the magnetization). Using the foregoing relations, we can write the density as
\begin{eqnarray}
\rho(\theta,t)=\frac{1}{2\pi}+\frac{1}{\pi}\sum_{n=1}^{+\infty}\left\lbrack M_x^{(n)}(t)\cos(n\theta)+M_y^{(n)}(t)\sin(n\theta)\right\rbrack.\nonumber\\
\label{gene4b}
\end{eqnarray}

\section{Fluctuation-dissipation theorem for the magnetization}
\label{sec_fdtm}

The distribution of the magnetization in the presence of a magnetic field is \cite{hmfmagn}:
\begin{eqnarray}
P(M_x)=\frac{1}{Z(\beta)} e^{-\beta N [F(M_x)-h M_x]},
\label{fdtm1}
\end{eqnarray}
where $F(M_x)$ is the free energy in the absence of a magnetic field (see Eq. (\ref{aace4})) and $Z(\beta)=\int  e^{-\beta N [F(M_x)-h M_x]}\, dM_x$ is the partition function. Identifying $\ln Z$ as the generating function of the connexe correlation functions, a classical calculation shows that
\begin{eqnarray}
\langle M_x\rangle=\frac{1}{\beta N}\frac{\partial\ln Z}{\partial h},\quad \langle (\Delta M_x)^2\rangle=\frac{1}{\beta^2 N^2}\frac{\partial^2\ln Z}{\partial h^2}.\qquad
\label{fdtm2}
\end{eqnarray}
This leads to the fluctuation-dissipation theorem
\begin{eqnarray}
\langle (\Delta M_x)^2\rangle=\frac{1}{\beta N}\frac{\partial \langle M_x\rangle}{\partial h}=\frac{\chi_M}{\beta N}.
\label{fdtm3}
\end{eqnarray}
This relation is valid for arbitrary $h$ and $T$. If we consider a weak field $h\rightarrow 0$ and expand the free energy close to $M_x=0$ in the homogeneous phase ($T>T_c$), we get
\begin{eqnarray}
P(M_x)\propto e^{-\beta N [\frac{1}{2}F''(0)M_x^2-h M_x]}.
\label{fdtm4}
\end{eqnarray}
From this expression, we immediately obtain $M_x=h/F''(0)$ and $\langle (\Delta M_x)^2\rangle=1/[\beta N F''(0)]$ with $F''(0)=2(T-T_c)$. This returns Eq. (\ref{fdtm3}) but the previous derivation is more general.

\section{Distribution of the density fluctuations in the homogeneous phase}
\label{sec_gene}

The equilibrium distribution of the density
is given by eq. (\ref{h5}) where $F[\rho]$ is the free
energy defined by Eq. (\ref{tsce1zero}). The
distribution of the density fluctuations about an equilibrium state is
\begin{eqnarray}
P[\delta\rho]\propto e^{-\beta N \delta^2F[\delta\rho]}.
\label{gene9}
\end{eqnarray}
Using Eq. (\ref{tsce4}), it can be rewritten as
\begin{eqnarray}
\label{gene10}
P[\delta\rho]\propto e^{-\beta N \left\lbrace {1\over 2}\int \delta\rho\delta\Phi d\theta+{1\over 2}T\int {(\delta\rho)^{2}\over \rho}d\theta\right\rbrace}.
\end{eqnarray}
If the equilibrium state is spatially homogeneous, using Eq. (\ref{dyn1a}) we find that the distribution of the different modes of the density fluctuations is given by
\begin{eqnarray}
P[\delta\hat{\rho}_n]\propto e^{-\beta N \sum_{n=1}^{+\infty}4\pi^2 (T+\hat{u}_n)|\delta\hat{\rho}_n|^2}.
\label{gene8}
\end{eqnarray}
From this distribution, we can compute the correlations of the fluctuations and we obtain Eq. (\ref{corr11b}). If we introduce an external field $\Psi(\theta)$, we have to replace $F[\rho]$ by $F[\rho]+\int\rho\Psi\, d\theta$ in the foregoing equations. The distribution of the different modes of the density fluctuations becomes
\begin{eqnarray}
P[\delta\hat{\rho}_n]\propto e^{-\beta N \left (\sum_{n=1}^{+\infty}4\pi^2 (T+\hat{u}_n)|\delta\hat{\rho}_n|^2+2\pi\sum_{n=-\infty}^{+\infty}\delta\hat{\rho}_n\hat{\Psi}_{-n}\right )}.\nonumber\\
\label{gene8b}
\end{eqnarray}
From this distribution, we can compute the change of density due to the external field. This leads to Eq. (\ref{de13b}). From Eqs. (\ref{de13b}) and (\ref{corr11b}) we obtain the fluctuation-dissipation theorem (\ref{fdt}).

We now consider the dynamical evolution of the fluctuations about a homogeneous state. As shown in Section \ref{sec_fmhp}, the Fourier components of the density fluctuations satisfy the equations
\begin{eqnarray}
\xi \frac{d \delta\hat{\rho}_n}{dt}+n^2(T+\hat{u}_n) \delta\hat{\rho}_n=i n \sqrt{\frac{\xi T}{\pi N}}\, \hat{R}_n(t).
\label{gene1}
\end{eqnarray}
These equations are valid in the homogeneous phase for $N\gg 1$ so that the fluctuations with respect to the homogeneous distribution $\rho=1/(2\pi)$ are small. Using Eq. (\ref{gene4b}), writing the equations satisfied by $M_x^{(n)}$ and $M_y^{(n)}$, and introducing the vector ${\bf M}_n=M_x^{(n)}+i M_y^{(n)}$, we get
\begin{eqnarray}
\xi\frac{d {\bf M}_n}{dt}+{n^2}(T+\hat{u}_n) {\bf M}_n=\sqrt{\frac{\xi Tn^2}{N}}{\bf R}_n(t),
\label{gene5}
\end{eqnarray}
where  ${\bf R}_n=R_{x}^{(n)}+iR_{y}^{(n)}$ with
$R_{x}^{(n)}=i\sqrt{\pi}(\hat{R}_n-\hat{R}_{-n})$ and
$R_{y}^{(n)}=-\sqrt{\pi}(\hat{R}_n+\hat{R}_{-n})$. This is a Gaussian white
noise with zero mean $\langle {\bf R}_n(t)\rangle={\bf 0}$ and variance $\langle
{R}_{n,i}(t)R_{m,j}(t')\rangle=\delta_{nm}\delta_{ij}\delta(t-t')$. We note that
${\bf M}_n=2\pi \delta\hat{\rho}_{-n}$ and ${\bf
R}_n=-2i\sqrt{\pi}\hat{R}_{-n}$. For fixed $n$, Eq. (\ref{gene5}) defines an
Ornstein-Uhlenbeck process. The corresponding 
Fokker-Planck equation for each mode is
\begin{eqnarray}
\xi\frac{\partial P_n}{\partial t}=\frac{\partial}{\partial {\bf M}_n}\cdot \left \lbrack\frac{Tn^2}{2N}\frac{\partial P_n}{\partial {\bf M}_n}+P_n (T+\hat{u}_n)n^2 {\bf M}_n\right\rbrack.\quad
\label{gene6}
\end{eqnarray}
This equation can be solved analytically as in Section \ref{sec_fpfluc} and we recover by this method the temporal correlations functions of Section \ref{sec_scse}. The stationary solution of the Fokker-Planck equation (\ref{gene6}) is
\begin{eqnarray}
P_n=\frac{\beta N (T+\hat{u}_n)}{\pi} e^{-\beta N (T+\hat{u}_n)|{\bf M}_n|^2}.
\label{gene7}
\end{eqnarray}
The complete distribution of the density fluctuations is obtained by taking the product of $P_n$ for the different  modes $n$. Recalling that $|{\bf M}_n|^2=4\pi^2 |\delta\hat{\rho}_n|^2$, we find that the stationary distribution of the density fluctuations is given by Eq. (\ref{gene8}).

\section{The two-body correlation function and the invalidity of the mean field approximation close to the critical point}
\label{sec_corr}

The two-body distribution function may be written as $P_2(\theta,\theta')=P_1(\theta)P_1(\theta')\left\lbrack 1+\frac{1}{N}h(\theta,\theta')+\frac{1}{N}\right\rbrack$ where $h(\theta,\theta')$ is the two-body correlation function. In the homogeneous phase, this relation becomes $P_2(\theta,\theta')=\rho^2\left\lbrack 1+\frac{1}{N}h(|\theta-\theta'|)+\frac{1}{N}\right\rbrack$ where $\rho=1/(2\pi)$ is the equilibrium density. The correlations of the density fluctuations $\langle \delta{\rho}(\theta)\delta{\rho}(\theta')\rangle$ are related to the two-body correlation function $h(|\theta-\theta'|)$ by (see, e.g., Appendix A of \cite{hb5}):
\begin{eqnarray}
\label{acorr1}
\langle \delta{\rho}(\theta)\delta{\rho}(\theta')\rangle=\frac{1}{N}\left\lbrack \rho\delta(\theta-\theta')+\rho^2 h(|\theta-\theta'|)\right\rbrack.
\end{eqnarray}
In the absence of interaction, we recover the well-known result $\langle \delta{\rho}(\theta)\delta{\rho}(\theta')\rangle=\frac{\rho}{N}\delta(\theta-\theta')$. The Fourier transform of Eq. (\ref{acorr1}) is
\begin{eqnarray}
\label{acorr2}
\langle \delta\hat{\rho}_{n}\delta\hat{\rho}_{n'}\rangle=\frac{1}{N}\rho^2(1+\hat{h}_n) \delta_{n,-n'}.
\end{eqnarray}
According to Eq. (\ref{acorr2}), the structure factor is related to the Fourier transform of the correlation function by
\begin{eqnarray}
\label{acorr3fe}
{S}_{n}=\frac{1}{2\pi} (1+\hat{h}_n).
\end{eqnarray}
Using Eq. (\ref{corr11b}), we find that the Fourier transform of the correlation function is
\begin{eqnarray}
\label{acorr3}
\hat{h}_{n}=-\frac{\hat{u}_n}{T+\hat{u}_n}.
\end{eqnarray}
If we neglect collective effects, we simply have ${h}=-{u}/T$. More generally, we can define an effective potential $u_{DH}$ whose Fourier transform is $\hat{u}_{DH}=\hat{u}_n/(1+\hat{u}_n/T)$. It can be viewed as a generalization of the Debye-H\"uckel potential in plasma physics. For the cosine potential (\ref{u}), whose Fourier transform is given by Eq. (\ref{jh4}),  we find that $\hat{h}_{n}=0$ for $n\neq\pm 1$ and
\begin{eqnarray}
\label{acorr4}
\hat{h}_{\pm 1}=\frac{1}{N}\frac{T_c}{T-T_c}.
\end{eqnarray}
The correlation function in physical space is therefore
\begin{eqnarray}
\label{acorr5}
h(\theta-\theta')=\frac{1}{N}\frac{1}{T-T_c}\cos(\theta-\theta').
\end{eqnarray}
This expression is valid at the order $1/N$. These results can also be obtained from the YBG hierarchy (see \cite{hb1} and Appendix A of \cite{hb5}). We note that the two-body correlation function diverges at the critical point $T_{c}$ where the homogeneous phase becomes unstable and the clustered phase appears. This implies that the mean-field approximation ceases to be valid close to the critical point.

Let us consider the relation between the energy and the temperature in the homogeneous phase. The exact expression of the energy, taking correlations into account, is
\begin{equation}
\label{acorr6}
E={T\over 2}+\frac{1}{2}\frac{N-1}{N}\int  P_2(\theta,\theta')u(\theta-\theta')\, d\theta d\theta'.
\end{equation}
Using the preceding results, we obtain
\begin{equation}
\label{acorr7}
E={T\over 2}+\frac{1}{2}-\frac{T_c}{2N(T-T_c)}.
\end{equation}
For fixed $T>T_c$ and $N\rightarrow +\infty$, we obtain the mean field result
\begin{equation}
\label{acorr8}
E={T\over 2}+\frac{1}{2}.
\end{equation}
However, finite $N$ effects modify the shape of the caloric curve in
the vicinity of the critical point. According to Eq. (\ref{acorr7}), the mean-field approximation is
valid when $N(T-T_c)\gg 1$. This condition requires that $N$ be larger and larger as $T$
approaches $T_c$.

Starting from Eq. (\ref{smf2}) and using the two-body correlation function (\ref{acorr5}), we can easily compute the variance of the magnetization. For example,
\begin{eqnarray}
\label{dim1}
\langle M_x^2\rangle&=&\frac{1}{N^2}\sum_{ij}\langle \cos\theta_i\cos\theta_j\rangle\nonumber\\
&=&\frac{1}{N^2}\sum_{i}\langle \cos^2\theta_i\rangle+\frac{1}{N^2}\sum_{i\neq j}\langle \cos\theta_i\cos\theta_j\rangle\nonumber\\
&=&\frac{1}{N}\int P_1\cos^2\theta\, d\theta\nonumber\\
&+&\frac{N(N-1)}{N^2}\int P_2(\theta,\theta')\cos\theta\cos\theta'\, d\theta d\theta'\nonumber\\
&=&\frac{1}{2N}\frac{1}{1-T_c/T}.
\end{eqnarray}
Similarly, we find that $\langle M_x M_y\rangle=0$ and $\langle M_x^2\rangle=\langle M_y^2\rangle=\langle M^2\rangle/2$ where $\langle M^2\rangle$ is given by Eq. (\ref{magn6a}).
This returns the results of Sections \ref{sec_efluc}, \ref{sec_scse}, and \ref{sec_fmhp}. The fluctuations of the magnetization  scale as $M\sim N^{-1/2}$ but they diverge as $(T-T_c)^{-1/2}$ at the critical temperature $T_c$. At high temperatures the variance of the fluctuation is simply given by $\langle M^2\rangle=1/N$. Indeed, the correlations are negligible and the distribution of the magnetization for $N\rightarrow +\infty$ can be directly obtained from the central limit theorem (CLT). This leads to the Gaussian distribution  (\ref{magn3}) with a variance  $\langle M^2\rangle=1/N$. Finally, using $F(\theta)=-\partial\Phi/\partial\theta=-M_x\sin\theta+M_y\cos\theta$, the spatial correlations of the force are given by
\begin{equation}
\label{corr4b}
\langle F(\theta)F(\theta')\rangle={1\over N}{1\over 2(1-T_c/T)}\cos(\theta-\theta').
\end{equation}

\section{Approximate analytical formulae for the magnetization}
\label{sec_formula}

In \cite{cvb} and in Sec. \ref{sec_meanclose} of this paper, we have established the following equation for the evolution of the magnetization close to the critical point $T_c$:
\begin{equation}
\label{formula1}
\xi\frac{dM}{dt}=(T_c-T)M-\frac{M^3}{4}.
\end{equation}
Its stable steady state returns the expression $M=2\sqrt{T_c-T}$ of the
magnetization close to the critical point (see Eq. (\ref{tp15})). The temporal
evolution of the magnetization is given by Eq. (\ref{mod13}).  These results are
valid for $T\rightarrow T_c^-$.

In a recent paper, Sonnenschein and Schimansky-Geier \cite{sonnen} have
proposed 
an approximate equation for the evolution of the magnetization. With our
notations it writes
\begin{equation}
\label{formula2}
\xi\frac{dM}{dt}=(T_c-T)M-\frac{M^5}{2}.
\end{equation}
This equation is based on a Gaussian approximation. It gives a good 
agreement with numerical simulations at sufficiently low temperatures but
becomes inaccurate close to the critical point. Its stable steady state is
$M=[2(T_c-T)]^{1/4}$ \cite{sonnen}. We can check that it reproduces the
asymptotic expansion (\ref{tp13}) of the magnetization for $T\rightarrow 0$. The
temporal evolution of the magnetization is given by \cite{sonnen}:
\begin{equation}
\label{formula2b}
M(t)=\frac{M}{\left\lbrack 1+\left ( \frac{M^4}{M_0^4}-1\right )e^{-4(T_{c}-T)t/\xi}\right\rbrack^{1/4}}.
\end{equation}

Interestingly, the two equations (\ref{formula1}) and (\ref{formula2}) appear to be complementary. We propose to unify them in a single equation
\begin{equation}
\label{formula3}
\xi\frac{dM}{dt}=(T_c-T)M-\left (\frac{T}{T_c}\right )^n \frac{M^3}{4}-\frac{M^5}{2},
\end{equation}
where $n$ is a fitting parameter. For $T\ll T_c$ we recover Eq. (\ref{formula2})
and for $T\rightarrow T_c$ we recover Eq. (\ref{formula1}) since $M\rightarrow
0$. The stable steady 
state of this equation has a simple analytial expression
\begin{equation}
\label{formula4}
M=\left\lbrack \sqrt{\frac{1}{16}\left (\frac{T}{T_c}\right )^{2n}+2T_c\left (1-\frac{T}{T_c}\right )}-\frac{1}{4}\left (\frac{T}{T_c}\right )^n\right\rbrack^{1/2}.
\end{equation}
We find that Eq. (\ref{formula4}) with $n=4$ gives an excellent agreement with
the exact value of the equilibrium magnetization for any temperature $T\le T_c$
(the analytical curve $M(T)$ is almost indistinguishable from the exact
numerical curve in Fig. \ref{beta}). On the other hand, considering a small
perturbation about a steady state of Eq. (\ref{formula3}), we find that the
perturbation evolves as $\delta M\propto e^{\omega_i t}$ with
\begin{equation}
\label{formula5}
\omega_i=\frac{1}{\xi}\left\lbrack T_c-T-\frac{3}{4}\left (\frac{T}{T_c}\right )^n M^2-\frac{5}{2}M^4 \right\rbrack.
\end{equation}
For $T\rightarrow T_c$ we recover the result $\omega_i=-2(T_c-T)/\xi$ [see Eq. (\ref{smf19})] and for $T\ll T_c$ we get $\omega_i=-4(T_c-T)/\xi$. This last expression does not give a very good agreement with the exact result shown in Fig. \ref{lambdaEulernew} for $T\ll T_c$ because Eq. (\ref{formula2}) remains an  approximation (the relaxation time is overestimated in the approach of \cite{sonnen}). Finally, Eq. (\ref{formula3}) can be solved analytically to give $t(M)$. We find that
\begin{equation}
\label{formula6}
\frac{M(t)^{2a}(M(t)^2-M_e^2)^b(M(t)^2-M_*^2)^c}{M_0^{2a}(M_0^2-M_e^2)^b(M_0^2-M_*^2)^c}=e^{-t/\xi},
\end{equation}
where $a=1/(M_e^2M_*^2)=-1/[2(T_c-T)]$, $b=1/[M_e^2(M_e^2-M_*^2)]$, and $c=1/[M_*^2(M_*^2-M_e^2)]$. We have introduced $M_e^2=M_+^2$ and $M_*^2=M_-^2$ where $M_{\pm}^2=\pm [\frac{1}{16}({T}/{T_c})^{2n}+2T_c (1-{T}/{T_c})]^{1/2}-\frac{1}{4}({T}/{T_c})^n$ are the roots of the r.h.s. of Eq. (\ref{formula3}) ($M_e$ is the equilibrium magnetization).

\end{document}